# Generalized Orbicular (m,n,o) T-Spherical Fuzzy Sets with Hamacher Aggregation Operators and Application to Multi-Criteria Group Decision Making

**Yasir Akhtar[1], Mehboob Ali[2], Miin-Shen Yang[1,*]**

[1]Department of Applied Mathematics, Chung Yuan Christian University, Chung-Li, Taoyuan 32023, Taiwan.
[2]Government College Gilgit, Gilgit-Baltistan, Pakistan.
[*]E-mail: msyang@cycu.edu.tw

## Abstract

This paper introduces a novel approach to enhance uncertainty representation, offering decision-makers a more comprehensive perspective for improved decision-making outcomes. We propose Generalized Orbicular (m,n,o) T-Spherical Fuzzy Set ($GO_{mno}$-TSFS), a flexible extension of existing fuzzy set models including Globular T-spherical fuzzy sets (G-TSFSs), T-spherical fuzzy sets (T-SFSs), (p,q,r) Spherical fuzzy sets, and (p,q) Quasirung orthopair fuzzy sets (QOFSs). The framework employs three adjustable parameters "m", "n", and "o" to finely tune the influence of membership degrees, allowing for adaptable weighting of various degrees of membership. By utilizing spheres to represent membership, indeterminacy, and non-membership levels, the model enhances accuracy in depicting vague, ambiguous, and imprecise data. Building upon the foundation of $GO_{mno}$-TSFSs, we introduce essential set operations and algebraic operations for $GO_{mno}$-TSF Values ($GO_{mno}$-TSFVs). Moreover, we also develop score functions, accuracy functions, and basic distance measures such as Hamming and Euclidean distances to further enhance the analytical capabilities of the framework. Additionally, we propose $GO_{mno}$-TSF Hamacher Weighted Averaging ($GO_{mno}$-TSFHWA) and $GO_{mno}$-TSFH Weighted Geometric ($GO_{mno}$-TSFHWG), aggregation operators tailored for our proposed sets. To demonstrate the practical applicability of our approach, we apply our proposed aggregation operators namely $GO_{mno}$-TSFHWA and $GO_{mno}$-TSFHWG to solve a Multi-Criteria Group Decision Making (MCGDM) problem, specifically for selecting the most suitable e-commerce online shopping platform from the top-rated options. Sensitivity analysis is also conducted to validate the reliability and efficacy of our results, affirming the utility and robustness of the proposed methodology in real-world decision-making scenarios.

## 1. Introduction

Fuzzy sets (FSs) [32] represent a concept within fuzzy logic, a logical framework designed to address uncertainty and imprecision. Unlike traditional set theory, which mandates items to either belong or not belong to a set, FSs allow for partial membership. Within an FS, every element receives a degree of membership (DoM), denoting its proximity to other elements. These DoMs range from 0, indicating non membership, to 1, indicating full membership, with intermediate values representing partial inclusion. FSs have various measures with applications in different fields [7, 25, 33]. Intuitionistic fuzzy sets (IFSs) [4] expand upon the concept of FS by incorporating not only DoM, but also a degree of non-membership (DoN). DoN signifies the absence of belonging of an element. IFSs have found application in various fields, including [15, 18, 29]. Although Atanassov's development of IFSs is esteemed, decision-makers, aggregation operators frequently encounter limitations in assigning values because of constraints on DoM and DoN. Afterward, Smarandache [26] proposed neutrosophic sets (NSs) to generalize IFSs so that NSs can characterize DoM, DoN, and also an indeterminacy membership, and Smarandache [27] then extended them to neutrosophic hypersoft sets (NHSSs)) in which Jafar et al. [16] and Saqlain et al. [24] gave the distance and similarity measures for NHSSs with application to renewable energy source selection and on the construction of NHSS-TOPSIS, and Rahman et al. [22] considered the fuzzy parameterized NHSS with application to solid waste management.

Furthermore, Cuong [8] introduced a novel concept termed a picture fuzzy set (PFS). PFS comprises three representative functions denoted by $\left(\ddot{\mathfrak{A}}\right)$ for DoM, $\left(\ddot{\Xi}\right)$ for degree of indeterminacy (DoI) and $\left(\ddot{\mathfrak{M}}\right)$ for DoN, with the stipulation that their collective sum falls within the range of [0, 1], expressed as $\text{sum}\left(\ddot{\mathfrak{A}},\ddot{\Xi},\ddot{\mathfrak{M}}\right)\in\left[0,1\right]$. The term $1-\text{sum}\left(\ddot{\mathfrak{A}},\ddot{\Xi},\ddot{\mathfrak{M}}\right)$ denotes the rejection level of an element within PFS. Since its inception, extensive research has been conducted on PFS, particularly focusing on its applications across various decision-making models in references [11,12]. Additional details and pertinent researches can be found in references [6,31].

Picture fuzzy sets adhere to the principles of both fuzzy sets and intuitionistic fuzzy sets, yet they encounter a notable constraint: the domain of picture fuzzy sets is constrained, thereby complicating the assignment of degrees (DoM, DoI, DoN) to characteristic functions autonomously.

Acknowledging the limitation in PFS, Mahmood et al. [20] introduced an innovative model termed the spherical fuzzy set (SFS), which functions as an extension of picture fuzzy sets. This extension broadens the domain of picture fuzzy sets, facilitating more adaptable degree assignments. The characteristic functions in SFSs are represented by $\ddot{\mathfrak{A}}, \ddot{\Xi},$ and $\ddot{\mathfrak{M}},$ under a specific constraint: while the sum of $\ddot{\mathfrak{A}}, \ddot{\Xi},$ and $\ddot{\mathfrak{M}},$ may exceed the unit interval, but their squares must remain within it, as indicated by $\operatorname{sum}\left(\ddot{\mathfrak{A}}^2, \ddot{\Xi}^2, \ddot{\mathfrak{M}}^2\right) \in [0,1].$ Furthermore, research focus on TOPSIS method, Dombi aggregation operators and multi-attribute decision making (DM) problems under SFSs are defined in [2,3]. The inclusion of this constraint broadened the scope of SFS beyond that of Picture fuzzy sets. However, when employing specific values such as $\ddot{\mathfrak{A}} = 0.50,$ $\ddot{\Xi} = 0.80,$ and $\ddot{\mathfrak{M}} = 0.75,$ even squaring alone fails to meet the criterion, as the total of these values (1.452) significantly surpasses the unit interval. To tackle such scenarios, Ullah et al. [28] introduced a further extension of SFSs termed as T-spherical fuzzy set (TSFS), which relaxes the constraints of SFS. In TSFSs, the triplet $\left(\ddot{\mathfrak{A}}, \ddot{\Xi}, \ddot{\mathfrak{M}}\right)$ is governed by the condition that the sum $\left(\ddot{\mathfrak{A}}^t, \ddot{\Xi}^t, \ddot{\mathfrak{M}}^t\right) \in [0,1],$ where $t \in \mathbb{Z}.$ By utilizing this new constraint, for instance, with the triplet (0.50, 0.80, 0.75) and $t = 4,$ the sum of the values yields 0.7885, fulfilling the requirement within the unit interval. This example underscores that TSFS serves as a generalization of IFS, PFS, and SFS. Some of the important applications of TSFSs in various domains can be found in [1, 13, 19]. According to Rahim et al. [21], traditional TSFSs require equal membership levels. To better reflect real situations and give decision-makers flexibility with membership grades, ($p, q$)-Spherical FSs are introduced. These sets satisfy the sum ($\ddot{\mathfrak{A}}^p(\ddot{k}), \ddot{\Xi}^r(\ddot{k}), \ddot{\mathfrak{M}}^q(\ddot{k})$) constraint, where $r = \operatorname{LCM}(p, q).$

Yang et al. [30] recently introduced Globular T-spherical fuzzy sets (G-TSFSs), which encapsulate the DoM, DoI and DoN, in terms of a sphere with a specific center and radius, satisfying $0 \leq \overset{\cdots}{\mathfrak{A}}_G^t(\overset{\cdots}{k}) + \overset{\cdots}{\Xi}_G^t(\overset{\cdots}{k}) + \overset{\cdots}{\mathfrak{M}}_G^t(\overset{\cdots}{k}) \leq 1$. They also proposed cosine similarity measures based on the central values and radius of G-TSFSs, along with Hamming and Euclidean distance measures. However, the G-TSFSs framework has notable limitations, especially in scenarios where decision-makers need to assign power to membership grades independently and differently, and where the sum of the DoM, DoI, and DoN raised to the same power 't' exceeds 1 *i.e.* $0 \leq \overset{\cdots}{\mathfrak{A}}_G^t(\overset{\cdots}{k}) + \overset{\cdots}{\Xi}_G^t(\overset{\cdots}{k}) + \overset{\cdots}{\mathfrak{M}}_G^t(\overset{\cdots}{k}) > 1$. For instance, if a decision-maker assigns DoM as 0.9, DoI as 0.8, and DoN as 0.7, the sum of these values exceeds 1 *i.e.* $0.85^4 + 0.8^4 + 0.7^4 = 1.1717$, violating the unity condition integral to G-TSFSs. Such situations highlight the inadequacy of the G-TSFS model and its associated principles in addressing complex decision-making problems.

This limitation in G-TSFSs prompts us to develop a new set-theoretic construct that overcomes these constraints. In response, we propose a novel framework called Generalized Orbicular (m, n, o) T-Spherical Fuzzy Sets (GO$_{mno}$-TSFSs). In this framework, the quadruplet $\left( \overset{\cdots}{\mathfrak{A}}, \overset{\cdots}{\Xi}, \overset{\cdots}{\mathfrak{M}}; \overset{\cdots}{\mathfrak{r}} \right)$ adheres to the condition that the sum of $\overset{\cdots}{\mathfrak{A}}^m, \overset{\cdots}{\Xi}^o$, and $\overset{\cdots}{\mathfrak{M}}^n$ satisfies unity condition, where $m, o$ and $n$ are integers such that $o = LCM(m,n)$. To illustrate, consider the previous example with values (0.85, 0.8, 0.7) and parameters $m = 4, n = 3, \ and \ o = 12$. The cumulative value yields 0.9337, fitting within the range [0, 1]. This example demonstrates that the GO$_{mno}$-TSF framework is a comprehensive extension of CIFS, CSFSs, TSFSs, and G-TSFSs, addressing the limitations of the existing models. The main contribution of the proposed research is outlined in the following:

1. Introduction of GO$_{mno}$-TSFSs as a novel approach to enhance uncertainty representation in decision-making processes.
2. Utilization of adjustable parameters "m", "n", and "o" to finely tune the influence of membership degrees, allowing for adaptable weighting of various degrees of membership and enhancing the accuracy of data portrayal.
3. Introduction of essential set operations and algebraic operations for GO$_{mno}$-TSF Values (GO$_{mno}$-TSFVs), enhancing the analytical capabilities of the framework.
4. Development of score functions, accuracy functions, and basic distance measures such as Hamming and Euclidean distances, further enriching the analytical toolkit for decision-making processes.

5. Proposal of GO_{mno}-TSFHWA and GO_{mno}-TSFHWG, aggregation operators tailored for the proposed sets, providing practical aggregation methods.

6. Application of the proposed approach in solving a MCGDM problem, specifically in selecting the most suitable e-commerce online shopping platform from top-rated options, demonstrating its practical applicability.

7. Validation of the proposed methodology through sensitivity analysis and comparative analysis, affirming its reliability and efficacy in real-world decision-making scenarios.

The structure of this paper is organized as follows:

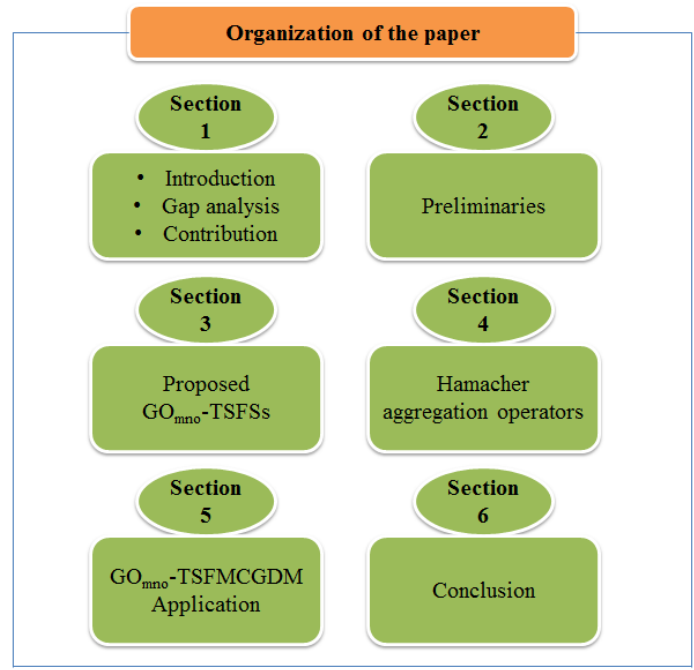

**Figure 1.** The structure of the proposed work.

## 2. Preliminaries

**Definition 1**. [30] A G-TSFSs $\ddot{R}_{\ddot{r}}$ in a universal set $\ddot{K}$ is defined by $\ddot{R}_{\ddot{r}} = \left\{ \langle \ddot{k}, \ddot{\mathfrak{A}}_{\ddot{R}}(\ddot{k}), \ddot{\Xi}_{\ddot{R}}(\ddot{k}), \ddot{\mathfrak{M}}\ddot{\mathfrak{r}}_{\ddot{R}}(\ddot{k}); \ddot{r} \rangle : \ddot{k} \in \ddot{K} \right\}$ where $\ddot{\mathfrak{A}}_{\ddot{R}}(\ddot{k}), \ddot{\Xi}_{\ddot{R}}(\ddot{k}), \ddot{\mathfrak{M}}\ddot{\mathfrak{r}}_{\ddot{R}}(\ddot{k}) \in [0,1]$ describe the DoM, DoI and DoN, respectively, with the condition $0 \leq \ddot{\mathfrak{A}}_{\ddot{R}}^{t}(\ddot{k}) + \ddot{\Xi}_{\ddot{R}}^{t}(\ddot{k}) + \ddot{\mathfrak{M}}\ddot{\mathfrak{r}}_{\ddot{R}}^{t}(\ddot{k}) \leq 1$ for $t$ being any positive integer, and $\ddot{r} \in [0,1]$ is the radius of the sphere centered on the point $\left\langle \ddot{\mathfrak{A}}_{\ddot{R}}(\ddot{\underline{k}}), \ddot{\underline{\Xi}}_{\ddot{R}}(\ddot{\underline{k}}), \ddot{\mathfrak{M}}_{\ddot{R}}(\ddot{\underline{k}}) \right\rangle$ on the globular space. The central point $\left\langle \ddot{\mathfrak{A}}_{\ddot{R}}(\ddot{\underline{k}}), \ddot{\underline{\Xi}}_{\ddot{R}}(\ddot{\underline{k}}), \ddot{\mathfrak{M}}_{\ddot{R}}(\ddot{\underline{k}}) \right\rangle$ is derived by using the values of DoM, DoI and DoN in TSFVs under consideration.

**Special Cases:** The special cases of G-TSFSs are illustrated in the following.

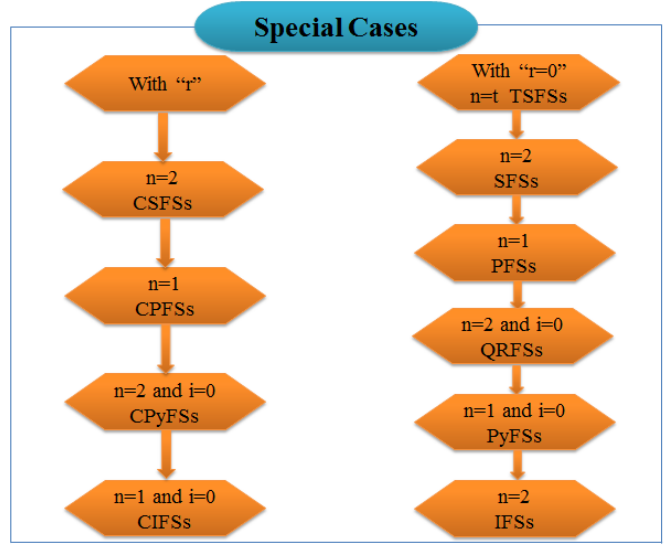

**Figure 2.** Specific cases of G-TSFSs are examined.

**Definition 2**. [21] A $p, q, r$ -SFSs $\ddot{R}$ on a universal set $K$ is defined by $\ddot{R} = \left\{ \langle \ddot{\underline{k}}, \ddot{\mathfrak{A}}_{\ddot{R}}(\ddot{\underline{k}}), \ddot{\underline{\Xi}}_{\ddot{R}}(\ddot{\underline{k}}), \ddot{\mathfrak{M}}_{\ddot{R}}(\ddot{\underline{k}}) \rangle : \ddot{\underline{k}} \in \ddot{K} \right\}$ where $\ddot{\mathfrak{A}}_R(\ddot{\underline{k}}), \ddot{\underline{\Xi}}_R(\ddot{\underline{k}}), \ddot{\mathfrak{M}}_R(\ddot{\underline{k}}) \in [0,1]$ describe the DoM, DoI and DoN, of an element $\ddot{\underline{k}} \in \ddot{K}$ such that $\left( \ddot{\mathfrak{A}}_{\ddot{R}}(\ddot{\underline{k}}) \right)^p + \left( \ddot{\underline{\Xi}}_{\ddot{R}}(\ddot{\underline{k}}) \right)^r + \left( \ddot{\mathfrak{M}}_{\ddot{R}}(\ddot{\underline{k}}) \right)^q$ (for all $p, q \geq 1$ and $r = LCM(p, q)$. A $p, q, r$ -spherical fuzzy number ( $p, q, r$ -SFN) can be expressed as a triplet $\left( \ddot{\mathfrak{A}}, \ddot{\underline{\Xi}}, \ddot{\mathfrak{M}} \right)$ such that $\ddot{\mathfrak{A}}, \ddot{\underline{\Xi}}, \ddot{\mathfrak{M}} \in [0,1]$ and $\ddot{\mathfrak{A}}^p + \ddot{\underline{\Xi}}^r + \ddot{\mathfrak{M}}^q \leq 1$ for all $p, q \geq 1$ and $r = LCM(p, q)$.

**Definition 3**. [17] A generalized IFSs $\ddot{R}$ on a universal set $\ddot{K}$ is defined by $\ddot{R} = \left\{ \langle \ddot{\underline{k}}, \ddot{\mathfrak{A}}_{\ddot{R}}(\ddot{\underline{k}}), \ddot{\underline{\Xi}}_{\ddot{R}}(\ddot{\underline{k}}), \ddot{\mathfrak{M}}_{\ddot{R}}(\ddot{\underline{k}}) \rangle : \ddot{\underline{k}} \in \ddot{K} \right\}$, where $\ddot{\mathfrak{A}}_{\ddot{R}}(\ddot{\underline{k}}), \ddot{\mathfrak{M}}_{\ddot{R}}(\ddot{\underline{k}}) \in [0,1]$ describe the DoM and DoN, respectively, with the condition $0 \leq \ddot{\mathfrak{A}}_{\ddot{R}}(\ddot{\underline{k}})^\wp + \ddot{\mathfrak{M}}_{\ddot{R}}(\ddot{\underline{k}})^\wp \leq 1$ for each $\ddot{\underline{k}} \in \ddot{K}$ and $\wp = n$ or $\frac{1}{n}$, $n = 1, 2, ..., N$.

### 3. Proposed Generalized Orbicular (m,n,o)-TSFSs

In this section, we introduce a cutting-edge concept known as Generalized Orbicular (m, n, o)-T-Spherical Fuzzy Set (GO$_{mno}$-TSFS). This concept represents a significant extension that encompasses various existing fuzzy set frameworks such as IFSs, CIFSs, PyFSs, CPyFSs, q-ROFSs, C-qROFS, PFSs, CPFSs, SFSs, CSFSs, TSFSs, and G-TSFSs. The GO$_{mno}$-TSFS allows decision makers to address uncertainty and impressions with a greater precision.

By enhancing the representation of uncertainty, this approach improves the accuracy and overall performance of the assessment mechanism. Thus, GO$_{mno}$-TSFSs provide decision makers with a more comprehensive viewpoint, enabling them to conduct more thorough evaluations during the decision-making process. Consequently, employing a more effective decision-making framework yields superior outcomes.

**Definition 4.** Let $\ddot{K}$ be a universal set. A Generalized Orbicular (m, n, o)-TSFS (GO$_{mno}$-TSFS), denoted by $\Theta_{\ddot{r}}$, defined over $\ddot{K}$ is mathematically articulated as:

$$\Theta_{\ddot{r}} = \left\{ \left\langle \ddot{k}, \ddot{\mathfrak{A}}_{\Theta}(\ddot{k}), \ddot{\Xi}_{\Theta}(\ddot{k}), \ddot{\mathfrak{M}}_{\Theta}(\ddot{k}); \ddot{r} \right\rangle :: \ddot{k} \in \ddot{K} \right\} \tag{1}$$

That satisfies $0 \leq \ddot{\mathfrak{A}}_{\Theta}^m(\ddot{k}) + \ddot{\Xi}_{\Theta}^o(\ddot{k}) + \ddot{\mathfrak{M}}_{\Theta}^n(\ddot{k}) \leq 1$ , $\forall m, n \geq 1$ and $o = LCM(m,n)$ with $m = t_1$ or $\dfrac{1}{t_1}$ , $o = t_2$ or $\dfrac{1}{t_2}$ , and $n = t_3$ or $\dfrac{1}{t_3}$ . Here $t_i (i = 1, 2, 3) \in \mathbb{Z}^+$ and $\ddot{\mathfrak{A}}_{\Theta}(\ddot{k}), \ddot{\Xi}_{\Theta}(\ddot{k}), \ddot{\mathfrak{M}}_{\Theta}(\ddot{k}) \in [0,1]$ signifies DoM, DoI, and DoN, respectively for each $\ddot{k} \in \ddot{K}$ . Whereas, $\ddot{r} \in [0,1]$ represents the radius of the sphere centered around the point $\left\langle \ddot{\mathfrak{A}}_{\Theta}(\ddot{k}), \ddot{\Xi}_{\Theta}(\ddot{k}), \ddot{\mathfrak{M}}_{\Theta}(\ddot{k}) \right\rangle$ on the orbicular space. The central point $\left\langle \ddot{\mathfrak{A}}_{\Theta}(\ddot{k}), \ddot{\Xi}_{\Theta}(\ddot{k}), \ddot{\mathfrak{M}}_{\Theta}(\ddot{k}) \right\rangle$ is achieved with the help of the values of DoM, DoI and DoN in (m, n, o)-SFVs under consideration.

In an (m, n, o)-SFS, every element is defined as a specific point within the spherical fuzzy interpretation framework. Conversely, in GO$_{mno}$-TSFS, each element is conceived as a sphere with a distinct center represented by $\left\langle \ddot{\mathfrak{A}}_\Theta(\ddot{k}), \ddot{\Xi}_\Theta(\ddot{k}), \ddot{\mathfrak{M}}_\Theta(\ddot{k}) \right\rangle$, along with an associated radius designated by $\ddot{r}$.

An (m, n, o)-SFS can be represented within the framework of GO$_{\text{mno}}$-TSFS as $\Theta_0 = \left\{ \left\langle \ddot{k}, \ddot{\mathfrak{A}}_\Theta(\ddot{k}), \ddot{\Xi}_\Theta(\ddot{k}), \ddot{\mathfrak{M}}_\Theta(\ddot{k}); 0 \right\rangle : \ddot{k} \in \ddot{K} \right\}$. Consequently, the concept of GO$_{\text{mno}}$-TSFSs extends the notion of (m, n, o)-SFSs as well as G-TSFSs. However, a GO$_{\text{mno}}$-TSFS with $\ddot{r} > 0$ necessitates a different approach than that employed for (m, n, o)-SFSs due to its unique characteristics. Hence, the rules governing GO$_{\text{mno}}$-TSFSs require reconstruction, with an expectation that the norms applied within GO$_{\text{mno}}$-TSFSs should exhibit greater inclusivity and effectiveness compared to those within (m, n, o)-SFSs.

**Definition 5.** Let $\hbar = \left\langle \ddot{\mathfrak{A}}_h, \ddot{\Xi}_h, \ddot{\mathfrak{M}}_h \right\rangle$ with $\ddot{\mathfrak{A}}_h, \ddot{\Xi}_h, \ddot{\mathfrak{M}}_h \in [0,1]$ be a (m, n, o)-SFV such that $\mathfrak{A}_h^m + \Xi_h^o + \mathfrak{M}_h^n \leq 1$ and let $r_h \in [0,1]$ be the radius of sphere centered on the point $\left\langle \ddot{\mathfrak{A}}_h, \ddot{\Xi}_h, \ddot{\mathfrak{M}}_h \right\rangle$. Then, the following quartet is called a GO$_{\text{mno}}$-TSF value (GO$_{\text{mno}}$-TSFV).

$$\hbar_{\ddot{r}} = \left\langle \ddot{\mathfrak{A}}_h, \ddot{\Xi}_h, \ddot{\mathfrak{M}}_h; \ddot{r} \right\rangle \tag{2}$$

**Remark:** As for the real numbers $\ddot{\mathfrak{A}}_h, \ddot{\Xi}_h, \ddot{\mathfrak{M}}_h \in [0,1]$ and $m = t_1$ or $\dfrac{1}{t_1}$, $o = t_2$ or $\dfrac{1}{t_2}$, and $n = t_3$ or $\dfrac{1}{t_3}$ with $t_i (i = 1, 2, 3) \in \mathbb{Z}^+$, we can establish the following conditions.

1. For $m, n, o \geq 1$, if $0 \leq \ddot{\mathfrak{A}}_h, \ddot{\Xi}_h, \ddot{\mathfrak{M}}_h \leq 1$ then $0 \leq \ddot{\mathfrak{A}}_h^m + \ddot{\Xi}_h^o + \ddot{\mathfrak{M}}_h^n \leq 1$. Thus, if $\Theta_{\ddot{r}} \in$ GO$_{\text{mno}}$-TSFS($\ddot{K}$) then $\Theta_{\ddot{r}} \in$ GO$_{\text{mno}}$-TSFS($mno, \ddot{K}$).

2. For $m, n, o < 1$, if $0 \leq \ddot{\mathfrak{A}}_h^m + \ddot{\Xi}_h^o + \ddot{\mathfrak{M}}_h^n \leq 1$ then $0 \leq \ddot{\mathfrak{A}}_h, \ddot{\Xi}_h, \ddot{\mathfrak{M}}_h \leq 1$. Thus, if $\Theta_{\ddot{r}} \in$ GO$_{\text{mno}}$-TSFS($mno, \ddot{K}$) then $\Theta_{\ddot{r}} \in$ GO$_{\text{mno}}$-TSFS($\ddot{K}$).

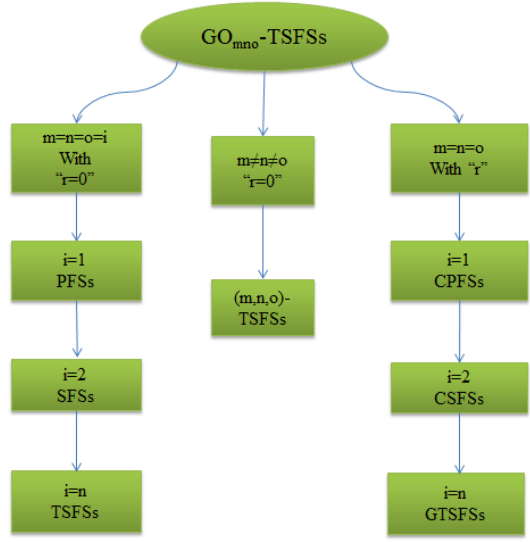

**Figure 3.** Special cases of GO$_{mno}$-TSFSs.

**Definition 6.** Let us have the two GO$_{mno}$-TSFSs $\underline{P}_{\breve{r}_1}$ and $\underline{Q}_{\breve{r}_2}$ in $K$ with

$$\underline{P}_{\breve{r}_1} = \left\{ \left\langle \breve{\breve{k}}, \breve{\mathfrak{A}}_P(\breve{\breve{k}}), \breve{\overline{\Xi}}_P(\breve{\breve{k}}), \breve{\mathfrak{M}}_P(\breve{\breve{k}}); \breve{r}_1 \right\rangle : \breve{\breve{k}} \in \breve{K} \right\} \text{ and } \underline{Q}_{\breve{r}_2} = \left\{ \left\langle \breve{\breve{k}}, \breve{\mathfrak{A}}_Q(\breve{\breve{k}}), \breve{\overline{\Xi}}_Q(\breve{\breve{k}}), \breve{\mathfrak{M}}_Q(\breve{\breve{k}}); \breve{r}_2 \right\rangle : \breve{\breve{k}} \in \breve{K} \right\} \text{ Then}$$

we define some basic set operations on GO$_{mno}$-TSFSs as:

(i) $\underline{P}_{\breve{r}_1} \subset \underline{Q}_{\breve{r}_2}$ iff $\breve{r}_1 \leq \breve{r}_2$ and $\breve{\mathfrak{A}}_P(\breve{\breve{k}}) \leq \breve{\mathfrak{A}}_Q(\breve{\breve{k}}), \breve{\overline{\Xi}}_P(\breve{\breve{k}}) \leq \breve{\overline{\Xi}}_Q(\breve{\breve{k}})$, and $\breve{\mathfrak{M}}_P(\breve{\breve{k}}) \geq \breve{\mathfrak{M}}_Q(\breve{\breve{k}})$

(ii) $\underline{P}_{\breve{r}_1} = \underline{Q}_{\breve{r}_2}$ iff $\breve{r}_1 = \breve{r}_2$ and $\breve{\mathfrak{A}}_P(\breve{\breve{k}}) = \breve{\mathfrak{A}}_Q(\breve{\breve{k}}), \breve{\overline{\Xi}}_P(\breve{\breve{k}}) = \breve{\overline{\Xi}}_Q(\breve{\breve{k}})$, and $\breve{\mathfrak{M}}_P(\breve{\breve{k}}) = \breve{\mathfrak{M}}_Q(\breve{\breve{k}})$

(iii) $\underline{P}_{\breve{r}_1}^{c} = \left\{ \left\langle \breve{\breve{k}}, \breve{\mathfrak{A}}_P(\breve{\breve{k}}), \breve{\overline{\Xi}}_P(\breve{\breve{k}}), \breve{\mathfrak{M}}_P(\breve{\breve{k}}); \breve{r}_1 \right\rangle : \breve{\breve{k}} \in \breve{K} \right\}$

(iv) $\underline{P}_{\breve{r}_1} \cup_{\min} \underline{Q}_{\breve{r}_2} = \left\{ \breve{\breve{k}}, \max(\breve{\mathfrak{A}}_P(\breve{\breve{k}}), \breve{\mathfrak{A}}_Q(\breve{\breve{k}})), \min(\breve{\overline{\Xi}}_P(\breve{\breve{k}}), \breve{\overline{\Xi}}_Q(\breve{\breve{k}})), \min(\breve{\mathfrak{M}}_P(\breve{\breve{k}}), \breve{\mathfrak{M}}_Q(\breve{\breve{k}})); \min(\breve{r}_1, \breve{r}_2) : \breve{\breve{k}} \in \breve{K} \right\}.$

(v) $\underline{P}_{\breve{r}_1} \cup_{\max} \underline{Q}_{\breve{r}_2} = \left\{ \breve{\breve{k}}, \max(\breve{\mathfrak{A}}_P(\breve{\breve{k}}), \breve{\mathfrak{A}}_Q(\breve{\breve{k}})), \min(\breve{\overline{\Xi}}_P(\breve{\breve{k}}), \breve{\overline{\Xi}}_Q(\breve{\breve{k}})), \min(\breve{\mathfrak{M}}_P(\breve{\breve{k}}), \breve{\mathfrak{M}}_Q(\breve{\breve{k}})); \max(\breve{r}_1, \breve{r}_2) : \breve{\breve{k}} \in \breve{K} \right\}.$

(vi) $\underline{P}_{\breve{r}_1} \cap_{\min} \underline{Q}_{\breve{r}_2} = \left\{ \breve{\breve{k}}, \min(\breve{\mathfrak{A}}_P(\breve{\breve{k}}), \breve{\mathfrak{A}}_Q(\breve{\breve{k}})), \min(\breve{\overline{\Xi}}_P(\breve{\breve{k}}), \breve{\overline{\Xi}}_Q(\breve{\breve{k}})), \max(\breve{\mathfrak{M}}_P(\breve{\breve{k}}), \breve{\mathfrak{M}}_Q(\breve{\breve{k}})); \min(\breve{r}_1, \breve{r}_2) : \breve{\breve{k}} \in \breve{K} \right\}.$

(vii) $\underline{P}_{\breve{r}_1} \cap_{\max} \underline{Q}_{\breve{r}_2} = \left\{ \breve{\breve{k}}, \min(\breve{\mathfrak{A}}_P(\breve{\breve{k}}), \breve{\mathfrak{A}}_Q(\breve{\breve{k}})), \min(\breve{\overline{\Xi}}_P(\breve{\breve{k}}), \breve{\overline{\Xi}}_Q(\breve{\breve{k}})), \max(\breve{\mathfrak{M}}_P(\breve{\breve{k}}), \breve{\mathfrak{M}}_Q(\breve{\breve{k}})); \max(\breve{r}_1, \breve{r}_2) : \breve{\breve{k}} \in \breve{K} \right\}.$

We give a numerical example to demonstrate the above fundamental operations in the following.

**Example 1:** Let the universal set be $\ddot{\underset{\sim}{K}} = \left\{ \ddot{\underset{\sim}{k}}_1, \ddot{\underset{\sim}{k}}_2, \ddot{\underset{\sim}{k}}_3 \right\}$, and the two GOmno-TSFSs on $\ddot{\underset{\sim}{K}}$ are:

$$\underset{\sim}{P}_{0.28} = \left\{ \langle \ddot{\underset{\sim}{k}}_1, 0.18, 0.34, 0.24; 0.28 \rangle, \langle \ddot{\underset{\sim}{k}}_2, 0.5, 0.38, 0.46; 0.28 \rangle, \langle \ddot{\underset{\sim}{k}}_3, 0.5, 0.23, 0.44; 0.28 \rangle, \langle \ddot{\underset{\sim}{k}}_4, 0.33, 0.35, 0.46; 0.28 \rangle \right\}$$

and

$$\underset{\sim}{Q}_{0.3} = \left\{ \langle \ddot{\underset{\sim}{k}}_1, 0.49, 0.28, 0.43; 0.3 \rangle, \langle \ddot{\underset{\sim}{k}}_2, 0.45, 0.41, 0.46; 0.3 \rangle, \langle \ddot{\underset{\sim}{k}}_3, 0.48, 0.41, 0.28; 0.3 \rangle, \langle \ddot{\underset{\sim}{k}}_4, 0.53, 0.18, 0.51; 0.3 \rangle \right\}.$$

Then, we have that $\underset{\sim}{P}_{0.28} \subset \underset{\sim}{Q}_{0.3}$. We also obtain

$$\underset{\sim}{P}_{0.28}^c = \left\{ \langle \ddot{\underset{\sim}{k}}_1, 0.24, 0.34, 0.18; 0.28 \rangle, \langle \ddot{\underset{\sim}{k}}_2, 0.46, 0.38, 0.5; 0.28 \rangle, \langle \ddot{\underset{\sim}{k}}_3, 0.44, 0.23, 0.5; 0.28 \rangle, \langle \ddot{\underset{\sim}{k}}_4, 0.46, 0.35, 0.33; 0.28 \rangle \right\};$$

$$\underset{\sim}{P}_{0.28} \cup_{\min} \underset{\sim}{Q}_{0.3} = \left\{ \langle \ddot{\underset{\sim}{k}}_1, 0.49, 0.28, 0.24; 0.28 \rangle, \langle \ddot{\underset{\sim}{k}}_2, 0.5, 0.38, 0.46; 0.28 \rangle, \langle \ddot{\underset{\sim}{k}}_3, 0.5, 0.23, 0.28; 0.28 \rangle, \langle \ddot{\underset{\sim}{k}}_4, 0.53, 0.18, 0.46; 0.28 \rangle \right\};$$

$$\underset{\sim}{P}_{0.28} \cup_{\max} \underset{\sim}{Q}_{0.3} = \left\{ \langle \ddot{\underset{\sim}{k}}_1, 0.49, 0.28, 0.24; 0.3 \rangle, \langle \ddot{\underset{\sim}{k}}_2, 0.5, 0.38, 0.46; 0.3 \rangle, \langle \ddot{\underset{\sim}{k}}_3, 0.5, 0.23, 0.28; 0.3 \rangle, \langle \ddot{\underset{\sim}{k}}_4, 0.53, 0.18, 0.46; 0.3 \rangle \right\};$$

$$\underset{\sim}{P}_{0.28} \cap_{\min} \underset{\sim}{Q}_{0.3} = \left\{ \langle \ddot{\underset{\sim}{k}}_1, 0.18, 0.28, 0.43; 0.28 \rangle, \langle \ddot{\underset{\sim}{k}}_2, 0.45, 0.38, 0.46; 0.28 \rangle, \langle \ddot{\underset{\sim}{k}}_3, 0.48, 0.23, 0.44; 0.28 \rangle, \langle \ddot{\underset{\sim}{k}}_4, 0.33, 0.18, 0.51; 0.28 \rangle \right\};$$

$$\underset{\sim}{P}_{0.28} \cap_{\max} \underset{\sim}{Q}_{0.3} = \left\{ \langle \ddot{\underset{\sim}{k}}_1, 0.18, 0.28, 0.43; 0.3 \rangle, \langle \ddot{\underset{\sim}{k}}_2, 0.45, 0.38, 0.46; 0.3 \rangle, \langle \ddot{\underset{\sim}{k}}_3, 0.48, 0.23, 0.44; 0.3 \rangle, \langle \ddot{\underset{\sim}{k}}_4, 0.33, 0.18, 0.51; 0.3 \rangle \right\}.$$

**Theorem 1.** Let $\underset{\sim}{P}_{\ddot{r}_1} = \left\{ \left\langle \ddot{\underset{\sim}{k}}, \ddot{\underset{\sim}{\mathfrak{A}}}_P(\ddot{\underset{\sim}{k}}), \ddot{\underset{\sim}{\Xi}}_P(\ddot{\underset{\sim}{k}}), \ddot{\underset{\sim}{\mathfrak{M}}}_P(\ddot{\underset{\sim}{k}}); \ddot{r}_1 \right\rangle : \ddot{\underset{\sim}{k}} \in \ddot{\underset{\sim}{K}} \right\}$ and

$\underset{\sim}{Q}_{\ddot{r}_2} = \left\{ \left\langle \ddot{\underset{\sim}{k}}, \ddot{\underset{\sim}{\mathfrak{A}}}_Q(\ddot{\underset{\sim}{k}}), \ddot{\underset{\sim}{\Xi}}_Q(\ddot{\underset{\sim}{k}}), \ddot{\underset{\sim}{\mathfrak{M}}}_Q(\ddot{\underset{\sim}{k}}); \ddot{r}_2 \right\rangle : \ddot{\underset{\sim}{k}} \in \ddot{\underset{\sim}{K}} \right\}$ be the two GO$_{\text{mno}}$-TSFSs in $\ddot{\underset{\sim}{K}}$. Then, we have the following results:

(1) $(\underset{\sim}{P}_{\ddot{r}_1} \cap_{\min} \underset{\sim}{Q}_{\ddot{r}_2})^c = \underset{\sim}{P}_{\ddot{r}_1}^c \cup_{\min} \underset{\sim}{Q}_{\ddot{r}_2}^c$

(2) $(\underset{\sim}{P}_{\ddot{r}_1} \cap_{\max} \underset{\sim}{Q}_{\ddot{r}_2})^c = \underset{\sim}{P}_{\ddot{r}_1}^c \cup_{\max} \underset{\sim}{Q}_{\ddot{r}_2}^c$

(3) $(\underset{\sim}{P}_{\ddot{r}_1} \cup_{\min} \underset{\sim}{Q}_{\ddot{r}_2})^c = \underset{\sim}{P}_{\ddot{r}_1}^c \cap_{\min} \underset{\sim}{Q}_{\ddot{r}_2}^c$

(4) $(\underset{\sim}{P}_{\ddot{r}_1} \cup_{\max} \underset{\sim}{Q}_{\ddot{r}_2})^c = \underset{\sim}{P}_{\ddot{r}_1}^c \cap_{\max} \underset{\sim}{Q}_{\ddot{r}_2}^c$.

In prior research, Atanassov [5] defined radius operations using the maximum and minimum functions within the unit interval [0,1]. In this work, we extend these operations over the same interval [0,1] and introduce four additional operations: $\otimes, \oplus, *$ and $\odot$ namely, algebraic product, algebraic sum, arithmetic mean, and geometric mean, respectively.

**Definition 7.** Let $\ddot{r_1}, \ddot{r_2} \in [0,1]$ and $m = t_1$ or $\dfrac{1}{t_1}$ , $o = t_2$ or $\dfrac{1}{t_2}$ , and $n = t_3$ or $\dfrac{1}{t_3}$ for all $t_i (i = 1,2,3) \in \mathbb{Z}^+$. The operations $\otimes, \oplus, *$ and $\odot$ on radius are defined respectively as:

$$\otimes(\ddot{r_1}, \ddot{r_2}) = (\ddot{r_1}\ddot{r_2}), \ \oplus(\ddot{r_1}, \ddot{r_2}) = \left(\ddot{r_1}^o + \ddot{r_2}^o - (\ddot{r_1}\ddot{r_2})^o\right)^{\frac{1}{o}}, \ *(\ddot{r_1}, \ddot{r_2}) = \left(\frac{\ddot{r_1}^o + \ddot{r_2}^o}{2}\right)^{\frac{1}{o}}, \ \odot(\ddot{r_1}, \ddot{r_2}) = \left(\sqrt[o]{\ddot{r_1}^o \ddot{r_2}^o}\right).$$

**Theorem 2.** The results of the operations defined in Definition 7 also belong to the unit interval, satisfying the fundamental requirement of the radius.

**Proof.** To prove the validity of these operations, we need to demonstrate that, for $\ddot{r_1}, \ddot{r_2} \in [0,1]$ and $m = t_1$ or $\dfrac{1}{t_1}, o = t_2$ or $\dfrac{1}{t_2}$, and $n = t_3$ or $\dfrac{1}{t_3}$ for all $t_i (i = 1,2,3) \in \mathbb{Z}^+$, the results of $\otimes, \oplus, *$, and $\odot$ to be within $[0,1]$. Let's begin with the operation $\otimes(\ddot{r_1}, \ddot{r_2})$. When $0 \le \ddot{r_1}, \ddot{r_2} \le 1$, it is evident that $0 \le \ddot{r_1}\ddot{r_2} \le 1$. Moving on to the operation $\oplus(\ddot{r_1}, \ddot{r_2})$, our aim is to prove $\ddot{r_1}^o + \ddot{r_2}^o - (\ddot{r_1}\ddot{r_2})^o \le 1$. Using the contradiction, suppose it is true for $\ddot{r_1}^o + \ddot{r_2}^o - (\ddot{r_1}\ddot{r_2})^o > 1$ such that, $\ddot{r_1}^o + \ddot{r_2}^o - (\ddot{r_1}\ddot{r_2})^o - 1 > 0$, $\ddot{r_1}^o + \ddot{r_2}^o - \ddot{r_1}^o\ddot{r_2}^o - 1 > 0$, $\ddot{r_1}^o - \ddot{r_1}^o\ddot{r_2}^o + \ddot{r_2}^o - 1 > 0$, $\ddot{r_1}^o\left(1 - \ddot{r_2}^o\right) - 1\left(1 - \ddot{r_2}^o\right) > 0$, $\left(\ddot{r_1}^o - 1\right)\left(1 - \ddot{r_2}^o\right) > 0$. For all $m = t_1$ or $\dfrac{1}{t_1}, o = t_2$ or $\dfrac{1}{t_2}$, and $n = t_3$ or $\dfrac{1}{t_3}$, it is obtained that $\left(\ddot{r_1}^o - 1\right)\left(1 - \ddot{r_2}^o\right) \le 0$, Therefore, it is contradicted, hence $0 \le \left(\ddot{r_1}^o + \ddot{r_2}^o - (\ddot{r_1}\ddot{r_2})^o\right)^{\frac{1}{o}} \le \left(1^o\right)^{\frac{1}{o}} = 1$ , $0 \le \left(\ddot{r_1}^o + \ddot{r_2}^o - (\ddot{r_1}\ddot{r_2})^o\right)^{\frac{1}{o}} \le 1$ . For the operation $*(\ddot{r_1}, \ddot{r_2})$ , since $\ddot{r_1}^o \le 1$ and $\ddot{r_2}^o \le 1$ then, $\dfrac{\ddot{r_1}^o + \ddot{r_2}^o}{2} \le 1$, so $\left(\dfrac{\ddot{r_1}^o + \ddot{r_2}^o}{2}\right)^{\frac{1}{o}} \le 1$. Lastly, for the operation $\odot(\ddot{r_1}, \ddot{r_2})$, it follows that $\ddot{r_1}^o \le 1$ and $\ddot{r_2}^o \le 1$, so $\sqrt{\ddot{r_1}^o\ddot{r_2}^o} \le 1$, and hence $\left(\sqrt{\ddot{r_1}^o\ddot{r_2}^o}\right)^{\frac{1}{o}} \le 1$. ∎

The operations defined in definition 7 that will affect the radius of GO$_{mno}$-TSFS. Next, we will define the general operations that apply to GO$_{mno}$-TSFS.

**Definition 8.** Let's consider $P_{\ddot{r}_1}, Q_{\ddot{r}_2} \in \text{GO}_{\text{mno}}\text{-TSFS}\left(\text{mno}, \ddot{K}\right)$, with $\ddot{r}_1, \ddot{r}_2 \in [0,1]$ and $m = t_1$ or $\dfrac{1}{t_1}$, $o = t_2$ or $\dfrac{1}{t_2}$, and $n = t_3$ or $\dfrac{1}{t_3}$ for all $t_i (i = 1, 2, 3) \in \mathbb{Z}^+$. For every $\ddot{k} \in \ddot{K}$, $\Upsilon \in \left(\min, \max, \otimes, \oplus, *, \odot\right)$ be the radius operators, the operations between $P_{\ddot{r}_1}$ and $Q_{\ddot{r}_2}$ can be defined as following:

(i) $P_{\ddot{r}_1} \cup_{\Upsilon} Q_{\ddot{r}_2} = \left\{ \ddot{k}, \max(\ddot{\mathfrak{A}}_P(\ddot{k}), \ddot{\mathfrak{A}}_Q(\ddot{k})), \min(\ddot{\Xi}_P(\ddot{k}), \ddot{\Xi}_Q(\ddot{k})), \min(\ddot{\mathfrak{M}}_P(\ddot{k}), \ddot{\mathfrak{M}}_Q(\ddot{k})); \Upsilon(\ddot{r}_1, \ddot{r}_2) : \ddot{k} \in \ddot{K} \right\}$.

(ii) $P_{\ddot{r}_1} \cap_{\Upsilon} Q_{\ddot{r}_2} = \left\{ \ddot{k}, \min(\ddot{\mathfrak{A}}_P(\ddot{k}), \ddot{\mathfrak{A}}_Q(\ddot{k})), \min(\ddot{\Xi}_P(\ddot{k}), \ddot{\Xi}_Q(\ddot{k})), \max(\ddot{\mathfrak{M}}_P(\ddot{k}), \ddot{\mathfrak{M}}_Q(\ddot{k})); \Upsilon(\ddot{r}_1, \ddot{r}_2) : \ddot{k} \in \ddot{K} \right\}$.

**Theorem 3.** For $P_{\ddot{r}_1}, Q_{\ddot{r}_2} \in \text{GO}_{\text{mno}}\text{-TSFS}$, $\mho\{\cap, \cup\}$ and $\Upsilon \in \left(\min, \max, \otimes, \oplus, *, \odot\right)$, it hold that $P_{\ddot{r}_1} \mho_{\Upsilon} Q_{\ddot{r}_2} \in \text{GO}_{\text{mno}}\text{-TSFS}$.

**Proof.** The proof for the radius has already been established in Theorem 2. To demonstrate this theorem: For operation $\cap_{\Upsilon}$ and $\cup_{\Upsilon}$, considering the case $P_{\ddot{r}_1} \cap_{\Upsilon} Q_{\ddot{r}_2}$ where $\max\left\{\ddot{\mathfrak{M}}_P\left(\ddot{k}\right), \ddot{\mathfrak{M}}_Q\left(\ddot{k}\right)\right\} = \ddot{\mathfrak{M}}_P\left(\ddot{k}\right)$ and $\max\left\{\ddot{\Xi}_P\left(\ddot{k}\right), \ddot{\Xi}_Q\left(\ddot{k}\right)\right\} = \ddot{\Xi}_P\left(\ddot{k}\right)$, we have,

$$0 \le \left(\ddot{\mathfrak{A}}_{P_{\ddot{r}_1} \cap_{\Upsilon} Q_{\ddot{r}_2}}\left(\ddot{k}\right)\right)^m + \left(\ddot{\Xi}_{P_{\ddot{r}_1} \cap_{\Upsilon} Q_{\ddot{r}_2}}\left(\ddot{k}\right)\right)^o + \left(\ddot{\mathfrak{M}}_{P_{\ddot{r}_1} \cap_{\Upsilon} Q_{\ddot{r}_2}}\left(\ddot{k}\right)\right)^n$$

$$= \left(\min\left\{\ddot{\mathfrak{A}}_P\left(\ddot{k}\right), \ddot{\mathfrak{A}}_Q\left(\ddot{k}\right)\right\}\right)^m + \left(\ddot{\Xi}_P\left(\ddot{k}\right)\right)^o + \left(\ddot{\mathfrak{M}}_P\left(\ddot{k}\right)\right)^n \le \left(\ddot{\mathfrak{A}}_P\left(\ddot{k}\right)\right)^m + \left(\ddot{\Xi}_P\left(\ddot{k}\right)\right)^o + \left(\ddot{\mathfrak{M}}_P\left(\ddot{k}\right)\right)^n \le 1.$$

If $\max\left\{\ddot{\mathfrak{M}}_P\left(\ddot{k}\right), \ddot{\mathfrak{M}}_Q\left(\ddot{k}\right)\right\} = \ddot{\mathfrak{M}}_Q\left(\ddot{k}\right)$ and $\max\left\{\ddot{\Xi}_P\left(\ddot{k}\right), \ddot{\Xi}_Q\left(\ddot{k}\right)\right\} = \ddot{\Xi}_Q\left(\ddot{k}\right)$, then similarly to the pervious proof we obtain $0 \le \left(\min\left\{\ddot{\mathfrak{A}}_P\left(\ddot{k}\right), \ddot{\mathfrak{A}}_Q\left(\ddot{k}\right)\right\}\right)^m + \left(\ddot{\Xi}_Q\left(\ddot{k}\right)\right)^o + \left(\ddot{\mathfrak{M}}_Q\left(\ddot{k}\right)\right)^n \le \left(\ddot{\mathfrak{A}}_Q\left(\ddot{k}\right)\right)^m + \left(\ddot{\Xi}_Q\left(\ddot{k}\right)\right)^o + \left(\ddot{\mathfrak{M}}_Q\left(\ddot{k}\right)\right)^n \le 1$. $P_{\ddot{r}_1} \cup_{\Upsilon} Q_{\ddot{r}_2}$ can also be proved using same method. ∎

**Definition 9.** Let's $\underset{\tilde{r_1}}{P}, \underset{\tilde{r_2}}{Q} \in \text{GO}_{\text{mno}}\text{-TSFS}\left(\text{mno}, \ddot{\ddot{K}}\right)$, with $\ddot{r_1}, \ddot{r_2} \in [0,1]$ and $m = t_1$ or $\dfrac{1}{t_1}$,

$o = t_2$ or $\dfrac{1}{t_2}$, and $n = t_3$ or $\dfrac{1}{t_3}$ for all $t_i(i=1,2,3) \in \mathbb{Z}^+$. For every $\ddot{\ddot{k}} \in \ddot{\ddot{K}}$, $\Upsilon \in \left(\min, \max, \otimes, \oplus, *, \odot\right)$

be the radius operators, the arithmetic mean, $\aleph_\Upsilon$ and geometric mean $\dagger_\Upsilon$ between $\underset{\tilde{r_1}}{P}$ and $\underset{\tilde{r_2}}{Q}$ can

be defined as following:

(1). $\underset{\tilde{r_1}}{P} \aleph_\Upsilon \underset{\tilde{r_2}}{Q} = \left\{ \ddot{\ddot{k}}, \left( \ddot{\ddot{\mathfrak{A}}}_P^m\left(\ddot{\ddot{k}}\right) + \ddot{\ddot{\mathfrak{A}}}_Q^m\left(\ddot{\ddot{k}}\right) \right)^{\frac{1}{m}}, \left( \ddot{\ddot{\Xi}}_P^o\left(\ddot{\ddot{k}}\right) + \ddot{\ddot{\Xi}}_Q^o\left(\ddot{\ddot{k}}\right) \right)^{\frac{1}{o}}, \left( \ddot{\ddot{\mathfrak{M}}}_P^n\left(\ddot{\ddot{k}}\right) + \ddot{\ddot{\mathfrak{M}}}_Q^n\left(\ddot{\ddot{k}}\right) \right)^{\frac{1}{n}} ; \Upsilon\left(\ddot{r_1}, \ddot{r_2}\right) : \ddot{\ddot{k}} \in \ddot{\ddot{K}} \right\}$

(2). $\underset{\tilde{r_1}}{P} \dagger_\Upsilon \underset{\tilde{r_2}}{Q} = \left\{ \ddot{\ddot{k}}, \left( \sqrt[m]{\ddot{\ddot{\mathfrak{A}}}_P^m\left(\ddot{\ddot{k}}\right) \ddot{\ddot{\mathfrak{A}}}_Q^m\left(\ddot{\ddot{k}}\right)} \right)^{\frac{1}{m}}, \left( \sqrt[o]{\ddot{\ddot{\Xi}}_P^o\left(\ddot{\ddot{k}}\right) \ddot{\ddot{\Xi}}_Q^o\left(\ddot{\ddot{k}}\right)} \right)^{\frac{1}{o}}, \left( \sqrt[n]{\ddot{\ddot{\mathfrak{M}}}_P^n\left(\ddot{\ddot{k}}\right) \ddot{\ddot{\mathfrak{M}}}_Q^n\left(\ddot{\ddot{k}}\right)} \right)^{\frac{1}{n}} ; \Upsilon\left(\ddot{r_1}, \ddot{r_2}\right) : \ddot{\ddot{k}} \in \ddot{\ddot{K}} \right\}.$

**Theorem 4.** The arithmetic mean $\aleph_\Upsilon$ and the geometric mean $\dagger_\Upsilon$ between two $\text{GO}_{\text{mno}}$-TSFSs in definition 9 are also $\text{GO}_{\text{mno}}$-TSFSs.

**Proof.** To prove these operations, we must show that for $\ddot{r_1}, \ddot{r_2} \in [0,1]$ and $m = t_1$ or $\dfrac{1}{t_1}$,

$o = t_2$ or $\dfrac{1}{t_2}$, and $n = t_3$ or $\dfrac{1}{t_3}$ for all $t_i(i=1,2,3) \in \mathbb{Z}^+$, the $\aleph_\Upsilon$ and $\dagger_\Upsilon$ satisfy the basic constraint of $\text{GO}_{\text{mno}}$-TSF framework. For operation $\aleph_\Upsilon$ we obtain,

$$0 \leq \left( \ddot{\ddot{\mathfrak{A}}}_{\underset{\tilde{r_1}}{P} \aleph_\Upsilon \underset{\tilde{r_2}}{Q}}\left(\ddot{\ddot{k}}\right) \right)^m + \left( \ddot{\ddot{\Xi}}_{\underset{\tilde{r_1}}{P} \aleph_\Upsilon \underset{\tilde{r_2}}{Q}}\left(\ddot{\ddot{k}}\right) \right)^o + \left( \ddot{\ddot{\mathfrak{M}}}_{\underset{\tilde{r_1}}{P} \aleph_\Upsilon \underset{\tilde{r_2}}{Q}}\left(\ddot{\ddot{k}}\right) \right)^n \leq 1$$

$$\leq \left( \frac{\ddot{\ddot{\mathfrak{A}}}_P^m\left(\ddot{\ddot{k}}\right) + \ddot{\ddot{\mathfrak{A}}}_Q^m\left(\ddot{\ddot{k}}\right)}{2} \right) + \left( \frac{\ddot{\ddot{\Xi}}_P^o\left(\ddot{\ddot{k}}\right) + \ddot{\ddot{\Xi}}_Q^o\left(\ddot{\ddot{k}}\right)}{2} \right) + \left( \frac{\ddot{\ddot{\mathfrak{M}}}_P^n\left(\ddot{\ddot{k}}\right) + \ddot{\ddot{\mathfrak{M}}}_Q^n\left(\ddot{\ddot{k}}\right)}{2} \right)$$

$$= \frac{\ddot{\ddot{\mathfrak{A}}}_P^m\left(\ddot{\ddot{k}}\right) + \ddot{\ddot{\mathfrak{A}}}_Q^m\left(\ddot{\ddot{k}}\right) + \ddot{\ddot{\Xi}}_P^o\left(\ddot{\ddot{k}}\right) + \ddot{\ddot{\Xi}}_Q^o\left(\ddot{\ddot{k}}\right) + \ddot{\ddot{\mathfrak{M}}}_P^n\left(\ddot{\ddot{k}}\right) + \ddot{\ddot{\mathfrak{M}}}_Q^n\left(\ddot{\ddot{k}}\right)}{2}$$

$$= \frac{\left( \ddot{\ddot{\mathfrak{A}}}_P^m\left(\ddot{\ddot{k}}\right) + \ddot{\ddot{\Xi}}_P^o\left(\ddot{\ddot{k}}\right) + \ddot{\ddot{\mathfrak{M}}}_P^n\left(\ddot{\ddot{k}}\right) \right) + \left( \mathfrak{A}_Q^m\left(\ddot{\ddot{k}}\right) + \ddot{\ddot{\Xi}}_Q^o\left(\ddot{\ddot{k}}\right) + \ddot{\ddot{\mathfrak{M}}}_Q^n\left(\ddot{\ddot{k}}\right) \right)}{2} \leq \frac{1+1}{2} = 1 \,.$$

Likewise for $\underline{P}_{\ddot{\vec{k}_1}} \dagger_\Upsilon \underline{Q}_{\ddot{\vec{k}_2}}$, we have,

$$0 \le \left(\dddot{\mathfrak{A}}_{\underline{P}_{\ddot{\vec{k}_1}}?_\Upsilon \underline{Q}_{\ddot{\vec{k}_2}}}\left(\ddot{\vec{k}}\right)\right)^m + \left(\dddot{\Xi}_{\underline{P}_{\ddot{\vec{k}_1}}} {}_\Upsilon {}_{\underline{Q}_{\ddot{\vec{k}_2}}}\left(\ddot{\vec{k}}\right)\right)^o + \left(\dddot{\mathfrak{M}}_{\underline{P}_{\ddot{\vec{k}_1}}?_\Upsilon \underline{Q}_{\ddot{\vec{k}_2}}}\left(\ddot{\vec{k}}\right)\right)^n \le 1$$

$$\le \sqrt{\dddot{\mathfrak{A}}_{\underline{P}}^m\left(\ddot{\vec{k}}\right)\dddot{\mathfrak{A}}_{\underline{Q}}^m\left(\ddot{\vec{k}}\right)} + \sqrt{\dddot{\Xi}_{\underline{P}}^o\left(\ddot{\vec{k}}\right)\dddot{\Xi}_{\underline{Q}}^o\left(\ddot{\vec{k}}\right)} + \sqrt{\dddot{\mathfrak{M}}_{\underline{P}}^n\left(\ddot{\vec{k}}\right)\dddot{\mathfrak{M}}_{\underline{Q}}^n\left(\ddot{\vec{k}}\right)}$$

$$\le \left(\frac{\dddot{\mathfrak{A}}_{\underline{P}}^m\left(\ddot{\vec{k}}\right)+\dddot{\mathfrak{A}}_{\underline{Q}}^m\left(\ddot{\vec{k}}\right)}{2}\right) + \left(\frac{\dddot{\Xi}_{\underline{P}}^o\left(\ddot{\vec{k}}\right)+\dddot{\Xi}_{\underline{Q}}^o\left(\ddot{\vec{k}}\right)}{2}\right) + \left(\frac{\dddot{\mathfrak{M}}_{\underline{P}}^n\left(\ddot{\vec{k}}\right)+\dddot{\mathfrak{M}}_{\underline{Q}}^n\left(\ddot{\vec{k}}\right)}{2}\right)$$

$$\le \frac{\left(\dddot{\mathfrak{A}}_{\underline{P}}^m\left(\ddot{\vec{k}}\right)+\dddot{\Xi}_{\underline{P}}^o\left(\ddot{\vec{k}}\right)+\dddot{\mathfrak{M}}_{\underline{P}}^n\left(\ddot{\vec{k}}\right)\right)+\left(\dddot{\mathfrak{A}}_{\underline{Q}}^m\left(\ddot{\vec{k}}\right)+\dddot{\Xi}_{\underline{Q}}^o\left(\ddot{\vec{k}}\right)+\dddot{\mathfrak{M}}_{\underline{Q}}^n\left(\ddot{\vec{k}}\right)\right)}{2} \le \frac{1+1}{2}=1$$ . It completes the proof. ■

### 3.1. Formation of GO$_{mno}$-TSF values

In this subsection, we outline the computational procedure for identifying the central point $\left\langle\dddot{\tilde{\mathfrak{A}}},\dddot{\tilde{\Xi}},\dddot{\tilde{\mathfrak{M}}}\right\rangle$ and radius $\ddot{\vec{r}}$ of a sphere based on the DoM, DoI, and DoN. Additionally, we illustrate the conversion process from (m,n,o)-TSFVs to GO$_{mno}$-TSFVs, using $\left\langle\dddot{\tilde{\mathfrak{A}}},\dddot{\tilde{\Xi}},\dddot{\tilde{\mathfrak{M}}}\right\rangle$ and $\ddot{\vec{r}}$ to depict GO$_{mno}$-TSFSs in relation to the sphere.

Let's denote a set of (m,n,o)-TSFVs as $\left\{\left\langle\dddot{\mathfrak{A}}_1,\dddot{\Xi}_1,\dddot{\mathfrak{M}}_1\right\rangle,\left\langle\dddot{\mathfrak{A}}_2,\dddot{\Xi}_2,\dddot{\mathfrak{M}}_3\right\rangle...,\left\langle\dddot{\mathfrak{A}}_{\ddot{\vec{l}}},\dddot{\Xi}_{\ddot{\vec{l}}},\dddot{\mathfrak{M}}_{\ddot{\vec{l}}}\right\rangle\right\}$. Then the central point, denoted as $\left\langle\dddot{\tilde{\mathfrak{A}}},\dddot{\tilde{\Xi}},\dddot{\tilde{\mathfrak{M}}}\right\rangle$, is defined as follows:

$$\left\langle\dddot{\tilde{\mathfrak{A}}},\dddot{\tilde{\Xi}},\dddot{\tilde{\mathfrak{M}}}\right\rangle = \left(\sqrt[m]{\frac{\sum_{i=1}^{\ddot{\vec{l}}}\dddot{\mathfrak{A}}_i^m}{\ddot{\vec{l}}}},\sqrt[o]{\frac{\sum_{i=1}^{\ddot{\vec{l}}}\dddot{\Xi}_i^o}{\ddot{\vec{l}}}},\sqrt[n]{\frac{\sum_{i=1}^{\ddot{\vec{l}}}\dddot{\mathfrak{M}}_i^n}{\ddot{\vec{l}}}}\right) \tag{3}$$

Subsequently, we can demonstrate that $\left\langle\dddot{\tilde{\mathfrak{A}}},\dddot{\tilde{\Xi}},\dddot{\tilde{\mathfrak{M}}};\ddot{\vec{r}}\right\rangle$ for $\ddot{\vec{r}}\in[0,1]$ transforms into a GO$_{mno}$-TSFV. Fundamentally, the sphere's radius $\ddot{\vec{r}}$ represents the greatest distance from the central point

$\left\langle \tilde{\tilde{\mathfrak{A}}}, \tilde{\tilde{\Xi}}, \tilde{\tilde{\mathfrak{M}}} \right\rangle$ to any point on its surface among the ensemble of (m,n,o)-TSFVs, forming a GO$_{mno}$-TSFV. This concept can be articulated mathematically as follows:

$$\ddot{r} = \min\left\{ \max_{1 \le i \le \ddot{l}} \sqrt{(\tilde{\tilde{\mathfrak{A}}}^m - \ddot{\mathfrak{A}}_i^m)^2 + (\tilde{\tilde{\Xi}}^o - \ddot{\Xi}_i^o)^2 + (\tilde{\tilde{\mathfrak{M}}}^n - \ddot{\mathfrak{M}}_i^n)^2}, 1 \right\} \tag{4}$$

**Theorem 5.** Let $\left\{ \left\langle \ddot{\mathfrak{A}}_1, \ddot{\Xi}_1, \ddot{\mathfrak{M}}_1 \right\rangle, \left\langle \ddot{\mathfrak{A}}_2, \ddot{\Xi}_2, \ddot{\mathfrak{M}}_3 \right\rangle ..., \left\langle \ddot{\mathfrak{A}}_{\ddot{l}}, \ddot{\Xi}_{\ddot{l}}, \ddot{\mathfrak{M}}_{\ddot{l}} \right\rangle \right\}$ denotes a set of (m,n,o)-TSFVs. Then $\left\langle \tilde{\tilde{\mathfrak{A}}}, \tilde{\tilde{\Xi}}, \tilde{\tilde{\mathfrak{M}}}; \ddot{r} \right\rangle$ is a GO$_{mno}$-TSFV where $\tilde{\tilde{\mathfrak{A}}} = \sqrt[m]{\dfrac{\sum_{i=1}^{\ddot{l}} \ddot{\mathfrak{A}}_i^m}{\ddot{l}}}$ , $\tilde{\tilde{\Xi}} = \sqrt[o]{\dfrac{\sum_{i=1}^{\ddot{l}} \ddot{\Xi}_i^o}{\ddot{l}}}$ , $\tilde{\tilde{\mathfrak{M}}} = \sqrt[n]{\dfrac{\sum_{i=1}^{\ddot{l}} \ddot{\mathfrak{M}}_i^n}{\ddot{l}}}$ and $\ddot{r}$ is defined as in Eq. (4).

**Proof**: Let's consider $\tilde{\tilde{\mathfrak{A}}}_h^m + \tilde{\tilde{\Xi}}_h^o + \tilde{\tilde{\mathfrak{M}}}_h^n$ with

$$\tilde{\tilde{\mathfrak{A}}}_h^m + \tilde{\tilde{\Xi}}_h^o + \tilde{\tilde{\mathfrak{M}}}_h^n = \left( \sqrt[m]{\sum_{i=1}^{\ddot{l}} \ddot{\mathfrak{A}}_i^m \Big/ \ddot{l}} \right)^m + \left( \sqrt[o]{\sum_{i=1}^{\ddot{l}} \ddot{\Xi}_i^o \Big/ \ddot{l}} \right)^o + \left( \sqrt[n]{\sum_{i=1}^{\ddot{l}} \ddot{\mathfrak{M}}_i^n \Big/ \ddot{l}} \right)^n$$

$$= \frac{\sum_{i=1}^{\ddot{l}} \ddot{\mathfrak{A}}_i^m}{\ddot{l}} + \frac{\sum_{i=1}^{\ddot{l}} \ddot{\Xi}_i^o}{\ddot{l}} + \frac{\sum_{i=1}^{\ddot{l}} \ddot{\mathfrak{M}}_i^n}{\ddot{l}} = \frac{\sum_{i=1}^{\ddot{l}} \ddot{\mathfrak{A}}_i^m + \sum_{i=1}^{\ddot{l}} \ddot{\Xi}_i^o + \sum_{i=1}^{\ddot{l}} \ddot{\mathfrak{M}}_i^n}{\ddot{l}}$$

$$= \frac{\sum_{i=1}^{\ddot{l}} \left( \ddot{\mathfrak{A}}_i^m + \ddot{\Xi}_i^o + \ddot{\mathfrak{M}}_i^n \right)}{\ddot{l}} \le \frac{\sum_{i=1}^{\ddot{l}} (1)}{\ddot{l}} = 1 \quad i.e. \; \tilde{\tilde{\mathfrak{A}}}_h^m + \tilde{\tilde{\Xi}}_h^o + \tilde{\tilde{\mathfrak{M}}}_h^n \le 1$$

Which is the basic constraint of a GO$_{mno}$-TSFV and further it is evident from eq. (4) that $0 \le \ddot{r} \le 1$. Hence, we can say that $\left\langle \tilde{\tilde{\mathfrak{A}}}, \tilde{\tilde{\Xi}}, \tilde{\tilde{\mathfrak{M}}}; \ddot{r} \right\rangle$ is a GO$_{mno}$-TSFV. ∎

**Definition 10.** Let $\tilde{P} = \left( \ddot{\mathfrak{A}}_{\tilde{P}}, \ddot{\Xi}_{\tilde{P}}, \ddot{\mathfrak{M}}_{\tilde{P}}; \ddot{r}_1 \right)$ and $\tilde{Q} = \left( \ddot{\mathfrak{A}}_{\tilde{Q}}, \ddot{\Xi}_{\tilde{Q}}, \ddot{\mathfrak{M}}_{\tilde{Q}}; \ddot{r}_2 \right)$ be two GO$_{mno}$-TSFVs, then

$$\tilde{\underset{\sim}{P}}^c = \left(\ddot{\mathfrak{A}}_{\tilde{P}}, \overset{\cdots}{\Xi}_{\tilde{P}}, \mathfrak{M}_{\tilde{P}}; \ddot{r}_1\right)$$

$$\tilde{\underset{\sim}{P}} \wedge \tilde{\underset{\sim}{Q}} = \left(\min(\ddot{\mathfrak{A}}_{\tilde{P}}, \ddot{\mathfrak{A}}_{\tilde{Q}}), \max(\overset{\cdots}{\Xi}_{\tilde{P}}, \overset{\cdots}{\Xi}_{\tilde{Q}}), \max(\mathfrak{M}_{\tilde{P}}, \mathfrak{M}_{\tilde{Q}}); \min(\ddot{r}_1, \ddot{r}_2)\right)$$

$$\tilde{\underset{\sim}{P}} \vee \tilde{\underset{\sim}{Q}} = \left(\max(\ddot{\mathfrak{A}}_{\tilde{P}}, \ddot{\mathfrak{A}}_{\tilde{Q}}), \min(\overset{\cdots}{\Xi}_{\tilde{P}}, \overset{\cdots}{\Xi}_{\tilde{Q}}), \min(\mathfrak{M}_{\tilde{P}}, \mathfrak{M}_{\tilde{Q}}); \max(\ddot{r}_1, \ddot{r}_2)\right)$$

$$\tilde{\underset{\sim}{P}} \oplus_{\min} \tilde{\underset{\sim}{Q}} = \left(\sqrt[m]{\ddot{\mathfrak{A}}_{\tilde{P}}^m + \ddot{\mathfrak{A}}_{\tilde{Q}}^m - \ddot{\mathfrak{A}}_{\tilde{P}}^m \ddot{\mathfrak{A}}_{\tilde{Q}}^m}, \overset{\cdots}{\Xi}_{\tilde{P}} \overset{\cdots}{\Xi}_{\tilde{Q}}, \mathfrak{M}_{\tilde{P}} \mathfrak{M}_{\tilde{Q}}; \min(\ddot{r}_1, \ddot{r}_2)\right)$$

$$\tilde{\underset{\sim}{P}} \oplus_{\max} \tilde{\underset{\sim}{Q}} = \left(\sqrt[m]{\ddot{\mathfrak{A}}_{\tilde{P}}^m + \ddot{\mathfrak{A}}_{\tilde{Q}}^m - \ddot{\mathfrak{A}}_{\tilde{P}}^m \ddot{\mathfrak{A}}_{\tilde{Q}}^m}, \overset{\cdots}{\Xi}_{\tilde{P}} \overset{\cdots}{\Xi}_{\tilde{Q}}, \mathfrak{M}_{\tilde{P}} \mathfrak{M}_{\tilde{Q}}; \max(\ddot{r}_1, \ddot{r}_2)\right)$$

$$\tilde{\underset{\sim}{P}} \otimes_{\min} \tilde{\underset{\sim}{Q}} = \left(\ddot{\mathfrak{A}}_{\tilde{P}} \ddot{\mathfrak{A}}_{\tilde{Q}}, \overset{\cdots}{\Xi}_{\tilde{P}} \overset{\cdots}{\Xi}_{\tilde{Q}}, \sqrt[n]{\mathfrak{M}_{\tilde{P}}^n + \mathfrak{M}_{\tilde{Q}}^n - \mathfrak{M}_{\tilde{P}}^n \mathfrak{M}_{\tilde{Q}}^n}; \min(\ddot{r}_1, \ddot{r}_2)\right)$$

$$\tilde{\underset{\sim}{P}} \otimes_{\max} \tilde{\underset{\sim}{Q}} = \left(\ddot{\mathfrak{A}}_{\tilde{P}} \ddot{\mathfrak{A}}_{\tilde{Q}}, \overset{\cdots}{\Xi}_{\tilde{P}} \overset{\cdots}{\Xi}_{\tilde{Q}}, \sqrt[n]{\mathfrak{M}_{\tilde{P}}^n + \mathfrak{M}_{\tilde{Q}}^n - \mathfrak{M}_{\tilde{P}}^n \mathfrak{M}_{\tilde{Q}}^n}; \max(\ddot{r}_1, \ddot{r}_2)\right)$$

$$\tilde{\underset{\sim}{\lambda}}_{\min} \tilde{\underset{\sim}{P}} = \left(\sqrt[m]{1 - \left(1 - \ddot{\mathfrak{A}}_{\tilde{P}}^m\right)^{\tilde{\lambda}}}, \overset{\cdots}{\Xi}_{\tilde{P}}^{\tilde{\lambda}}, \mathfrak{M}_{\tilde{P}}^{\tilde{\lambda}}; \ddot{r}_1^{\tilde{\lambda}}\right)$$

$$\tilde{\underset{\sim}{\lambda}}_{\max} \tilde{\underset{\sim}{P}} = \left(\sqrt[m]{1 - \left(1 - \ddot{\mathfrak{A}}_{\tilde{P}}^m\right)^{\tilde{\lambda}}}, \overset{\cdots}{\Xi}_{\tilde{P}}^{\tilde{\lambda}}, \mathfrak{M}_{\tilde{P}}^{\tilde{\lambda}}; \sqrt[o]{1 - \left(1 - \ddot{r}_1^o\right)^{\tilde{\lambda}}}\right)$$

$$\tilde{\underset{\sim}{P}}^{\tilde{\lambda}_{\min}} = \left(\ddot{\mathfrak{A}}_{\tilde{P}}^{\tilde{\lambda}}, \overset{\cdots}{\Xi}_{\tilde{P}}^{\tilde{\lambda}}, \sqrt[n]{1 - \left(1 - \mathfrak{M}_{\tilde{P}}^n\right)^{\tilde{\lambda}}}; \ddot{r}_1^{\tilde{\lambda}}\right)$$

$$\tilde{\underset{\sim}{P}}^{\tilde{\lambda}_{\max}} = \left(\ddot{\mathfrak{A}}_{\tilde{P}}^{\tilde{\lambda}}, \overset{\cdots}{\Xi}_{\tilde{P}}^{\tilde{\lambda}}, \sqrt[n]{1 - \left(1 - \mathfrak{M}_{\tilde{P}}^n\right)^{\tilde{\lambda}}}; \sqrt[o]{1 - \left(1 - \ddot{r}_1^o\right)^{\tilde{\lambda}}}\right)$$

**Theorem 6.** Let $\tilde{\underset{\sim}{P}} = \left(\ddot{\mathfrak{A}}_{\tilde{P}}, \overset{\cdots}{\Xi}_{\tilde{P}}, \mathfrak{M}_{\tilde{P}}; \ddot{r}_1\right)$ and $\tilde{\underset{\sim}{Q}} = \left(\ddot{\mathfrak{A}}_{\tilde{Q}}, \overset{\cdots}{\Xi}_{\tilde{Q}}, \mathfrak{M}_{\tilde{Q}}; \ddot{r}_2\right)$ be two GO$_{mno}$-TSFVs, $0 \le \tilde{\underset{\sim}{\lambda}}, \tilde{\underset{\sim}{\lambda}}_1, \tilde{\underset{\sim}{\lambda}}_2 \le 1$, then

(1) $\tilde{\underset{\sim}{P}} \oplus \tilde{\underset{\sim}{Q}} = \tilde{\underset{\sim}{Q}} \oplus \tilde{\underset{\sim}{P}}$;

(2) $\tilde{\underset{\sim}{P}} \otimes \tilde{\underset{\sim}{Q}} = \tilde{\underset{\sim}{Q}} \otimes \tilde{\underset{\sim}{P}}$;

(3) $\tilde{\underset{\sim}{\lambda}}\left(\tilde{\underset{\sim}{P}} \oplus \tilde{\underset{\sim}{Q}}\right) = \tilde{\underset{\sim}{\lambda}}\tilde{\underset{\sim}{P}} \oplus \tilde{\underset{\sim}{\lambda}}\tilde{\underset{\sim}{Q}}$;

(4) $\left(\tilde{\underset{\sim}{P}} \otimes \tilde{\underset{\sim}{Q}}\right)^{\tilde{\underset{\sim}{\lambda}}} = \tilde{\underset{\sim}{P}}^{\tilde{\underset{\sim}{\lambda}}} \otimes \tilde{\underset{\sim}{Q}}^{\tilde{\underset{\sim}{\lambda}}}$;

(5) $\tilde{\underset{\sim}{\lambda}}_1 \tilde{\underset{\sim}{P}} \oplus \tilde{\underset{\sim}{\lambda}}_2 \tilde{\underset{\sim}{P}} = \left(\tilde{\underset{\sim}{\lambda}}_1 \oplus \tilde{\underset{\sim}{\lambda}}_2\right)\tilde{\underset{\sim}{P}}$;

(6) $\tilde{\underset{\sim}{P}}^{\tilde{\underset{\sim}{\lambda}}_1} \otimes \tilde{\underset{\sim}{P}}^{\tilde{\underset{\sim}{\lambda}}_2} = \tilde{\underset{\sim}{P}}^{\tilde{\underset{\sim}{\lambda}}_1 + \tilde{\underset{\sim}{\lambda}}_2}$;

(7) $\left(\tilde{\underset{\sim}{P}}^{\tilde{\underset{\sim}{\lambda}}_1}\right)^{\tilde{\underset{\sim}{\lambda}}_2} = \tilde{\underset{\sim}{P}}^{\tilde{\underset{\sim}{\lambda}}_1 \tilde{\underset{\sim}{\lambda}}_2}$;

### 3.2. Some basic measures and ranking of GO$_{mno}$-TSFSs

In this subsection, we present fundamental tools designed specifically for GO$_{mno}$-TSFVs and subsequently devise a method for assessing them. This evaluation procedure facilitates comparison and prioritization among GO$_{mno}$-TSFVs, considering their unique properties and attributes.

**Definition 11.** Let $\hat{h} = \left\langle \ddot{\underset{\sim}{\mathfrak{A}}}_{\hbar}, \ddot{\underset{\sim}{\Xi}}_{\hbar}, \ddot{\underset{\sim}{\mathfrak{M}}}_{\hbar}; \ddot{r}_{\underset{\sim}{\sim}} \right\rangle$ be a GO$_{mno}$-TSFV. Then we define its score function (SF) and accuracy function (AF) in the following.

Score function: $\vec{\bar{\Psi}}(\hbar) = \dfrac{1}{2}\left(\ddot{\underset{\sim}{\mathfrak{A}}}_{\hbar}^{m} - \ddot{\underset{\sim}{\Xi}}_{\hbar}^{o} - \ddot{\underset{\sim}{\mathfrak{M}}}_{\hbar}^{n} + \ddot{r}_{\hbar}\left(2\eta - 1\right)\right)$      (5)

Here, $\vec{\bar{\Psi}}(\hbar) \in [-1,1]$, $r_{\hbar} \in [0,1]$, and $\eta \in [0,1]$

Accuracy function: $\vec{\bar{\xi}}(\hbar) = \ddot{\underset{\sim}{\mathfrak{A}}}_{\hbar}^{m} + \ddot{\underset{\sim}{\Xi}}_{\hbar}^{o} + \ddot{\underset{\sim}{\mathfrak{M}}}_{\hbar}^{n}$      (6)

Where, $\vec{\bar{\xi}}(\hbar) \in [0,1]$.

**Theorem 7.** (Monotonicity feature)

Let $\alpha = \left\langle \ddot{\underset{\sim}{\mathfrak{A}}}_{\alpha}, \ddot{\underset{\sim}{\Xi}}_{\alpha}, \ddot{\underset{\sim}{\mathfrak{M}}}_{\alpha}; \ddot{r}_{1} \right\rangle$ and $\beta = \left\langle \ddot{\underset{\sim}{\mathfrak{A}}}_{\beta}, \ddot{\underset{\sim}{\Xi}}_{\beta}, \ddot{\underset{\sim}{\mathfrak{M}}}_{\beta}; \ddot{r}_{2} \right\rangle$ be two GO$_{mno}$-TSFVs with $\vec{\bar{\Psi}}(\alpha)$ and $\vec{\bar{\Psi}}(\beta)$ be their corresponding SF. Then the score function $\vec{\bar{\Psi}}(\alpha)$ is monotonically increasing function with respect to $\ddot{\mathfrak{A}}_\alpha$ and $\ddot{r}_i$ while it monotonically decreasing function with respect to $\ddot{\Xi}_\alpha$ and $\ddot{\mathfrak{M}}_\alpha$.

**Proof.** (a) As we have, SF is $\vec{\ddot{\Psi}}(\alpha) = \frac{1}{2}\left(\ddot{\mathfrak{A}}_\alpha^m - \ddot{\Xi}_\alpha^o - \ddot{\mathfrak{M}}_\alpha^n + \ddot{r}_i(2\eta-1)\right)$, thus we have

$$\frac{\partial \vec{\ddot{\Psi}}(\alpha)}{\partial \ddot{\mathfrak{A}}_\alpha} = \frac{1}{2}\left(m\ddot{\mathfrak{A}}_\alpha^{m-1}\right) = \frac{m\ddot{\mathfrak{A}}_\alpha^{m-1}}{2}. \text{ So } \ddot{\mathfrak{A}}_\alpha \in [0,1] \text{ and } \frac{\partial \vec{\ddot{\Psi}}(\alpha)}{\partial \ddot{\mathfrak{A}}_\alpha} \geq 0, \text{ hence the SF is monotonically}$$

increasing function with respect to $\ddot{\mathfrak{A}}_\alpha$. Similarly $\frac{\partial}{\partial \ddot{r}_i}\left(\vec{\ddot{\Psi}}(\alpha)\right) = \frac{1}{2}(2\eta-1) = \left(\frac{2\eta-1}{2}\right) \geq 0$. Hence

$\ddot{r}_i \in [0,1]$ and $\frac{\partial}{\partial \ddot{r}_i}\left(\vec{\ddot{\Psi}}(\alpha)\right) \geq 0$, so the SF is monotonically increasing function with respect to $\ddot{r}_i$.

Now for $\ddot{\mathfrak{A}}_\alpha$ and $\ddot{\mathfrak{M}}_\alpha$, we can prove that the SF is monotonically decreasing function with respect to $\ddot{\mathfrak{A}}_\alpha$ and $\ddot{\mathfrak{M}}_\alpha$. We have that $\frac{\partial}{\partial \ddot{\Xi}_\alpha}\left(\vec{\ddot{\Psi}}(\alpha)\right) = \frac{1}{2}\left(0 + o\ddot{\Xi}_\alpha^{o-1} + 0 + 0\right) = -\frac{o\ddot{\Xi}_\alpha^{o-1}}{2} \leq 0$

Hence $\ddot{\Xi}_\alpha \in [0,1]$ and $\frac{\partial}{\partial \ddot{\Xi}_\alpha}\left(\vec{\ddot{\Psi}}(\alpha)\right) \leq 0$. Thus the SF is monotonically decreasing function with respect to $\ddot{\Xi}_\alpha$. Similarly, $\frac{\partial}{\partial \ddot{\mathfrak{M}}_\alpha}\left(\vec{\ddot{\Psi}}(\alpha)\right) = \frac{1}{2}\left(0 + 0 + n\ddot{\mathfrak{M}}_\alpha^{n-1} + 0\right) = -\frac{n\ddot{\mathfrak{M}}_\alpha^{n-1}}{2} \leq 0$. Therefore, the SF

decreases monotonically with respect to $\ddot{\mathfrak{M}}_\alpha$. ∎

**Theorem 8.** (Boundedness feature).

Let $\alpha = \langle \ddot{\mathfrak{A}}_\alpha, \ddot{\Xi}_\alpha, \ddot{\mathfrak{M}}_\alpha; \ddot{r}_i \rangle$ be a GO$_{mno}$-TSFV. Then the range of SF is: $-1 \leq \vec{\ddot{\Psi}}(\alpha) \leq 1$.

**Proof.** It is known that $\vec{\ddot{\Psi}}(\alpha) = \frac{1}{2}\left(\ddot{\mathfrak{A}}_\alpha^m - \ddot{\Xi}_\alpha^o - \ddot{\mathfrak{M}}_\alpha^n + \ddot{r}_i(2\eta-1)\right)$, where $\vec{\ddot{\Psi}}(\alpha) \in [-1,1]$. In order to prove this theorem, we have following intuitive cases to be considered;

Case 1: If $\ddot{\mathfrak{A}} = 1, \ddot{\Xi} = \ddot{\mathfrak{M}} = 0$, $\ddot{r}_i = 1$ and $\eta = 1$ then $\vec{\ddot{\Psi}}(\alpha) = 1$.

Case 2: If $\ddot{\mathfrak{A}} = 1, \ddot{\Xi} = \ddot{\mathfrak{M}} = 0$, $\ddot{r}_i = 1$ and $\eta = 0$ then $\vec{\ddot{\Psi}}(\alpha) = 0$.

Case 3: If $\ddot{\mathfrak{A}} = \ddot{\overline{\overline{\Xi}}} = 0, \ddot{\mathfrak{M}} = 1$, $\ddot{r} = 1$ and $\eta = 0$ then $\vec{\overline{\overline{\Psi}}}(\alpha) = -1$.

Hence it is proved that $-1 \le \vec{\overline{\overline{\Psi}}}(\alpha) \le 1$. ∎

**Theorem 9.** (Symmetry feature).

Let $\alpha = \langle \ddot{\mathfrak{A}}_\alpha, \ddot{\overline{\overline{\Xi}}}_\alpha, \ddot{\mathfrak{M}}_\alpha; \ddot{r}_1 \rangle$ and $\beta = \left( \ddot{\mathfrak{A}}_\beta, \ddot{\overline{\overline{\Xi}}}_\beta, \ddot{\mathfrak{M}}_\beta; \ddot{r}_2 \right)$ be two GO$_{mno}$-TSFVs with $\vec{\overline{\overline{\Psi}}}(\alpha)$ and $\vec{\overline{\overline{\Psi}}}(\beta)$ be their corresponding SF, while $\alpha^c = \langle \ddot{\mathfrak{M}}_\alpha, \ddot{\overline{\overline{\Xi}}}_\alpha, \ddot{\mathfrak{A}}_\alpha; \ddot{r}_1 \rangle$ and $\beta^c = \langle \ddot{\mathfrak{M}}_\beta, \ddot{\overline{\overline{\Xi}}}_\beta, \ddot{\mathfrak{A}}_\beta; \ddot{r}_2 \rangle$ be their respective complements. Where, $\vec{\overline{\overline{\Psi}}}(\alpha^c)$ and $\vec{\overline{\overline{\Psi}}}(\beta^c)$ be the complements of their corresponding score functions respectively. Then, $\vec{\overline{\overline{\Psi}}}(\alpha) \le \vec{\overline{\overline{\Psi}}}(\beta)$ iff $\vec{\overline{\overline{\Psi}}}(\alpha^c) \ge \vec{\overline{\overline{\Psi}}}(\beta^c)$.

**Proof.** As we have $\vec{\overline{\overline{\Psi}}}(\alpha) = \frac{1}{2} \left( \ddot{\mathfrak{A}}_\alpha^m - \ddot{\overline{\overline{\Xi}}}_\alpha^o - \ddot{\mathfrak{M}}_\alpha^n + \ddot{r}_1 (2\eta - 1) \right)$ and $\vec{\overline{\overline{\Psi}}}(\beta) = \frac{1}{2} \left( \ddot{\mathfrak{A}}_\beta^m - \ddot{\overline{\overline{\Xi}}}_\beta^o - \ddot{\mathfrak{M}}_\beta^n + \ddot{r}_2 (2\eta - 1) \right)$. Consider $\vec{\overline{\overline{\Psi}}}(\alpha) \le \vec{\overline{\overline{\Psi}}}(\beta)$ iff $\frac{1}{2} \left( \ddot{\mathfrak{A}}_\alpha^m - \ddot{\overline{\overline{\Xi}}}_\alpha^o - \ddot{\mathfrak{M}}_\alpha^n + \ddot{r}_1 (2\eta - 1) \right) \le \frac{1}{2} \left( \ddot{\mathfrak{A}}_\beta^m - \ddot{\overline{\overline{\Xi}}}_\beta^o - \ddot{\mathfrak{M}}_\beta^n + \ddot{r}_2 (2\eta - 1) \right)$. By taking the complements of memberships grades, we have $\vec{\overline{\overline{\Psi}}}(\alpha) \le \vec{\overline{\overline{\Psi}}}(\beta)$ iff $\frac{1}{2} \left( \ddot{\mathfrak{A}}_\alpha^m - \ddot{\overline{\overline{\Xi}}}_\alpha^o - \ddot{\mathfrak{M}}_\alpha^n + \ddot{r}_1 (2\eta - 1) \right) \ge \frac{1}{2} \left( \ddot{\mathfrak{A}}_\beta^m - \ddot{\overline{\overline{\Xi}}}_\beta^o - \ddot{\mathfrak{M}}_\beta^n + \ddot{r}_2 (2\eta - 1) \right)$. Thus, $\vec{\overline{\overline{\Psi}}}(\alpha) \le \vec{\overline{\overline{\Psi}}}(\beta)$ iff $\vec{\overline{\overline{\Psi}}}(\alpha^c) \ge \vec{\overline{\overline{\Psi}}}(\beta^c)$. ∎

**Theorem 10.** (Monotonicity feature)

An AF $\vec{\overline{\overline{\xi}}}(\alpha)$ is monotonically increasing functions with respect to $\ddot{\mathfrak{A}}_\alpha, \ddot{\overline{\overline{\Xi}}}_\alpha, \ddot{\mathfrak{M}}_\alpha$ and $\ddot{r}_1$.

**Proof.** As with theorem 7, the proof is similar. ∎

**Theorem 11.** (Symmetry feature)

Let $\alpha = \langle \ddot{\mathfrak{A}}_\alpha, \ddot{\Xi}_\alpha, \ddot{\mathfrak{M}}_\alpha; \ddot{r}_1 \rangle$ and $\beta = \langle \ddot{\mathfrak{A}}_\beta, \ddot{\Xi}_\beta, \ddot{\mathfrak{M}}_\beta; \ddot{r}_2 \rangle$ be two GO$_{mno}$-TSFVs with $\ddot{\bar{\bar{\xi}}}(\alpha)$ and $\ddot{\bar{\bar{\xi}}}(\beta)$ be their corresponding AF, while $\alpha^c = \langle \ddot{\mathfrak{M}}_\alpha, \ddot{\Xi}_\alpha, \ddot{\mathfrak{A}}_\alpha; \ddot{r}_1 \rangle$ and $\beta^c = \langle \ddot{\mathfrak{M}}_\beta, \ddot{\Xi}_\beta, \ddot{\mathfrak{A}}_\beta; \ddot{r}_2 \rangle$ be their respective complements. Then, $\ddot{\bar{\bar{\xi}}}(\alpha) = \ddot{\bar{\bar{\xi}}}(\beta)$

**Proof.** As with theorem 9, the proof is similar. ∎

**Definition 12.** Let $\hat{h}$ and $\hat{\lambda}$ denote two GO$_{mno}$-TSFVs with $\hat{h} = \langle \ddot{\mathfrak{A}}_{\hat{h}}, \ddot{\Xi}_{\hat{h}}, \ddot{\mathfrak{M}}_{\hat{h}}; \ddot{r}_1 \rangle$ and $\hat{\lambda} = \langle \ddot{\mathfrak{A}}_{\hat{\lambda}}, \ddot{\Xi}_{\hat{\lambda}}, \ddot{\mathfrak{M}}_{\hat{\lambda}}; \ddot{r}_2 \rangle$ as their mathematical forms respectively. Subsequently, we establish a ranking system for GO$_{mno}$-TSFVs based on the recently introduced score function defined in (5) and the AF described in (6) as follows.

- If $\ddot{\bar{\Psi}}(\hat{h}) > \ddot{\bar{\Psi}}(\hat{\lambda})$, then $\hat{h} > \hat{\lambda}$

- If $\ddot{\bar{\Psi}}(\hat{h}) < \ddot{\bar{\Psi}}(\hat{\lambda})$, then $\hat{h} < \hat{\lambda}$

- If $\ddot{\bar{\Psi}}(\hat{h}) = \ddot{\bar{\Psi}}(\hat{\lambda})$, then

    - If $\ddot{\bar{\bar{\xi}}}(\hat{h}) > \ddot{\bar{\bar{\xi}}}(\hat{\lambda})$, then $\hat{h} > \hat{\lambda}$

    - If $\ddot{\bar{\bar{\xi}}}(\hat{h}) < \ddot{\bar{\bar{\xi}}}(\hat{\lambda})$, then $\hat{h} < \hat{\lambda}$

    - If $\ddot{\bar{\bar{\xi}}}(\hat{h}) = \ddot{\bar{\bar{\xi}}}(\hat{\lambda})$, then $\hat{h} \approx \hat{\lambda}$.

We note that in Definition 4, we introduced a GO$_{mno}$-TSFS $\Theta_r$ in terms of $\ddot{\underbar{k}}$ as $\Theta_{\ddot{r}} = \left\{ \langle \ddot{\underbar{k}}, \ddot{\mathfrak{A}}_\Theta(\ddot{\underbar{k}}), \ddot{\Xi}_\Theta(\ddot{\underbar{k}}), \ddot{\mathfrak{M}}_\Theta(\ddot{\underbar{k}}); \ddot{r} \rangle : \ddot{\underbar{k}} \in \ddot{\mathcal{K}} \right\}$. However, we can define a more versatile type of GO$_{mno}$-TSFS $\Theta_{\ddot{r}} = \left\{ \langle \ddot{\underbar{k}}, \ddot{\mathfrak{A}}_\Theta(\ddot{\underbar{k}}), \ddot{\Xi}_\Theta(\ddot{\underbar{k}}), \ddot{\mathfrak{M}}_\Theta(\ddot{\underbar{k}}); \ddot{r}(\ddot{\underbar{k}}) \rangle : \ddot{\underbar{k}} \in \ddot{\mathcal{K}} \right\}$ in which the radius $\ddot{r}$ is allowed to vary with $\ddot{\underbar{k}}$. Naturally, if $\ddot{r}$ remains constant (a fixed value) for all $\ddot{\underbar{k}}$, then it converges to the same GO$_{mno}$-TSFS as defined in definition 4.

Next, we proceed to define distance measures for these GO$_{mno}$-TSFSs. Let's assume we have two GO$_{mno}$-TSFSs, $\Theta_{\ddot{r}_1} = \left\{ \langle \ddot{\underbar{k}}, \ddot{\mathfrak{A}}_\Theta(\ddot{\underbar{k}}), \ddot{\Xi}_\Theta(\ddot{\underbar{k}}), \ddot{\mathfrak{M}}_\Theta(\ddot{\underbar{k}}); \ddot{r}_1(\ddot{\underbar{k}}) \rangle : \ddot{\underbar{k}} \in \ddot{\mathcal{K}} \right\}$ and $\Theta_{\ddot{r}_2} = \left\{ \langle \ddot{\underbar{k}}, \ddot{\mathfrak{A}}_\Theta(\ddot{\underbar{k}}), \ddot{\Xi}_\Theta(\ddot{\underbar{k}}), \ddot{\mathfrak{M}}_\Theta(\ddot{\underbar{k}}); \ddot{r}_2(\ddot{\underbar{k}}) \rangle : \ddot{\underbar{k}} \in \ddot{\mathcal{K}} \right\}$, with radii $\ddot{r}_1(\ddot{\underbar{k}})$ and $\ddot{r}_2(\ddot{\underbar{k}})$ respectively.

**Definition 13**. The Hamming distance between two $GO_{mno}$-TSFSs, denoted as $\Theta_{\ddot{F_1}}$ and $\Theta_{\ddot{F_2}}$, under each $\ddot{\ddot{k}} \in \ddot{\ddot{K}}$ is mathematically defined as:

$$\Pi_1(\Theta_{\ddot{F_1}}, \Theta_{\ddot{F_2}}) = \frac{1}{2}\left(\frac{1}{2}\left(\left|\ddot{\mathfrak{A}}^m_{\Theta_{\ddot{F_1}}}(\ddot{\ddot{k}}) - \ddot{\mathfrak{A}}^m_{\Theta_{\ddot{F_2}}}(\ddot{\ddot{k}})\right| + \left|\ddot{\ddot{\Xi}}^o_{\Theta_{\ddot{F_1}}}(\ddot{\ddot{k}}) - \ddot{\ddot{\Xi}}^o_{\Theta_{\ddot{F_2}}}(\ddot{\ddot{k}})\right| + \left|\mathfrak{M}^n_{\Theta_{\ddot{F_1}}}(\ddot{\ddot{k}}) - \mathfrak{M}^n_{\Theta_{\ddot{F_2}}}(\ddot{\ddot{k}})\right|\right) + \left|\ddot{r_1}(\ddot{\ddot{k}}) - \ddot{r_2}(\ddot{\ddot{k}})\right|\right) \quad (7)$$

**Definition 14**. For the two $GO_{mno}$-TSFSs, denoted as $\Theta_{\ddot{F_1}}$ and $\Theta_{\ddot{F_2}}$, under each $\ddot{\ddot{k}} \in \ddot{\ddot{K}}$, we define the Euclidean distance as:

$$\Omega_1(\Theta_{\ddot{F_1}}, \Theta_{\ddot{F_2}}) = \frac{1}{2}\left(\sqrt{\frac{1}{2}\left((\ddot{\mathfrak{A}}^m_{\Theta_{\ddot{F_1}}}(\ddot{\ddot{k}}) - \ddot{\mathfrak{A}}^m_{\Theta_{\ddot{F_2}}}(\ddot{\ddot{k}}))^2 + (\ddot{\ddot{\Xi}}^o_{\Theta_{\ddot{F_1}}}(\ddot{\ddot{k}}) - \ddot{\ddot{\Xi}}^o_{\Theta_{\ddot{F_2}}}(\ddot{\ddot{k}}))^2 + (\mathfrak{M}^n_{\Theta_{\ddot{F_1}}}(\ddot{\ddot{k}}) - \mathfrak{M}^n_{\Theta_{\ddot{F_2}}}(\ddot{\ddot{k}}))^2\right)} + \left|\ddot{r_1}(\ddot{\ddot{k}}) - \ddot{r_2}(\ddot{\ddot{k}})\right|\right) \quad (8)$$

**Definition 15**. The definition of the normalized Hamming distance between two $GO_{mno}$-TSFSs $\Theta_{\ddot{F_1}}$ and $\Theta_{\ddot{F_2}}$ in $\ddot{\ddot{K}} = \left\{\ddot{\ddot{k}}_1, \ddot{\ddot{k}}_2, ..., \ddot{\ddot{k}}_l\right\}$, is as follows:

$$\Pi_2(\Theta_{\ddot{F_1}}, \Theta_{\ddot{F_2}}) = \frac{1}{2l}\left(\sum_{i=1}^{l}\left(\frac{1}{2}\left(\left|\ddot{\mathfrak{A}}^m_{\Theta_{\ddot{F_1}}}(\ddot{\ddot{k}}_i) - \ddot{\mathfrak{A}}^m_{\Theta_{\ddot{F_2}}}(\ddot{\ddot{k}}_i)\right| + \left|\ddot{\ddot{\Xi}}^o_{\Theta_{\ddot{F_1}}}(\ddot{\ddot{k}}_i) - \ddot{\ddot{\Xi}}^o_{\Theta_{\ddot{F_2}}}(\ddot{\ddot{k}}_i)\right| + \left|\mathfrak{M}^n_{\Theta_{\ddot{F_1}}}(\ddot{\ddot{k}}_i) - \mathfrak{M}^n_{\Theta_{\ddot{F_2}}}(\ddot{\ddot{k}}_i)\right|\right) + \left|\ddot{r_1}(\ddot{\ddot{k}}_i) - \ddot{r_2}(\ddot{\ddot{k}}_i)\right|\right)\right) \quad (9)$$

**Definition 16**. We define the normalized Euclidean distance between two $GO_{mno}$-TSFSs $\Theta_{\ddot{F_1}}$ and $\Theta_{\ddot{F_2}}$ in $\ddot{\ddot{K}} = \left\{\ddot{\ddot{k}}_1, \ddot{\ddot{k}}_2, ..., \ddot{\ddot{k}}_l\right\}$, as:

$$\Omega_2(\Theta_{\ddot{F_1}}, \Theta_{\ddot{F_2}}) = \frac{1}{2}\left(\sqrt{\frac{1}{2l}\sum_{i=1}^{l}\left((\ddot{\mathfrak{A}}^m_{\Theta_{\ddot{F_1}}}(\ddot{\ddot{k}}_i) - \ddot{\mathfrak{A}}^m_{\Theta_{\ddot{F_2}}}(\ddot{\ddot{k}}_i))^2 + (\ddot{\ddot{\Xi}}^o_{\Theta_{\ddot{F_1}}}(\ddot{\ddot{k}}_i) - \ddot{\ddot{\Xi}}^o_{\Theta_{\ddot{F_2}}}(\ddot{\ddot{k}}_i))^2 + (\mathfrak{M}^n_{\Theta_{\ddot{F_1}}}(\ddot{\ddot{k}}_i) - \mathfrak{M}^n_{\Theta_{\ddot{F_2}}}(\ddot{\ddot{k}}_i))^2\right)} + \frac{1}{l}\sum_{i=1}^{l}\left|\ddot{r_1}(\ddot{\ddot{k}}_i) - \ddot{r_2}(\ddot{\ddot{k}}_i)\right|\right) \quad (10)$$

## 4. HAMCHER AOs for $GO_{mno}$-TSFSs:

### 4.1. Hamacher operations:

T-norms $\left(\underline{T}\right)$ and t-conorms $\left(\underline{\underline{T}}\right)$ are fundamental concepts in FS theory, employed to define generalized union and intersection operations for fuzzy [17]. Roychowdhury and Wang [23] provided the formal definitions and conditions for $\underline{T}$ and $\underline{\underline{T}}$. Building upon $\underline{T}$ and $\underline{\underline{T}}$, Deschrijver and Kerre [9] introduced a generalized union and intersection for IFS. Furthermore, Hamacher [14] proposed a more comprehensive set of $\underline{T}$ and $\underline{\underline{T}}$ operations. The Hamacher $\underline{T}$ and $\underline{\underline{T}}$, which include the Hamacher product and Hamacher sum, respectively, serve as examples of these generalized set operations. The definitions of these Hamacher operations are as follows:

Suppose $d^*, e^*$ as the element of real number, where $\varsigma > 0$.

The Hamacher product denoted as $\otimes$ serves as an example of a $\underline{T}$ operation, while the Hamacher sum denoted as $\oplus$ functions as an example of a $\underset{\sim}{T}$ operation. We have:

$$\underline{T}\left(d^*, e^*\right) = d^* \otimes e^* = \left( \frac{d^* e^*}{\varsigma + \left(1 - \varsigma\right)\left(d^* + e^* - d^* e^*\right)} \right) \tag{11}$$

$$\underset{\sim}{T}\left(d^*, e^*\right) = d^* \oplus e^* = \left( \frac{d^* + e^* - d^* e^* - \left(1 - \varsigma\right) d^* e^*}{1 - \left(1 - \varsigma\right) d^* e^*} \right) \tag{12}$$

In particular, when $\varsigma = 1$ Hamcher's $\underline{T}$ and $\underset{\sim}{T}$ are reduced to

$$\underline{T}\left(d^*, e^*\right) = d^* \otimes e^* = d^* e^*, \tag{13}$$

$$\underset{\sim}{T}\left(d^*, e^*\right) = d^* \oplus e^* = d^* + e^* - d^* e^*, \tag{14}$$

In particular, these are the algebraic $\underline{T}$ and $\underset{\sim}{T}$, respectively. Specifically, when $\varsigma = 2$, the Hamacher $\underline{T}$ and $\underset{\sim}{T}$ will be equivalent to

$$\underline{T}\left(d^*, e^*\right) = d^* \otimes e^* = \frac{d^* e^*}{1 - \left(1 - d^*\right)\left(1 - e^*\right)}, \tag{15}$$

$$\underset{\sim}{T}\left(d^*, e^*\right) = d^* \oplus e^* = \frac{d^* + e^*}{1 + d^* e^*}, \tag{16}$$

Hamacher $\underline{T}$ and $\underset{\sim}{T}$, are respectively known as [9].

### 4.2. Hamcher operations for GO$_{mno}$-TSFSs:

In this section, we evaluate the following operators, GO$_{mno}$-TSF Hamacher weighted averaging (GO$_{mno}$-TSFHWA), and GO$_{mno}$-TSF Hamacher weighted geometric (GO$_{mno}$-TSFHWG), operators, Inspired by Eqs. 11-12, we define the $\underset{\sim}{T}\ H$ and $\underset{\sim}{T}\ S$ as the Hamacher product $\underset{\sim}{H}$ and Hamacher sum $\underset{\sim}{S}$ respectively. Consequently, the generalized intersection and union of two GO$_{mno}$-TSFSs, denoted as A and B, are represented by the Hamacher product $\left(\underline{a}_1 \otimes \underline{a}_2\right)$ and Hamacher sum $\left(\underline{a}_1 \oplus \underline{a}_2\right)$ on two GO$_{mno}$-TSFSs $\underline{a}_1$ and $\underline{a}_2$ respectively, outlined as follows.

$$
\underset{\sim}{\tilde{\underline{\tau}}} \oplus \underset{\sim}{\tilde{\underline{\iota}}} = \left(
\begin{array}{cc}
\sqrt[m]{\dfrac{\ddot{\mathfrak{A}}_{\tilde{\tau}}^m + \ddot{\mathfrak{A}}_{\tilde{\iota}}^m - \ddot{\mathfrak{A}}_{\tilde{\tau}}^m \ddot{\mathfrak{A}}_{\tilde{\iota}}^m - \left(1-\varsigma\right)\ddot{\mathfrak{A}}_{\tilde{\tau}}^m \ddot{\mathfrak{A}}_{\tilde{\iota}}^m}{1-\left(1-\varsigma\right)\ddot{\mathfrak{A}}_{\tilde{\tau}}^m \ddot{\mathfrak{A}}_{\tilde{\iota}}^m}}, & \dfrac{\ddot{\Xi}_{\tilde{\tau}}\ddot{\Xi}_{\tilde{\iota}}}{\sqrt[o]{\varsigma-\left(1-\varsigma\right)\left(\ddot{\Xi}_{\tilde{\tau}}^o + \ddot{\Xi}_{\tilde{\iota}}^o - \ddot{\Xi}_{\tilde{\tau}}^o \ddot{\Xi}_{\tilde{\iota}}^o\right)}}, \\[20pt]
\dfrac{\ddot{\mathfrak{M}}_{\tilde{\tau}}\ddot{\mathfrak{M}}_{\tilde{\iota}}}{\sqrt[n]{\varsigma-\left(1-\varsigma\right)\left(\ddot{\mathfrak{M}}_{\tilde{\tau}}^n + \ddot{\mathfrak{M}}_{\tilde{\iota}}^n - \ddot{\mathfrak{M}}_{\tilde{\tau}}^n \ddot{\mathfrak{M}}_{\tilde{\iota}}^n\right)}}, & \sqrt[o]{\dfrac{\ddot{r}_{\tilde{\tau}}^o + \ddot{r}_{\tilde{\iota}}^o - \ddot{r}_{\tilde{\tau}}^o \ddot{r}_{\tilde{\iota}}^o - \left(1-\varsigma\right)\ddot{r}_{\tilde{\tau}}^o \ddot{r}_{\tilde{\iota}}^o}{1-\left(1-\varsigma\right)\ddot{r}_{\tilde{\tau}}^o \ddot{r}_{\tilde{\iota}}^o}}
\end{array}
\right) \tag{17}
$$

$$
\underset{\sim}{\tilde{\underline{\tau}}} \otimes \underset{\sim}{\tilde{\underline{\iota}}} = \left(
\begin{array}{cc}
\dfrac{\ddot{\mathfrak{A}}_{\tilde{\tau}}\ddot{\mathfrak{A}}_{\tilde{\iota}}}{\sqrt[m]{\varsigma-\left(1-\varsigma\right)\left(\ddot{\mathfrak{A}}_{\tilde{\tau}}^m + \ddot{\mathfrak{A}}_{\tilde{\iota}}^m - \ddot{\mathfrak{A}}_{\tilde{\tau}}^m \ddot{\mathfrak{A}}_{\tilde{\iota}}^m\right)}}, & \sqrt[o]{\dfrac{\ddot{\Xi}_{\tilde{\tau}}^o + \ddot{\Xi}_{\tilde{\iota}}^o - \ddot{\Xi}_{\tilde{\tau}}^o \ddot{\Xi}_{\tilde{\iota}}^o - \left(1-\varsigma\right)\ddot{\Xi}_{\tilde{\tau}}^o \ddot{\Xi}_{\tilde{\iota}}^o}{1-\left(1-\varsigma\right)\ddot{\Xi}_{\tilde{\tau}}^o \ddot{\Xi}_{\tilde{\iota}}^o}}, \\[20pt]
\sqrt[n]{\dfrac{\ddot{\mathfrak{M}}_{\tilde{\tau}}^n + \ddot{\mathfrak{M}}_{\tilde{\iota}}^n - \ddot{\mathfrak{M}}_{\tilde{\tau}}^n \ddot{\mathfrak{M}}_{\tilde{\iota}}^n - \left(1-\varsigma\right)\ddot{\mathfrak{M}}_{\tilde{\tau}}^n \ddot{\mathfrak{M}}_{\tilde{\iota}}^n}{1-\left(1-\varsigma\right)\ddot{\mathfrak{M}}_{\tilde{\tau}}^n \ddot{\mathfrak{M}}_{\tilde{\iota}}^n}}, & \dfrac{\ddot{r}_{\tilde{\tau}}\ddot{r}_{\tilde{\iota}}}{\sqrt[o]{\varsigma-\left(1-\varsigma\right)\left(\ddot{r}_{\tilde{\tau}}^o + \ddot{r}_{\tilde{\iota}}^o - \ddot{r}_{\tilde{\tau}}^o \ddot{r}_{\tilde{\iota}}^o\right)}}
\end{array}
\right) \tag{18}
$$

$$
\underset{\sim}{\delta \tilde{\underline{\tau}}} = \left(
\begin{array}{cc}
\sqrt[m]{\dfrac{\left(1+\left(\varsigma-1\right)\ddot{\mathfrak{A}}_{\tilde{\tau}}^m\right)^{\delta} - \left(1-\ddot{\mathfrak{A}}_{\tilde{\tau}}^m\right)^{\delta}}{\left(1+\left(\varsigma-1\right)\ddot{\mathfrak{A}}_{\tilde{\tau}}^m\right)^{\delta} + \left(\varsigma-1\right)\left(1-\ddot{\mathfrak{A}}_{\tilde{\tau}}^m\right)^{\delta}}}, & \dfrac{\sqrt[o]{\varsigma}\ddot{\Xi}_{\tilde{\tau}}^{\delta}}{\sqrt[o]{\left(1+\left(\varsigma-1\right)\left(1-\ddot{\Xi}_{\tilde{\tau}}^o\right)\right)^{\delta} + \left(\varsigma-1\right)\left(\ddot{\Xi}_{\tilde{\tau}}^o\right)^{\delta}}}, \\[20pt]
\dfrac{\sqrt[n]{\varsigma}\ddot{\mathfrak{M}}_{\tilde{\tau}}^{\delta}}{\sqrt[n]{\left(1+\left(\varsigma-1\right)\left(1-\ddot{\mathfrak{M}}_{\tilde{\tau}}^n\right)\right)^{\delta} + \left(\varsigma-1\right)\left(\ddot{\mathfrak{M}}_{\tilde{\tau}}^n\right)^{\delta}}}, & \sqrt[o]{\dfrac{\left(1+\left(\varsigma-1\right)\ddot{r}_{\tilde{\tau}}^o\right)^{\delta} - \left(1-\ddot{r}_{\tilde{\tau}}^o\right)^{\delta}}{\left(1+\left(\varsigma-1\right)\ddot{r}_{\tilde{\tau}}^o\right)^{\delta} + \left(\varsigma-1\right)\left(1-\ddot{r}_{\tilde{\tau}}^o\right)^{\delta}}}
\end{array}
\right), \delta \in \left[0,1\right] \tag{19}
$$

$$
\underset{\sim}{\tilde{\underline{\tau}}}^{\delta} = \left(
\begin{array}{cc}
\dfrac{\sqrt[m]{\varsigma}\ddot{\mathfrak{A}}_{\tilde{\tau}}^{\delta}}{\sqrt[m]{\left(1+\left(\varsigma-1\right)\left(1-\ddot{\mathfrak{A}}_{\tilde{\tau}}^m\right)\right)^{\delta} + \left(\varsigma-1\right)\left(\ddot{\mathfrak{A}}_{\tilde{\tau}}^m\right)^{\delta}}}, & \sqrt[o]{\dfrac{\left(1+\left(\varsigma-1\right)\ddot{\Xi}_{\tilde{\tau}}^o\right)^{\delta} - \left(1-\ddot{\Xi}_{\tilde{\tau}}^o\right)^{\delta}}{\left(1+\left(\varsigma-1\right)\ddot{\Xi}_{\tilde{\tau}}^o\right)^{\delta} + \left(\varsigma-1\right)\left(1-\ddot{\Xi}_{\tilde{\tau}}^o\right)^{\delta}}}, \\[20pt]
\sqrt[n]{\dfrac{\left(1+\left(\varsigma-1\right)\ddot{\mathfrak{M}}_{\tilde{\tau}}^n\right)^{\delta} - \left(1-\ddot{\mathfrak{M}}_{\tilde{\tau}}^n\right)^{\delta}}{\left(1+\left(\varsigma-1\right)\ddot{\mathfrak{M}}_{\tilde{\tau}}^n\right)^{\delta} + \left(\varsigma-1\right)\left(1-\ddot{\mathfrak{M}}_{\tilde{\tau}}^n\right)^{\delta}}}, & \dfrac{\sqrt[o]{\varsigma}\ddot{r}_{\tilde{\tau}}^{\delta}}{\sqrt[o]{\left(1+\left(\varsigma-1\right)\left(1-\ddot{r}_{\tilde{\tau}}^o\right)\right)^{\delta} + \left(\varsigma-1\right)\left(\ddot{r}_{\tilde{\tau}}^o\right)^{\delta}}}
\end{array}
\right), \delta \in \left[0,1\right] \tag{20}
$$

### 4.3. GO$_{mno}$-TSF Hamacher weighted averaging operator

**Definition 17.** Assuming $\tilde{z}_i = \left\langle \ddot{\mathfrak{A}}_{\tilde{z}_i}, \ddot{\Xi}_{\tilde{z}_i}, \ddot{\mathfrak{M}}_{\tilde{z}_i}; \ddot{r}_{\tilde{z}_i} \right\rangle (i=1,2,...,n)$ be the assembly of GO$_{mno}$-TSFVs. The GO$_{mno}$-TSF Hamacher weighted averaging (GO$_{mno}$-TSFHWA) operator is defined as a mapping $\tilde{P}^n \rightarrow \tilde{P}$, Characterized as:

$$\text{GO}_{mno}\text{-TSFHWA}_{\tilde{\omega}}\left(\tilde{z}_1, \tilde{z}_2, ..., \tilde{z}_n\right) = \oplus_{i=1}^n \left(\tilde{\omega}_i \tilde{z}_i\right), \tag{21}$$

Where $\tilde{\omega} = \left(\tilde{\omega}_1, \tilde{\omega}_2, ..., \tilde{\omega}_n\right)$ represent the weight vector of $\tilde{z}_i (i=1,2,...,n)$ and $\tilde{\omega}_i > 0, \sum_{i=1}^n \tilde{\omega}_i = 1$.

**Theorem 12:** The aggregation value obtained using the GO$_{mno}$-TSFHWA operator is also a GO$_{mno}$-TSFV, where

$$\text{GO}_{mno}\text{-TSFHWA}_{\tilde{\omega}}\left(\tilde{z}_1, \tilde{z}_2, ..., \tilde{z}_n\right) = \left( \begin{array}{cc} \sqrt[m]{\dfrac{\prod_{i=1}^n\left(1+(\varsigma-1)\ddot{\mathfrak{A}}_{\tilde{z}_i}^m\right)^{\tilde{\omega}_i} - \prod_{i=1}^n\left(1-\ddot{\mathfrak{A}}_{\tilde{z}_i}^m\right)^{\tilde{\omega}_i}}{\prod_{i=1}^n\left(1+(\varsigma-1)\ddot{\mathfrak{A}}_{\tilde{z}_i}^m\right)^{\tilde{\omega}_i} + (\varsigma-1)\prod_{i=1}^n\left(1-\ddot{\mathfrak{A}}_{\tilde{z}_i}^m\right)^{\tilde{\omega}_i}}}, & \dfrac{\sqrt[n]{\varsigma}\prod_{i=1}^n \ddot{\Xi}_{\tilde{z}_i}^{\tilde{\omega}_i}}{\sqrt[n]{\prod_{i=1}^n\left(1+(\varsigma-1)\left(1-\ddot{\Xi}_{\tilde{z}_i}^n\right)\right)^{\tilde{\omega}_i} + (\varsigma-1)\prod_{i=1}^n\left(\ddot{\Xi}_{\tilde{z}_i}^n\right)^{\tilde{\omega}_i}}}, \\ \dfrac{\sqrt[n]{\varsigma}\prod_{i=1}^n \ddot{\mathfrak{M}}_{\tilde{z}_i}^{\tilde{\omega}_i}}{\sqrt[n]{\prod_{i=1}^n\left(1+(\varsigma-1)\left(1-\ddot{\mathfrak{M}}_{\tilde{z}_i}^n\right)\right)^{\tilde{\omega}_i} + (\varsigma-1)\prod_{i=1}^n\left(\ddot{\mathfrak{M}}_{\tilde{z}_i}^n\right)^{\tilde{\omega}_i}}}, & \sqrt[n]{\dfrac{\prod_{i=1}^n\left(1+(\varsigma-1)\ddot{r}_{\tilde{z}_i}^n\right)^{\tilde{\omega}_i} - \prod_{i=1}^n\left(1-\ddot{r}_{\tilde{z}_i}^n\right)^{\tilde{\omega}_i}}{\prod_{i=1}^n\left(1+(\varsigma-1)\ddot{r}_{\tilde{z}_i}^n\right)^{\tilde{\omega}_i} + (\varsigma-1)\prod_{i=1}^n\left(1-\ddot{r}_{\tilde{z}_i}^n\right)^{\tilde{\omega}_i}}} \end{array} \right) \tag{22}$$

Here $\tilde{\omega} = \left(\tilde{\omega}_1, \tilde{\omega}_2, ..., \tilde{\omega}_n\right)$ represent the weight vector of $\tilde{z}_i (i=1,2,...,n)$ and $\tilde{\omega}_i > 0, \sum_{i=1}^n \tilde{\omega}_i = 1$.

**Proof:** We proof Eq. (22) by mathematical induction on $n$.

(a) When $n=2$ we have $\text{GO}_{mno}\text{-TSFHWA}_{\tilde{\omega}}\left(\tilde{z}_1, \tilde{z}_2\right) = \tilde{\omega}_1 \tilde{z}_1 \oplus \tilde{\omega}_2 \tilde{z}_2$. By Theorem 3, we can see that both $\tilde{\omega}_1 \tilde{z}_1$ and $\tilde{\omega}_2 \tilde{z}_2$ are GO$_{mno}$-TSFVs, and the value of $\tilde{\omega}_1 \tilde{z}_1 \oplus \tilde{\omega}_2 \tilde{z}_2$ is also a GO$_{mno}$-TSFV. From the operational laws of GO$_{mno}$-TSFV, we have

$$\tilde{\omega}_1\tilde{\underline{\tau}}_1 = \left( \begin{array}{c} \sqrt[m]{\dfrac{\left(1+\left(\varsigma-1\right)\ddot{\ddot{\mathfrak{A}}}_{\tilde{\underline{\tau}}_1}^m\right)^{\tilde{\omega}_1}-\left(1-\ddot{\ddot{\mathfrak{A}}}_{\tilde{\underline{\tau}}_1}^m\right)^{\tilde{\omega}_1}}{\left(1+\left(\varsigma-1\right)\ddot{\ddot{\mathfrak{A}}}_{\tilde{\underline{\tau}}_1}^m\right)^{\tilde{\omega}_1}+\left(\varsigma-1\right)\left(1-\ddot{\ddot{\mathfrak{A}}}_{\tilde{\underline{\tau}}_1}^m\right)^{\tilde{\omega}_1}}}, \quad \dfrac{\sqrt[o]{\varsigma}\ddot{\ddot{\Xi}}_{\tilde{\underline{\tau}}_1}^{\tilde{\omega}_1}}{\sqrt[o]{\left(1+\left(\varsigma-1\right)\left(1-\ddot{\ddot{\Xi}}_{\tilde{\underline{\tau}}_1}^o\right)\right)^{\tilde{\omega}_1}+\left(\varsigma-1\right)\left(\ddot{\ddot{\Xi}}_{\tilde{\underline{\tau}}_1}^o\right)^{\tilde{\omega}_1}}}, \\[3em] \dfrac{\sqrt[n]{\varsigma}\ddot{\ddot{\mathfrak{M}}}_{\tilde{\underline{\tau}}_1}^{\tilde{\omega}_1}}{\sqrt[n]{\left(1+\left(\varsigma-1\right)\left(1-\ddot{\ddot{\mathfrak{M}}}_{\tilde{\underline{\tau}}_1}^n\right)\right)^{\tilde{\omega}_1}+\left(\varsigma-1\right)\left(\ddot{\ddot{\mathfrak{M}}}_{\tilde{\underline{\tau}}_1}^n\right)^{\tilde{\omega}_1}}}, \quad \sqrt[o]{\dfrac{\left(1+\left(\varsigma-1\right)\ddot{\ddot{\imath}}_{\tilde{\underline{\tau}}_1}^o\right)^{\tilde{\omega}_1}-\left(1-\ddot{\ddot{\imath}}_{\tilde{\underline{\tau}}_1}^o\right)^{\tilde{\omega}_1}}{\left(1+\left(\varsigma-1\right)\ddot{\ddot{\imath}}_{\tilde{\underline{\tau}}_1}^o\right)^{\tilde{\omega}_1}+\left(\varsigma-1\right)\left(1-\ddot{\ddot{\imath}}_{\tilde{\underline{\tau}}_1}^o\right)^{\tilde{\omega}_1}}} \end{array} \right),$$

$$\tilde{\omega}_2\tilde{\underline{\tau}}_2 = \left( \begin{array}{c} \sqrt[m]{\dfrac{\left(1+\left(\varsigma-1\right)\ddot{\ddot{\mathfrak{A}}}_{\tilde{\underline{\tau}}_2}^m\right)^{\tilde{\omega}_2}-\left(1-\ddot{\ddot{\mathfrak{A}}}_{\tilde{\underline{\tau}}_2}^m\right)^{\tilde{\omega}_2}}{\left(1+\left(\varsigma-1\right)\ddot{\ddot{\mathfrak{A}}}_{\tilde{\underline{\tau}}_2}^m\right)^{\tilde{\omega}_2}+\left(\varsigma-1\right)\left(1-\ddot{\ddot{\mathfrak{A}}}_{\tilde{\underline{\tau}}_2}^m\right)^{\tilde{\omega}_2}}}, \quad \dfrac{\sqrt[o]{\varsigma}\ddot{\ddot{\Xi}}_{\tilde{\underline{\tau}}_2}^{\tilde{\omega}_2}}{\sqrt[o]{\left(1+\left(\varsigma-1\right)\left(1-\ddot{\ddot{\Xi}}_{\tilde{\underline{\tau}}_2}^o\right)\right)^{\tilde{\omega}_2}+\left(\varsigma-1\right)\left(\ddot{\ddot{\Xi}}_{\tilde{\underline{\tau}}_2}^o\right)^{\tilde{\omega}_2}}}, \\[3em] \dfrac{\sqrt[n]{\varsigma}\ddot{\ddot{\mathfrak{M}}}_{\tilde{\underline{\tau}}_2}^{\tilde{\omega}_2}}{\sqrt[n]{\left(1+\left(\varsigma-1\right)\left(1-\ddot{\ddot{\mathfrak{M}}}_{\tilde{\underline{\tau}}_2}^n\right)\right)^{\tilde{\omega}_2}+\left(\varsigma-1\right)\left(\ddot{\ddot{\mathfrak{M}}}_{\tilde{\underline{\tau}}_2}^n\right)^{\tilde{\omega}_2}}}, \quad \sqrt[o]{\dfrac{\left(1+\left(\varsigma-1\right)\ddot{\ddot{\imath}}_{\tilde{\underline{\tau}}_2}^o\right)^{\tilde{\omega}_2}-\left(1-\ddot{\ddot{\imath}}_{\tilde{\underline{\tau}}_2}^o\right)^{\tilde{\omega}_2}}{\left(1+\left(\varsigma-1\right)\ddot{\ddot{\imath}}_{\tilde{\underline{\tau}}_2}^o\right)^{\tilde{\omega}_2}+\left(\varsigma-1\right)\left(1-\ddot{\ddot{\imath}}_{\tilde{\underline{\tau}}_2}^o\right)^{\tilde{\omega}_2}}} \end{array} \right).$$

Then, $\mathrm{GO_{mno}TSFHWA}_{\tilde{\omega}}\left(\tilde{\underline{\tau}}_1,\tilde{\underline{\tau}}_2\right)=\tilde{\omega}_1\tilde{\underline{\tau}}_1\oplus\tilde{\omega}_2\tilde{\underline{\tau}}_2$

$$= \left( \begin{array}{c} \sqrt[m]{\dfrac{\left(1+\left(\varsigma-1\right)\ddot{\ddot{\mathfrak{A}}}_{\tilde{\underline{\tau}}_1}^m\right)^{\tilde{\omega}_1}-\left(1-\ddot{\ddot{\mathfrak{A}}}_{\tilde{\underline{\tau}}_1}^m\right)^{\tilde{\omega}_1}}{\left(1+\left(\varsigma-1\right)\ddot{\ddot{\mathfrak{A}}}_{\tilde{\underline{\tau}}_1}^m\right)^{\tilde{\omega}_1}+\left(\varsigma-1\right)\left(1-\ddot{\ddot{\mathfrak{A}}}_{\tilde{\underline{\tau}}_1}^m\right)^{\tilde{\omega}_1}}}, \quad \dfrac{\sqrt[o]{\varsigma}\ddot{\ddot{\Xi}}_{\tilde{\underline{\tau}}_1}^{\tilde{\omega}_1}}{\sqrt[o]{\left(1+\left(\varsigma-1\right)\left(1-\ddot{\ddot{\Xi}}_{\tilde{\underline{\tau}}_1}^o\right)\right)^{\tilde{\omega}_1}+\left(\varsigma-1\right)\left(\ddot{\ddot{\Xi}}_{\tilde{\underline{\tau}}_1}^o\right)^{\tilde{\omega}_1}}}, \\[3em] \dfrac{\sqrt[n]{\varsigma}\ddot{\ddot{\mathfrak{M}}}_{\tilde{\underline{\tau}}_1}^{\tilde{\omega}_1}}{\sqrt[n]{\left(1+\left(\varsigma-1\right)\left(1-\ddot{\ddot{\mathfrak{M}}}_{\tilde{\underline{\tau}}_1}^n\right)\right)^{\tilde{\omega}_1}+\left(\varsigma-1\right)\left(\ddot{\ddot{\mathfrak{M}}}_{\tilde{\underline{\tau}}_1}^n\right)^{\tilde{\omega}_1}}}, \quad \sqrt[o]{\dfrac{\left(1+\left(\varsigma-1\right)\ddot{\ddot{\imath}}_{\tilde{\underline{\tau}}_1}^o\right)^{\tilde{\omega}_1}-\left(1-\ddot{\ddot{\imath}}_{\tilde{\underline{\tau}}_1}^o\right)^{\tilde{\omega}_1}}{\left(1+\left(\varsigma-1\right)\ddot{\ddot{\imath}}_{\tilde{\underline{\tau}}_1}^o\right)^{\tilde{\omega}_1}+\left(\varsigma-1\right)\left(1-\ddot{\ddot{\imath}}_{\tilde{\underline{\tau}}_1}^o\right)^{\tilde{\omega}_1}}} \end{array} \right) \oplus$$

$$\left( \begin{array}{c} \sqrt[m]{\dfrac{\left(1+\left(\varsigma-1\right)\ddot{\ddot{\mathfrak{A}}}_{\tilde{\underline{\tau}}_2}^m\right)^{\tilde{\omega}_2}-\left(1-\ddot{\ddot{\mathfrak{A}}}_{\tilde{\underline{\tau}}_2}^m\right)^{\tilde{\omega}_2}}{\left(1+\left(\varsigma-1\right)\ddot{\ddot{\mathfrak{A}}}_{\tilde{\underline{\tau}}_2}^m\right)^{\tilde{\omega}_2}+\left(\varsigma-1\right)\left(1-\ddot{\ddot{\mathfrak{A}}}_{\tilde{\underline{\tau}}_2}^m\right)^{\tilde{\omega}_2}}}, \quad \dfrac{\sqrt[o]{\varsigma}\ddot{\ddot{\Xi}}_{\tilde{\underline{\tau}}_2}^{\tilde{\omega}_2}}{\sqrt[o]{\left(1+\left(\varsigma-1\right)\left(1-\ddot{\ddot{\Xi}}_{\tilde{\underline{\tau}}_2}^o\right)\right)^{\tilde{\omega}_2}+\left(\varsigma-1\right)\left(\ddot{\ddot{\Xi}}_{\tilde{\underline{\tau}}_2}^o\right)^{\tilde{\omega}_2}}}, \\[3em] \dfrac{\sqrt[n]{\varsigma}\ddot{\ddot{\mathfrak{M}}}_{\tilde{\underline{\tau}}_2}^{\tilde{\omega}_2}}{\sqrt[n]{\left(1+\left(\varsigma-1\right)\left(1-\ddot{\ddot{\mathfrak{M}}}_{\tilde{\underline{\tau}}_2}^n\right)\right)^{\tilde{\omega}_2}+\left(\varsigma-1\right)\left(\ddot{\ddot{\mathfrak{M}}}_{\tilde{\underline{\tau}}_2}^n\right)^{\tilde{\omega}_2}}}, \quad \sqrt[o]{\dfrac{\left(1+\left(\varsigma-1\right)\ddot{\ddot{\imath}}_{\tilde{\underline{\tau}}_2}^o\right)^{\tilde{\omega}_2}-\left(1-\ddot{\ddot{\imath}}_{\tilde{\underline{\tau}}_2}^o\right)^{\tilde{\omega}_2}}{\left(1+\left(\varsigma-1\right)\ddot{\ddot{\imath}}_{\tilde{\underline{\tau}}_2}^o\right)^{\tilde{\omega}_2}+\left(\varsigma-1\right)\left(1-\ddot{\ddot{\imath}}_{\tilde{\underline{\tau}}_2}^o\right)^{\tilde{\omega}_2}}} \end{array} \right)$$

$$= \left( \begin{array}{cc} \sqrt[m]{\dfrac{\prod_{i=1}^{2}\left(1+(\varsigma-1)\ddddot{\mathfrak{A}}_{\bar{\xi}_i}^{m}\right)^{\tilde{\omega}_i} - \prod_{i=1}^{2}\left(1-\ddddot{\mathfrak{A}}_{\bar{\xi}_i}^{m}\right)^{\tilde{\omega}_i}}{\prod_{i=1}^{2}\left(1+(\varsigma-1)\ddddot{\mathfrak{A}}_{\bar{\xi}_i}^{m}\right)^{\tilde{\omega}_i} + (\varsigma-1)\prod_{i=1}^{2}\left(1-\ddddot{\mathfrak{A}}_{\bar{\xi}_i}^{m}\right)^{\tilde{\omega}_i}}}, & \dfrac{\sqrt[o]{\varsigma}\prod_{i=1}^{2}\ddddot{\Xi}_{\bar{\xi}_i}^{\tilde{\omega}_i}}{\sqrt[o]{\prod_{i=1}^{2}\left(1+(\varsigma-1)\left(1-\ddddot{\Xi}_{\bar{\xi}_i}^{o}\right)\right)^{\tilde{\omega}_i} + (\varsigma-1)\prod_{i=1}^{2}\left(\ddddot{\Xi}_{\bar{\xi}_i}^{o}\right)^{\tilde{\omega}_i}}}, \\[3em] \dfrac{\sqrt[n]{\varsigma}\prod_{i=1}^{2}\ddddot{\mathfrak{N}}_{\bar{\xi}_i}^{\tilde{\omega}_i}}{\sqrt[n]{\prod_{i=1}^{2}\left(1+(\varsigma-1)\left(1-\ddddot{\mathfrak{N}}_{\bar{\xi}_i}^{n}\right)\right)^{\tilde{\omega}_i} + (\varsigma-1)\prod_{i=1}^{2}\left(\ddddot{\mathfrak{N}}_{\bar{\xi}_i}^{n}\right)^{\tilde{\omega}_i}}}, & \sqrt[o]{\dfrac{\prod_{i=1}^{2}\left(1+(\varsigma-1)\ddot{r}_{\bar{\xi}_i}^{o}\right)^{\tilde{\omega}_i} - \prod_{i=1}^{2}\left(1-\ddot{r}_{\bar{\xi}_i}^{o}\right)^{\tilde{\omega}_i}}{\prod_{i=1}^{2}\left(1+(\varsigma-1)\ddot{r}_{\bar{\xi}_i}^{o}\right)^{\tilde{\omega}_i} + (\varsigma-1)\prod_{i=1}^{2}\left(1-\ddot{r}_{\bar{\xi}_i}^{o}\right)^{\tilde{\omega}_i}}} \end{array} \right)$$

(b) Assume that $n = \tilde{l}$, Equation (22) holds, i.e.

$$\text{GO}_{mno}\text{-TSFHWA}_{\tilde{\omega}}\left(\tilde{\underline{\xi}}_1, \tilde{\underline{\xi}}_2, ..., \tilde{\underline{\xi}}_{\tilde{l}}\right) = \tilde{\omega}_1\tilde{\underline{\xi}}_1 \oplus \tilde{\omega}_2\tilde{\underline{\xi}}_2 \oplus ... \oplus \tilde{\omega}_{\tilde{l}}\tilde{\underline{\xi}}_{\tilde{l}}$$

$$= \left( \begin{array}{cc} \sqrt[m]{\dfrac{\prod_{i=1}^{\tilde{l}}\left(1+(\varsigma-1)\ddddot{\mathfrak{A}}_{\bar{\xi}_i}^{m}\right)^{\tilde{\omega}_i} - \prod_{i=1}^{\tilde{l}}\left(1-\ddddot{\mathfrak{A}}_{\bar{\xi}_i}^{m}\right)^{\tilde{\omega}_i}}{\prod_{i=1}^{\tilde{l}}\left(1+(\varsigma-1)\ddddot{\mathfrak{A}}_{\bar{\xi}_i}^{m}\right)^{\tilde{\omega}_i} + (\varsigma-1)\prod_{i=1}^{\tilde{l}}\left(1-\ddddot{\mathfrak{A}}_{\bar{\xi}_i}^{m}\right)^{\tilde{\omega}_i}}}, & \dfrac{\sqrt[o]{\varsigma}\prod_{i=1}^{\tilde{l}}\ddddot{\Xi}_{\bar{\xi}_i}^{\tilde{\omega}_i}}{\sqrt[o]{\prod_{i=1}^{\tilde{l}}\left(1+(\varsigma-1)\left(1-\ddddot{\Xi}_{\bar{\xi}_i}^{o}\right)\right)^{\tilde{\omega}_i} + (\varsigma-1)\prod_{i=1}^{\tilde{l}}\left(\ddddot{\Xi}_{\bar{\xi}_i}^{o}\right)^{\tilde{\omega}_i}}}, \\[3em] \dfrac{\sqrt[n]{\varsigma}\prod_{i=1}^{\tilde{l}}\ddddot{\mathfrak{N}}_{\bar{\xi}_i}^{\tilde{\omega}_i}}{\sqrt[n]{\prod_{i=1}^{\tilde{l}}\left(1+(\varsigma-1)\left(1-\ddddot{\mathfrak{N}}_{\bar{\xi}_i}^{n}\right)\right)^{\tilde{\omega}_i} + (\varsigma-1)\prod_{i=1}^{\tilde{l}}\left(\ddddot{\mathfrak{N}}_{\bar{\xi}_i}^{n}\right)^{\tilde{\omega}_i}}}, & \sqrt[o]{\dfrac{\prod_{i=1}^{\tilde{l}}\left(1+(\varsigma-1)\ddot{r}_{\bar{\xi}_i}^{o}\right)^{\tilde{\omega}_i} - \prod_{i=1}^{\tilde{l}}\left(1-\ddot{r}_{\bar{\xi}_i}^{o}\right)^{\tilde{\omega}_i}}{\prod_{i=1}^{\tilde{l}}\left(1+(\varsigma-1)\ddot{r}_{\bar{\xi}_i}^{o}\right)^{\tilde{\omega}_i} + (\varsigma-1)\prod_{i=1}^{\tilde{l}}\left(1-\ddot{r}_{\bar{\xi}_i}^{o}\right)^{\tilde{\omega}_i}}} \end{array} \right)$$

And the aggregation value is also a GO$_{mno}$-TSFV, and when $n = \tilde{l} + 1$, by the operational laws of GO$_{mno}$-TSFV, we have

$$\text{GO}_{mno}\text{-TSFHWA}_{\tilde{\omega}}\left(\tilde{\underline{\xi}}_1, \tilde{\underline{\xi}}_2, ..., \tilde{\underline{\xi}}_{\tilde{l}}\right) = \tilde{\omega}_1\tilde{\underline{\xi}}_1 \oplus \tilde{\omega}_2\tilde{\underline{\xi}}_2 \oplus ... \oplus \tilde{\omega}_{\tilde{l}}\tilde{\underline{\xi}}_{\tilde{l}} \oplus \tilde{\omega}_{\tilde{l}+1}\tilde{\underline{\xi}}_{\tilde{l}+1}$$

$$= \left( \begin{array}{cc} \sqrt[m]{\dfrac{\prod_{i=1}^{\tilde{l}}\left(1+(\varsigma-1)\ddddot{\mathfrak{A}}_{\bar{\xi}_i}^{m}\right)^{\tilde{\omega}_i} - \prod_{i=1}^{\tilde{l}}\left(1-\ddddot{\mathfrak{A}}_{\bar{\xi}_i}^{m}\right)^{\tilde{\omega}_i}}{\prod_{i=1}^{\tilde{l}}\left(1+(\varsigma-1)\ddddot{\mathfrak{A}}_{\bar{\xi}_i}^{m}\right)^{\tilde{\omega}_i} + (\varsigma-1)\prod_{i=1}^{\tilde{l}}\left(1-\ddddot{\mathfrak{A}}_{\bar{\xi}_i}^{m}\right)^{\tilde{\omega}_i}}}, & \dfrac{\sqrt[o]{\varsigma}\prod_{i=1}^{\tilde{l}}\ddddot{\Xi}_{\bar{\xi}_i}^{\tilde{\omega}_i}}{\sqrt[o]{\prod_{i=1}^{\tilde{l}}\left(1+(\varsigma-1)\left(1-\ddddot{\Xi}_{\bar{\xi}_i}^{o}\right)\right)^{\tilde{\omega}_i} + (\varsigma-1)\prod_{i=1}^{\tilde{l}}\left(\ddddot{\Xi}_{\bar{\xi}_i}^{o}\right)^{\tilde{\omega}_i}}}, \\[3em] \dfrac{\sqrt[n]{\varsigma}\prod_{i=1}^{\tilde{l}}\ddddot{\mathfrak{N}}_{\bar{\xi}_i}^{\tilde{\omega}_i}}{\sqrt[n]{\prod_{i=1}^{\tilde{l}}\left(1+(\varsigma-1)\left(1-\ddddot{\mathfrak{N}}_{\bar{\xi}_i}^{n}\right)\right)^{\tilde{\omega}_i} + (\varsigma-1)\prod_{i=1}^{\tilde{l}}\left(\ddddot{\mathfrak{N}}_{\bar{\xi}_i}^{n}\right)^{\tilde{\omega}_i}}}, & \sqrt[o]{\dfrac{\prod_{i=1}^{\tilde{l}}\left(1+(\varsigma-1)\ddot{r}_{\bar{\xi}_i}^{o}\right)^{\tilde{\omega}_i} - \prod_{i=1}^{\tilde{l}}\left(1-\ddot{r}_{\bar{\xi}_i}^{o}\right)^{\tilde{\omega}_i}}{\prod_{i=1}^{\tilde{l}}\left(1+(\varsigma-1)\ddot{r}_{\bar{\xi}_i}^{o}\right)^{\tilde{\omega}_i} + (\varsigma-1)\prod_{i=1}^{\tilde{l}}\left(1-\ddot{r}_{\bar{\xi}_i}^{o}\right)^{\tilde{\omega}_i}}} \end{array} \right) \oplus$$

$$\left( \sqrt[m]{\frac{\left(1+\left(\varsigma-1\right)\ddot{\mathfrak{A}}_{\tilde{\varepsilon}_{\tilde{l}+1}}^{m}\right)^{\tilde{\omega}_{\tilde{l}+1}}-\left(1-\ddot{\mathfrak{A}}_{\tilde{\varepsilon}_{\tilde{l}+1}}^{m}\right)^{\tilde{\omega}_{\tilde{l}+1}}}{\left(1+\left(\varsigma-1\right)\ddot{\mathfrak{A}}_{\tilde{\varepsilon}_{\tilde{l}+1}}^{m}\right)^{\tilde{\omega}_{\tilde{l}+1}}+\left(\varsigma-1\right)\left(1-\ddot{\mathfrak{A}}_{\tilde{\varepsilon}_{\tilde{l}+1}}^{m}\right)^{\tilde{\omega}_{\tilde{l}+1}}}}, \frac{\sqrt[o]{\varsigma}\,\ddot{\ddot{\Xi}}_{\tilde{\varepsilon}_{\tilde{l}+1}}^{\tilde{\omega}_{\tilde{l}+1}}}{\sqrt[o]{\left(1+\left(\varsigma-1\right)\left(1-\ddot{\ddot{\Xi}}_{\tilde{\varepsilon}_{\tilde{l}+1}}^{o}\right)\right)^{\tilde{\omega}_{\tilde{l}+1}}+\left(\varsigma-1\right)\left(\ddot{\ddot{\Xi}}_{\tilde{\varepsilon}_{\tilde{l}+1}}^{o}\right)^{\tilde{\omega}_{\tilde{l}+1}}}}, \right.$$
$$\left. \frac{\sqrt[n]{\gamma}\,\ddot{\ddot{\mathfrak{M}}}_{\tilde{\varepsilon}_{\tilde{l}+1}}^{\tilde{\omega}_{\tilde{l}+1}}}{\sqrt[n]{\left(1+\left(\varsigma-1\right)\left(1-\ddot{\ddot{\mathfrak{M}}}_{\tilde{\varepsilon}_{\tilde{l}+1}}^{n}\right)\right)^{\tilde{\omega}_{\tilde{l}+1}}+\left(\varsigma-1\right)\left(\ddot{\ddot{\mathfrak{M}}}_{\tilde{\varepsilon}_{\tilde{l}+1}}^{n}\right)^{\tilde{\omega}_{\tilde{l}+1}}}}, \sqrt[o]{\frac{\left(1+\left(\varsigma-1\right)\ddot{r}_{\tilde{\varepsilon}_{\tilde{l}+1}}^{o}\right)^{\tilde{\omega}_{\tilde{l}+1}}-\left(1-\ddot{r}_{\tilde{\varepsilon}_{\tilde{l}+1}}^{o}\right)^{\tilde{\omega}_{\tilde{l}+1}}}{\left(1+\left(\varsigma-1\right)\ddot{r}_{\tilde{\varepsilon}_{\tilde{l}+1}}^{o}\right)^{\tilde{\omega}_{\tilde{l}+1}}+\left(\varsigma-1\right)\left(1-\ddot{r}_{\tilde{\varepsilon}_{\tilde{l}+1}}^{o}\right)^{\tilde{\omega}_{\tilde{l}+1}}}} \right)$$

$$= \left( \sqrt[m]{\frac{\prod_{i=1}^{\tilde{l}+1}\left(1+\left(\varsigma-1\right)\ddot{\mathfrak{A}}_{\tilde{\varepsilon}_{i}}^{m}\right)^{\tilde{\omega}_{i}}-\prod_{i=1}^{\tilde{l}+1}\left(1-\ddot{\mathfrak{A}}_{\tilde{\varepsilon}_{i}}^{m}\right)^{\tilde{\omega}_{i}}}{\prod_{i=1}^{\tilde{l}+1}\left(1+\left(\varsigma-1\right)\ddot{\mathfrak{A}}_{\tilde{\varepsilon}_{i}}^{m}\right)^{\tilde{\omega}_{i}}+\left(\varsigma-1\right)\prod_{i=1}^{\tilde{l}+1}\left(1-\ddot{\mathfrak{A}}_{\tilde{\varepsilon}_{i}}^{m}\right)^{\tilde{\omega}_{i}}}}, \frac{\sqrt[o]{\varsigma}\,\prod_{i=1}^{\tilde{l}+1}\ddot{\ddot{\Xi}}_{\tilde{\varepsilon}_{i}}^{\tilde{\omega}_{i}}}{\sqrt[o]{\prod_{i=1}^{\tilde{l}+1}\left(1+\left(\varsigma-1\right)\left(1-\ddot{\ddot{\Xi}}_{\tilde{\varepsilon}_{i}}^{o}\right)\right)^{\tilde{\omega}_{i}}+\left(\varsigma-1\right)\prod_{i=1}^{\tilde{l}+1}\left(\ddot{\ddot{\Xi}}_{\tilde{\varepsilon}_{i}}^{o}\right)^{\tilde{\omega}_{i}}}}, \right.$$
$$\left. \frac{\sqrt[n]{\varsigma}\,\prod_{i=1}^{\tilde{l}+1}\ddot{\ddot{\mathfrak{M}}}_{\tilde{\varepsilon}_{i}}^{\tilde{\omega}_{i}}}{\sqrt[n]{\prod_{i=1}^{\tilde{l}+1}\left(1+\left(\varsigma-1\right)\left(1-\ddot{\ddot{\mathfrak{M}}}_{\tilde{\varepsilon}_{i}}^{n}\right)\right)^{\tilde{\omega}_{i}}+\left(\varsigma-1\right)\prod_{i=1}^{\tilde{l}+1}\left(\ddot{\ddot{\mathfrak{M}}}_{\tilde{\varepsilon}_{i}}^{n}\right)^{\tilde{\omega}_{i}}}}, \sqrt[o]{\frac{\prod_{i=1}^{\tilde{l}+1}\left(1+\left(\varsigma-1\right)\ddot{r}_{\tilde{\varepsilon}_{i}}^{o}\right)^{\tilde{\omega}_{i}}-\prod_{i=1}^{\tilde{l}+1}\left(1-\ddot{r}_{\tilde{\varepsilon}_{i}}^{o}\right)^{\tilde{\omega}_{i}}}{\prod_{i=1}^{\tilde{l}+1}\left(1+\left(\varsigma-1\right)\ddot{r}_{\tilde{\varepsilon}_{i}}^{o}\right)^{\tilde{\omega}_{i}}+\left(\varsigma-1\right)\prod_{i=1}^{\tilde{l}+1}\left(1-\ddot{r}_{\tilde{\varepsilon}_{i}}^{o}\right)^{\tilde{\omega}_{i}}}} \right)$$

By this, the aggregated value is also a GO$_{mno}$-TSFV. Therefore, when $n=\tilde{l}+1$, Eq. (22) holds, the following applies. Thus, by (a) and (b), we know that Eq. (22) holds for all $n$. We have completed the proof. ∎

We can also obtain the following GO$_{mno}$-TSFHWA properties that can be similarly proven.

**(1) Idempotency:** If $\tilde{\underline{\varepsilon}}_i\left(i=1,2,...,n\right)$ are identical. i.e. $\tilde{\underline{\varepsilon}}_i=\tilde{\underline{\varepsilon}}\;\forall i$ , then
$$\text{GO}_{mno}\text{-TSFHWA}_{\tilde{\omega}}\left(\tilde{\underline{\varepsilon}}_1,\tilde{\underline{\varepsilon}}_2,...,\tilde{\underline{\varepsilon}}_n\right)=\tilde{\underline{\varepsilon}}.$$

**(2) Boundedness:** Let $\tilde{\underline{\varepsilon}}_i\left(i=1,2,...,n\right)$ be a collection of GO$_{mno}$-TSFVs, and let
$\tilde{\underline{\varepsilon}}^{-}=\min_{1\le i\le n}\left(\tilde{\underline{\varepsilon}}_i\right),\tilde{\underline{\varepsilon}}^{+}=\max_{1\le i\le n}\left(\tilde{\underline{\varepsilon}}_i\right).$ Then ,
$$\tilde{\underline{\varepsilon}}^{-}\le\text{GO}_{mno}\text{-TSFHWA}_{\tilde{\omega}}\left(\tilde{\underline{\varepsilon}}_1,\tilde{\underline{\varepsilon}}_2,...,\tilde{\underline{\varepsilon}}_n\right)\le\tilde{\underline{\varepsilon}}^{+}.$$

**(3) Monotonicity:** Let $\tilde{\underline{\varepsilon}}_i\left(i=1,2,...,n\right)$ and $\tilde{\underline{\varepsilon}}_i'\left(i=1,2,...,n\right)$ be two sets of GO$_{mno}$-TSFVs, if $\tilde{\underline{\varepsilon}}_i\le\tilde{\underline{\varepsilon}}_i',\;\forall i$ , then
$$\text{GO}_{mno}\text{-TSFHWA}_{\tilde{\omega}}\left(\tilde{\underline{\varepsilon}}_1,\tilde{\underline{\varepsilon}}_2,...,\tilde{\underline{\varepsilon}}_n\right)\le\text{GO}_{mno}\text{-TSFHWA}_{\tilde{\omega}}\left(\tilde{\underline{\varepsilon}}_1',\tilde{\underline{\varepsilon}}_2',...,\tilde{\underline{\varepsilon}}_n'\right).$$

### 4.4. GO$_{mno}$-TSF Hamacher weighted geometric operator

This section introduces Hamacher weighted geometric operator that incorporate GO$_{mno}$-TSF information. That is known as GO$_{mno}$-TSF Hamacher weighted geometric (GO$_{mno}$-TSFHWG) operator.

**Definition 18.** Assuming $\tilde{\tau}_i = \left\langle \ddot{\mathfrak{A}}_{\tilde{\tau}_i}, \ddot{\Xi}_{\tilde{\tau}_i}, \ddot{\mathfrak{M}}_{\tilde{\tau}_i}; \ddot{r}_{\tilde{\tau}_i} \right\rangle \left( i = 1, 2, ..., n \right)$ be the collection of GO$_{mno}$-TSFVs. The GO$_{mno}$-TSFHWG operator is defined as a mapping $\tilde{P}^n \rightarrow \tilde{P}$, characterized as:

$$\text{GO}_{mno}\text{-TSFHWG}_{\tilde{\omega}}\left(\tilde{\tau}_1, \tilde{\tau}_2, ..., \tilde{\tau}_n\right) = \otimes_{i=1}^{n} \tilde{\tau}_i^{\tilde{\omega}_i} \tag{23}$$

where $\tilde{\omega} = \left(\tilde{\omega}_1, \tilde{\omega}_2, ..., \tilde{\omega}_n\right)$ represent the weight vector of $\tilde{\tau}_i \left( i = 1, 2, ..., n \right)$ and $\tilde{\omega}_i > 0, \sum_{i=1}^{n} \tilde{\omega}_i = 1$.

We can derive the following results from Definition 18 and Theorem 6.

**Theorem 13.** The aggregated value obtained by using GO$_{mno}$-TSFHWG operator is also a GO$_{mno}$-TSFV, where

$$\text{GO}_{mno}\text{-TSFHWG}_{\tilde{\omega}}\left(\tilde{\tau}_1, \tilde{\tau}_2, ..., \tilde{\tau}_n\right) = \left( \begin{array}{cc} \dfrac{\sqrt{\varsigma}\prod\limits_{i=1}^{n}\ddot{\mathfrak{A}}_{\tilde{\tau}_i}^{*\tilde{\omega}_i}}{\sqrt{\prod\limits_{i=1}^{n}\left(1+\left(\varsigma-1\right)\left(1-\ddot{\mathfrak{A}}_{\tilde{\tau}_i}^{*}\right)\right)^{\tilde{\omega}_i}+\left(\varsigma-1\right)\prod\limits_{i=1}^{n}\left(\ddot{\mathfrak{A}}_{\tilde{\tau}_i}^{*}\right)^{\tilde{\omega}_i}}}, & \sqrt{\dfrac{\prod\limits_{i=1}^{n}\left(1+\left(\varsigma-1\right)\ddot{\Xi}_{\tilde{\tau}_i}^{*}\right)^{\tilde{\omega}_i}-\prod\limits_{i=1}^{n}\left(1-\ddot{\Xi}_{\tilde{\tau}_i}^{*}\right)^{\tilde{\omega}_i}}{\prod\limits_{i=1}^{n}\left(1+\left(\varsigma-1\right)\ddot{\Xi}_{\tilde{\tau}_i}^{*}\right)^{\tilde{\omega}_i}+\left(\gamma-1\right)\prod\limits_{i=1}^{n}\left(1-\ddot{\Xi}_{\tilde{\tau}_i}^{*}\right)^{\tilde{\omega}_i}}}, \\[4ex] \sqrt{\dfrac{\prod\limits_{i=1}^{n}\left(1+\left(\varsigma-1\right)\ddot{\mathfrak{M}}_{\tilde{\tau}_i}^{*}\right)^{\tilde{\omega}_i}-\prod\limits_{i=1}^{n}\left(1-\ddot{\mathfrak{M}}_{\tilde{\tau}_i}^{*}\right)^{\tilde{\omega}_i}}{\prod\limits_{i=1}^{n}\left(1+\left(\varsigma-1\right)\ddot{\mathfrak{M}}_{\tilde{\tau}_i}^{*}\right)^{\tilde{\omega}_i}+\left(\varsigma-1\right)\prod\limits_{i=1}^{n}\left(1-\ddot{\mathfrak{M}}_{\tilde{\tau}_i}^{*}\right)^{\tilde{\omega}_i}}}, & \dfrac{\sqrt{\varsigma}\prod\limits_{i=1}^{n}\ddot{r}_{\tilde{\tau}_i}^{*\tilde{\omega}_i}}{\sqrt{\prod\limits_{i=1}^{n}\left(1+\left(\varsigma-1\right)\left(1-\ddot{r}_{\tilde{\tau}_i}^{*}\right)\right)^{\tilde{\omega}_i}+\left(\varsigma-1\right)\prod\limits_{i=1}^{n}\left(\ddot{r}_{\tilde{\tau}_i}^{*}\right)^{\tilde{\omega}_i}}} \end{array} \right) \tag{24}$$

Here $\tilde{\omega} = \left(\tilde{\omega}_1, \tilde{\omega}_2, ..., \tilde{\omega}_n\right)$ represent the weight vector of $\tilde{\tau}_i \left( i = 1, 2, ..., n \right)$ and $\tilde{\omega}_i > 0, \sum_{i=1}^{n} \tilde{\omega}_i = 1$.

**Proof:** We have Equation. (24) by calculated induction on.

(a) When $n = 2$, we have $\text{GO}_{mno}\text{-TSFHWG}_{\tilde{\omega}}\left(\tilde{\tau}_1, \tilde{\tau}_2\right) = \tilde{\tau}_1^{\tilde{\omega}_1} \otimes \tilde{\tau}_2^{\tilde{\omega}_2}$. By Theorem 3, we can see that both $\tilde{\tau}_1^{\tilde{\omega}_1}$ and $\tilde{\tau}_2^{\tilde{\omega}_2}$ are GO$_{mno}$-TSFVs, and the value of $\tilde{\tau}_1^{\tilde{\omega}_1} \otimes \tilde{\tau}_2^{\tilde{\omega}_2}$ is also a GO$_{mno}$-TSFV. From the operational laws of GO$_{mno}$-TSFVs, we have

$$
\tilde{\underline{\mathcal{L}}}_1^{\tilde{\omega}_1} = \left(
\begin{array}{cc}
\dfrac{\sqrt[m]{\varsigma}\,\ddot{\tilde{\mathfrak{A}}}_{\tilde{\mathcal{L}}_1}^{\tilde{\omega}_1}}{\sqrt[m]{\left(1+\left(\varsigma-1\right)\left(1-\ddot{\tilde{\mathfrak{A}}}_{\tilde{\mathcal{L}}_1}^{m}\right)\right)^{\tilde{\omega}_1}+\left(\varsigma-1\right)\left(\ddot{\tilde{\mathfrak{A}}}_{\tilde{\mathcal{L}}_1}^{m}\right)^{\tilde{\omega}_1}}}, &
\sqrt[o]{\dfrac{\left(1+\left(\varsigma-1\right)\dddot{\Xi}_{\tilde{\mathcal{L}}_1}^{o}\right)^{\tilde{\omega}_1}-\left(1-\dddot{\Xi}_{\tilde{\mathcal{L}}_1}^{o}\right)^{\tilde{\omega}_1}}{\left(1+\left(\varsigma-1\right)\dddot{\Xi}_{\tilde{\mathcal{L}}_1}^{o}\right)^{\tilde{\omega}_1}+\left(\varsigma-1\right)\left(1-\dddot{\Xi}_{\tilde{\mathcal{L}}_1}^{o}\right)^{\tilde{\omega}_1}}}, \\[4mm]
\sqrt[n]{\dfrac{\left(1+\left(\varsigma-1\right)\dddot{\mathfrak{M}}_{\tilde{\mathcal{L}}_1}^{n}\right)^{\tilde{\omega}_1}-\left(1-\dddot{\mathfrak{M}}_{\tilde{\mathcal{L}}_1}^{n}\right)^{\tilde{\omega}_1}}{\left(1+\left(\varsigma-1\right)\dddot{\mathfrak{M}}_{\tilde{\mathcal{L}}_1}^{n}\right)^{\tilde{\omega}_1}+\left(\varsigma-1\right)\left(1-\dddot{\mathfrak{M}}_{\tilde{\mathcal{L}}_1}^{n}\right)^{\tilde{\omega}_1}}}, &
\dfrac{\sqrt[q]{\varsigma}\,\ddot{\dot{\imath}}_{\tilde{\mathcal{L}}_1}^{\tilde{\omega}_1}}{\sqrt[q]{\left(1+\left(\varsigma-1\right)\left(1-\ddot{\dot{\imath}}_{\tilde{\mathcal{L}}_1}^{o}\right)\right)^{\tilde{\omega}_1}+\left(\varsigma-1\right)\left(\ddot{\dot{\imath}}_{\tilde{\mathcal{L}}_1}^{o}\right)^{\tilde{\omega}_1}}}
\end{array}
\right),
$$

$$
\tilde{\underline{\mathcal{L}}}_2^{\tilde{\omega}_2} = \left(
\begin{array}{cc}
\dfrac{\sqrt[m]{\varsigma}\,\ddot{\tilde{\mathfrak{A}}}_{\tilde{\mathcal{L}}_2}^{\tilde{\omega}_2}}{\sqrt[m]{\left(1+\left(\varsigma-1\right)\left(1-\ddot{\tilde{\mathfrak{A}}}_{\tilde{\mathcal{L}}_2}^{m}\right)\right)^{\tilde{\omega}_2}+\left(\varsigma-1\right)\left(\ddot{\tilde{\mathfrak{A}}}_{\tilde{\mathcal{L}}_2}^{m}\right)^{\tilde{\omega}_2}}}, &
\sqrt[o]{\dfrac{\left(1+\left(\varsigma-1\right)\dddot{\Xi}_{\tilde{\mathcal{L}}_2}^{o}\right)^{\tilde{\omega}_2}-\left(1-\dddot{\Xi}_{\tilde{\mathcal{L}}_2}^{o}\right)^{\tilde{\omega}_2}}{\left(1+\left(\varsigma-1\right)\dddot{\Xi}_{\tilde{\mathcal{L}}_2}^{o}\right)^{\tilde{\omega}_2}+\left(\varsigma-1\right)\left(1-\dddot{\Xi}_{\tilde{\mathcal{L}}_2}^{o}\right)^{\tilde{\omega}_2}}}, \\[4mm]
\sqrt[n]{\dfrac{\left(1+\left(\varsigma-1\right)\dddot{\mathfrak{M}}_{\tilde{\mathcal{L}}_2}^{n}\right)^{\tilde{\omega}_2}-\left(1-\dddot{\mathfrak{M}}_{\tilde{\mathcal{L}}_2}^{n}\right)^{\tilde{\omega}_2}}{\left(1+\left(\varsigma-1\right)\dddot{\mathfrak{M}}_{\tilde{\mathcal{L}}_2}^{n}\right)^{\tilde{\omega}_2}+\left(\varsigma-1\right)\left(1-\dddot{\mathfrak{M}}_{\tilde{\mathcal{L}}_2}^{n}\right)^{\tilde{\omega}_2}}}, &
\dfrac{\sqrt[q]{\varsigma}\,\ddot{\dot{\imath}}_{\tilde{\mathcal{L}}_2}^{\tilde{\omega}_2}}{\sqrt[q]{\left(1+\left(\varsigma-1\right)\left(1-\ddot{\dot{\imath}}_{\tilde{\mathcal{L}}_2}^{o}\right)\right)^{\tilde{\omega}_2}+\left(\varsigma-1\right)\left(\ddot{\dot{\imath}}_{\tilde{\mathcal{L}}_2}^{o}\right)^{\tilde{\omega}_2}}}
\end{array}
\right).
$$

$$
\text{GO}_{\text{mno}}\text{-TSFHWG}_{\tilde{\omega}}\left(\tilde{\underline{\mathcal{L}}}_1,\tilde{\underline{\mathcal{L}}}_2\right)=\tilde{\underline{\mathcal{L}}}_1^{\tilde{\omega}_1}\otimes\tilde{\underline{\mathcal{L}}}_2^{\tilde{\omega}_2}
$$

$$
= \left(
\begin{array}{cc}
\dfrac{\sqrt[m]{\varsigma}\,\ddot{\tilde{\mathfrak{A}}}_{\tilde{\mathcal{L}}_1}^{\tilde{\omega}_1}}{\sqrt[m]{\left(1+\left(\varsigma-1\right)\left(1-\ddot{\tilde{\mathfrak{A}}}_{\tilde{\mathcal{L}}_1}^{m}\right)\right)^{\tilde{\omega}_1}+\left(\varsigma-1\right)\left(\ddot{\tilde{\mathfrak{A}}}_{\tilde{\mathcal{L}}_1}^{m}\right)^{\tilde{\omega}_1}}}, &
\sqrt[o]{\dfrac{\left(1+\left(\varsigma-1\right)\dddot{\Xi}_{\tilde{\mathcal{L}}_1}^{o}\right)^{\tilde{\omega}_1}-\left(1-\dddot{\Xi}_{\tilde{\mathcal{L}}_1}^{o}\right)^{\tilde{\omega}_1}}{\left(1+\left(\varsigma-1\right)\dddot{\Xi}_{\tilde{\mathcal{L}}_1}^{o}\right)^{\tilde{\omega}_1}+\left(\varsigma-1\right)\left(1-\dddot{\Xi}_{\tilde{\mathcal{L}}_1}^{o}\right)^{\tilde{\omega}_1}}}, \\[4mm]
\sqrt[n]{\dfrac{\left(1+\left(\varsigma-1\right)\dddot{\mathfrak{M}}_{\tilde{\mathcal{L}}_1}^{n}\right)^{\tilde{\omega}_1}-\left(1-\dddot{\mathfrak{M}}_{\tilde{\mathcal{L}}_1}^{n}\right)^{\tilde{\omega}_1}}{\left(1+\left(\varsigma-1\right)\dddot{\mathfrak{M}}_{\tilde{\mathcal{L}}_1}^{n}\right)^{\tilde{\omega}_1}+\left(\varsigma-1\right)\left(1-\dddot{\mathfrak{M}}_{\tilde{\mathcal{L}}_1}^{n}\right)^{\tilde{\omega}_1}}}, &
\dfrac{\sqrt[q]{\varsigma}\,\ddot{\dot{\imath}}_{\tilde{\mathcal{L}}_1}^{\tilde{\omega}_1}}{\sqrt[q]{\left(1+\left(\varsigma-1\right)\left(1-\ddot{\dot{\imath}}_{\tilde{\mathcal{L}}_1}^{o}\right)\right)^{\tilde{\omega}_1}+\left(\varsigma-1\right)\left(\ddot{\dot{\imath}}_{\tilde{\mathcal{L}}_1}^{o}\right)^{\tilde{\omega}_1}}}
\end{array}
\right)\otimes
$$

$$
\left(
\begin{array}{cc}
\dfrac{\sqrt[m]{\varsigma}\,\ddot{\tilde{\mathfrak{A}}}_{\tilde{\mathcal{L}}_2}^{\tilde{\omega}_2}}{\sqrt[m]{\left(1+\left(\varsigma-1\right)\left(1-\ddot{\tilde{\mathfrak{A}}}_{\tilde{\mathcal{L}}_2}^{m}\right)\right)^{\tilde{\omega}_2}+\left(\varsigma-1\right)\left(\ddot{\tilde{\mathfrak{A}}}_{\tilde{\mathcal{L}}_2}^{m}\right)^{\tilde{\omega}_2}}}, &
\sqrt[o]{\dfrac{\left(1+\left(\varsigma-1\right)\dddot{\Xi}_{\tilde{\mathcal{L}}_2}^{o}\right)^{\tilde{\omega}_2}-\left(1-\dddot{\Xi}_{\tilde{\mathcal{L}}_2}^{o}\right)^{\tilde{\omega}_2}}{\left(1+\left(\varsigma-1\right)\dddot{\Xi}_{\tilde{\mathcal{L}}_2}^{o}\right)^{\tilde{\omega}_2}+\left(\varsigma-1\right)\left(1-\dddot{\Xi}_{\tilde{\mathcal{L}}_2}^{o}\right)^{\tilde{\omega}_2}}}, \\[4mm]
\sqrt[n]{\dfrac{\left(1+\left(\varsigma-1\right)\dddot{\mathfrak{M}}_{\tilde{\mathcal{L}}_2}^{n}\right)^{\tilde{\omega}_2}-\left(1-\dddot{\mathfrak{M}}_{\tilde{\mathcal{L}}_2}^{n}\right)^{\tilde{\omega}_2}}{\left(1+\left(\varsigma-1\right)\dddot{\mathfrak{M}}_{\tilde{\mathcal{L}}_2}^{n}\right)^{\tilde{\omega}_2}+\left(\varsigma-1\right)\left(1-\dddot{\mathfrak{M}}_{\tilde{\mathcal{L}}_2}^{n}\right)^{\tilde{\omega}_2}}}, &
\dfrac{\sqrt[q]{\varsigma}\,\ddot{\dot{\imath}}_{\tilde{\mathcal{L}}_2}^{\tilde{\omega}_2}}{\sqrt[q]{\left(1+\left(\varsigma-1\right)\left(1-\ddot{\dot{\imath}}_{\tilde{\mathcal{L}}_2}^{o}\right)\right)^{\tilde{\omega}_2}+\left(\varsigma-1\right)\left(\ddot{\dot{\imath}}_{\tilde{\mathcal{L}}_2}^{o}\right)^{\tilde{\omega}_2}}}
\end{array}
\right)
$$

$$= \left( \frac{\sqrt[m]{\varsigma} \prod\limits_{i=1}^{2} \ddot{\mathfrak{A}}_{\tilde{\varsigma}_i}^{\tilde{\omega}_i}}{\sqrt[m]{\prod\limits_{i=1}^{2} \left(1 + (\varsigma - 1)\left(1 - \ddot{\mathfrak{A}}_{\tilde{\varsigma}_i}^m\right)\right)^{\tilde{\omega}_i} + (\varsigma - 1)\prod\limits_{i=1}^{2} \left(\ddot{\mathfrak{A}}_{\tilde{\varsigma}_i}^m\right)^{\tilde{\omega}_i}}}, \sqrt[o]{\frac{\prod\limits_{i=1}^{2} \left(1 + (\varsigma - 1)\ddot{\Xi}_{\tilde{\varsigma}_i}^o\right)^{\tilde{\omega}_i} - \prod\limits_{i=1}^{2} \left(1 - \ddot{\Xi}_{\tilde{\varsigma}_i}^o\right)^{\tilde{\omega}_i}}{\prod\limits_{i=1}^{2} \left(1 + (\varsigma - 1)\ddot{\Xi}_{\tilde{\varsigma}_i}^o\right)^{\tilde{\omega}_i} + (\varsigma - 1)\prod\limits_{i=1}^{2} \left(1 - \ddot{\Xi}_{\tilde{\varsigma}_i}^o\right)^{\tilde{\omega}_i}}}, \right.$$

$$\left. \sqrt[n]{\frac{\prod\limits_{i=1}^{2} \left(1 + (\varsigma - 1)\ddot{\mathfrak{M}}_{\tilde{\varsigma}_i}^n\right)^{\tilde{\omega}_i} - \prod\limits_{i=1}^{2} \left(1 - \ddot{\mathfrak{M}}_{\tilde{\varsigma}_i}^n\right)^{\tilde{\omega}_i}}{\prod\limits_{i=1}^{2} \left(1 + (\varsigma - 1)\ddot{\mathfrak{M}}_{\tilde{\varsigma}_i}^n\right)^{\tilde{\omega}_i} + (\varsigma - 1)\prod\limits_{i=1}^{2} \left(1 - \ddot{\mathfrak{M}}_{\tilde{\varsigma}_i}^n\right)^{\tilde{\omega}_i}}}, \frac{\sqrt[o]{\varsigma} \prod\limits_{i=1}^{2} \ddot{r}_{\tilde{\varsigma}_i}^{\tilde{\omega}_i}}{\sqrt[o]{\prod\limits_{i=1}^{2} \left(1 + (\varsigma - 1)\left(1 - \ddot{r}_{\tilde{\varsigma}_i}^o\right)\right)^{\tilde{\omega}_i} + (\varsigma - 1)\prod\limits_{i=1}^{2} \left(\ddot{r}_{\tilde{\varsigma}_i}^o\right)^{\tilde{\omega}_i}}} \right)$$

(b) Assume that $n = \tilde{l}$ Equation (24) holds, i.e.,

$$\text{GO}_{\text{mno}}\text{-TSFHWG}_\omega \left(\tilde{\varsigma}_1, \tilde{\varsigma}_2, ..., \tilde{\varsigma}_{\tilde{l}}\right) = \tilde{\varsigma}_1^{\tilde{\omega}_1} \otimes \tilde{\varsigma}_2^{\tilde{\omega}_2} \otimes ... \otimes \tilde{\varsigma}_{\tilde{l}}^{\tilde{\omega}_{\tilde{l}}}$$

$$= \left( \frac{\sqrt[m]{\varsigma} \prod\limits_{i=1}^{\tilde{l}} \ddot{\mathfrak{A}}_{\tilde{\varsigma}_i}^{\tilde{\omega}_i}}{\sqrt[m]{\prod\limits_{i=1}^{\tilde{l}} \left(1 + (\varsigma - 1)\left(1 - \ddot{\mathfrak{A}}_{\tilde{\varsigma}_i}^m\right)\right)^{\tilde{\omega}_i} + (\varsigma - 1)\prod\limits_{i=1}^{\tilde{l}} \left(\ddot{\mathfrak{A}}_{\tilde{\varsigma}_i}^m\right)^{\tilde{\omega}_i}}}, \sqrt[o]{\frac{\prod\limits_{i=1}^{\tilde{l}} \left(1 + (\varsigma - 1)\ddot{\Xi}_{\tilde{\varsigma}_i}^o\right)^{\tilde{\omega}_i} - \prod\limits_{i=1}^{\tilde{l}} \left(1 - \ddot{\Xi}_{\tilde{\varsigma}_i}^o\right)^{\tilde{\omega}_i}}{\prod\limits_{i=1}^{\tilde{l}} \left(1 + (\varsigma - 1)\ddot{\Xi}_{\tilde{\varsigma}_i}^o\right)^{\tilde{\omega}_i} + (\varsigma - 1)\prod\limits_{i=1}^{\tilde{l}} \left(1 - \ddot{\Xi}_{\tilde{\varsigma}_i}^o\right)^{\tilde{\omega}_i}}}, \right.$$

$$\left. \sqrt[n]{\frac{\prod\limits_{i=1}^{\tilde{l}} \left(1 + (\varsigma - 1)\ddot{\mathfrak{M}}_{\tilde{\varsigma}_i}^n\right)^{\tilde{\omega}_i} - \prod\limits_{i=1}^{\tilde{l}} \left(1 - \ddot{\mathfrak{M}}_{\tilde{\varsigma}_i}^n\right)^{\tilde{\omega}_i}}{\prod\limits_{i=1}^{\tilde{l}} \left(1 + (\varsigma - 1)\ddot{\mathfrak{M}}_{\tilde{\varsigma}_i}^n\right)^{\tilde{\omega}_i} + (\varsigma - 1)\prod\limits_{i=1}^{\tilde{l}} \left(1 - \ddot{\mathfrak{M}}_{\tilde{\varsigma}_i}^n\right)^{\tilde{\omega}_i}}}, \frac{\sqrt[o]{\varsigma} \prod\limits_{i=1}^{\tilde{l}} \ddot{r}_{\tilde{\varsigma}_i}^{\tilde{\omega}_i}}{\sqrt[o]{\prod\limits_{i=1}^{\tilde{l}} \left(1 + (\varsigma - 1)\left(1 - \ddot{r}_{\tilde{\varsigma}_i}^o\right)\right)^{\tilde{\omega}_i} + (\varsigma - 1)\prod\limits_{i=1}^{\tilde{l}} \left(\ddot{r}_{\tilde{\varsigma}_i}^o\right)^{\tilde{\omega}_i}}} \right)$$

And the aggregation value is a $\text{GO}_{\text{mno}}$-TSFV, then when $n = \tilde{l} + 1$, by the operational laws of $\text{GO}_{\text{mno}}$-TSFV, we have

$$\text{GO}_{\text{mno}}\text{-TSFHWG}_\omega \left(\tilde{\varsigma}_1, \tilde{\varsigma}_2, ..., \tilde{\varsigma}_{\tilde{l}}\right) = \tilde{\varsigma}_1^{\tilde{\omega}_1} \otimes \tilde{\varsigma}_2^{\tilde{\omega}_2} \otimes ... \otimes \tilde{\varsigma}_{\tilde{l}}^{\tilde{\omega}_{\tilde{l}}} \otimes \tilde{\varsigma}_{\tilde{l}+1}^{\tilde{\omega}_{\tilde{l}+1}}$$

$$= \left( \begin{array}{cc} \dfrac{\sqrt[m]{\varsigma} \prod\limits_{i=1}^{\tilde{l}} \overset{\cdots}{\mathfrak{A}}_{\tilde{\tau}_i}^{\tilde{\omega}_i}}{\sqrt[m]{\prod\limits_{i=1}^{\tilde{l}} \left(1+\left(\varsigma-1\right)\left(1-\overset{\cdots}{\mathfrak{A}}_{\tilde{\tau}_i}^m\right)\right)^{\tilde{\omega}_i} + \left(\varsigma-1\right) \prod\limits_{i=1}^{\tilde{l}} \left(\overset{\cdots}{\mathfrak{A}}_{\tilde{\tau}_i}^m\right)^{\tilde{\omega}_i}}}, & \sqrt[o]{\dfrac{\prod\limits_{i=1}^{\tilde{l}} \left(1+\left(\varsigma-1\right)\overset{\cdots}{\Xi}_{\tilde{\tau}_i}^o\right)^{\tilde{\omega}_i} - \prod\limits_{i=1}^{\tilde{l}} \left(1-\overset{\cdots}{\Xi}_{\tilde{\tau}_i}^o\right)^{\tilde{\omega}_i}}{\prod\limits_{i=1}^{\tilde{l}} \left(1+\left(\varsigma-1\right)\overset{\cdots}{\Xi}_{\tilde{\tau}_i}^o\right)^{\tilde{\omega}_i} + \left(\varsigma-1\right) \prod\limits_{i=1}^{\tilde{l}} \left(1-\overset{\cdots}{\Xi}_{\tilde{\tau}_i}^o\right)^{\tilde{\omega}_i}}}, \\[3em] \dfrac{\sqrt[n]{\prod\limits_{i=1}^{\tilde{l}} \left(1+\left(\varsigma-1\right)\overset{\cdots}{\mathfrak{M}}_{\tilde{\tau}_i}^n\right)^{\tilde{\omega}_i} - \prod\limits_{i=1}^{\tilde{l}} \left(1-\overset{\cdots}{\mathfrak{M}}_{\tilde{\tau}_i}^n\right)^{\tilde{\omega}_i}}}{\sqrt[n]{\prod\limits_{i=1}^{\tilde{l}} \left(1+\left(\varsigma-1\right)\overset{\cdots}{\mathfrak{M}}_{\tilde{\tau}_i}^n\right)^{\tilde{\omega}_i} + \left(\varsigma-1\right) \prod\limits_{i=1}^{\tilde{l}} \left(1-\overset{\cdots}{\mathfrak{M}}_{\tilde{\tau}_i}^n\right)^{\tilde{\omega}_i}}}, & \dfrac{\sqrt[o]{\varsigma} \prod\limits_{i=1}^{\tilde{l}} \overset{\cdots}{r}_{\tilde{\tau}_i}^{\tilde{\omega}_i}}{\sqrt[o]{\prod\limits_{i=1}^{\tilde{l}} \left(1+\left(\varsigma-1\right)\left(1-\overset{\cdots}{r}_{\tilde{\tau}_i}^o\right)\right)^{\tilde{\omega}_i} + \left(\varsigma-1\right) \prod\limits_{i=1}^{\tilde{l}} \left(\overset{\cdots}{r}_{\tilde{\tau}_i}^o\right)^{\tilde{\omega}_i}}} \end{array} \right) \otimes$$

$$\left( \begin{array}{cc} \dfrac{\sqrt[m]{\varsigma} \, \overset{\cdots}{\mathfrak{A}}_{\tilde{\tau}_{\tilde{l}+1}}^{\tilde{\omega}_{\tilde{l}+1}}}{\sqrt[m]{\left(1+\left(\varsigma-1\right)\left(1-\overset{\cdots}{\mathfrak{A}}_{\tilde{\tau}_{\tilde{l}+1}}^m\right)\right)^{\tilde{\omega}_{\tilde{l}+1}} + \left(\varsigma-1\right)\left(\overset{\cdots}{\mathfrak{A}}_{\tilde{\tau}_{\tilde{l}+1}}^m\right)^{\tilde{\omega}_{\tilde{l}+1}}}}, & \sqrt[o]{\dfrac{\left(1+\left(\varsigma-1\right)\overset{\cdots}{\Xi}_{\tilde{\tau}_{\tilde{l}+1}}^o\right)^{\tilde{\omega}_{\tilde{l}+1}} - \left(1-\overset{\cdots}{\Xi}_{\tilde{\tau}_{\tilde{l}+1}}^o\right)^{\tilde{\omega}_{\tilde{l}+1}}}{\left(1+\left(\varsigma-1\right)\overset{\cdots}{\Xi}_{\tilde{\tau}_{\tilde{l}+1}}^o\right)^{\tilde{\omega}_{\tilde{l}+1}} + \left(\varsigma-1\right)\left(1-\overset{\cdots}{\Xi}_{\tilde{\tau}_{\tilde{l}+1}}^o\right)^{\tilde{\omega}_{\tilde{l}+1}}}}, \\[3em] \dfrac{\sqrt[n]{\left(1+\left(\varsigma-1\right)\overset{\cdots}{\mathfrak{M}}_{\tilde{\tau}_{\tilde{l}+1}}^n\right)^{\tilde{\omega}_{\tilde{l}+1}} - \left(1-\overset{\cdots}{\mathfrak{M}}_{\tilde{\tau}_{\tilde{l}+1}}^n\right)^{\tilde{\omega}_{\tilde{l}+1}}}}{\sqrt[n]{\left(1+\left(\varsigma-1\right)\overset{\cdots}{\mathfrak{M}}_{\tilde{\tau}_{\tilde{l}+1}}^n\right)^{\tilde{\omega}_{\tilde{l}+1}} + \left(\varsigma-1\right)\left(1-\overset{\cdots}{\mathfrak{M}}_{\tilde{\tau}_{\tilde{l}+1}}^n\right)^{\tilde{\omega}_{\tilde{l}+1}}}}, & \dfrac{\sqrt[o]{\varsigma} \, \overset{\cdots}{r}_{\tilde{\tau}_{\tilde{l}+1}}^{\tilde{\omega}_{\tilde{l}+1}}}{\sqrt[o]{\left(1+\left(\varsigma-1\right)\left(1-\overset{\cdots}{r}_{\tilde{\tau}_{\tilde{l}+1}}^o\right)\right)^{\tilde{\omega}_{\tilde{l}+1}} + \left(\varsigma-1\right)\left(\overset{\cdots}{r}_{\tilde{\tau}_{\tilde{l}+1}}^o\right)^{\tilde{\omega}_{\tilde{l}+1}}}} \end{array} \right)$$

$$= \left( \begin{array}{cc} \dfrac{\sqrt[m]{\varsigma} \prod\limits_{i=1}^{\tilde{l}+1} \overset{\cdots}{\mathfrak{A}}_{\tilde{\tau}_i}^{\tilde{\omega}_i}}{\sqrt[m]{\prod\limits_{i=1}^{\tilde{l}+1} \left(1+\left(\varsigma-1\right)\left(1-\overset{\cdots}{\mathfrak{A}}_{\tilde{\tau}_i}^m\right)\right)^{\tilde{\omega}_i} + \left(\varsigma-1\right) \prod\limits_{i=1}^{\tilde{l}+1} \left(\overset{\cdots}{\mathfrak{A}}_{\tilde{\tau}_i}^m\right)^{\tilde{\omega}_i}}}, & \sqrt[o]{\dfrac{\prod\limits_{i=1}^{\tilde{l}+1} \left(1+\left(\varsigma-1\right)\overset{\cdots}{\Xi}_{\tilde{\tau}_i}^o\right)^{\tilde{\omega}_i} - \prod\limits_{i=1}^{\tilde{l}+1} \left(1-\overset{\cdots}{\Xi}_{\tilde{\tau}_i}^o\right)^{\tilde{\omega}_i}}{\prod\limits_{i=1}^{\tilde{l}+1} \left(1+\left(\varsigma-1\right)\overset{\cdots}{\Xi}_{\tilde{\tau}_i}^o\right)^{\tilde{\omega}_i} + \left(\varsigma-1\right) \prod\limits_{i=1}^{\tilde{l}+1} \left(1-\overset{\cdots}{\Xi}_{\tilde{\tau}_i}^o\right)^{\tilde{\omega}_i}}}, \\[3em] \dfrac{\sqrt[n]{\prod\limits_{i=1}^{\tilde{l}+1} \left(1+\left(\varsigma-1\right)\overset{\cdots}{\mathfrak{M}}_{\tilde{\tau}_i}^n\right)^{\tilde{\omega}_i} - \prod\limits_{i=1}^{\tilde{l}+1} \left(1-\overset{\cdots}{\mathfrak{M}}_{\tilde{\tau}_i}^n\right)^{\tilde{\omega}_i}}}{\sqrt[n]{\prod\limits_{i=1}^{\tilde{l}+1} \left(1+\left(\varsigma-1\right)\overset{\cdots}{\mathfrak{M}}_{\tilde{\tau}_i}^n\right)^{\tilde{\omega}_i} + \left(\varsigma-1\right) \prod\limits_{i=1}^{\tilde{l}+1} \left(1-\overset{\cdots}{\mathfrak{M}}_{\tilde{\tau}_i}^n\right)^{\tilde{\omega}_i}}}, & \dfrac{\sqrt[o]{\varsigma} \prod\limits_{i=1}^{\tilde{l}+1} \overset{\cdots}{r}_{\tilde{\tau}_i}^{\tilde{\omega}_i}}{\sqrt[o]{\prod\limits_{i=1}^{\tilde{l}+1} \left(1+\left(\varsigma-1\right)\left(1-\overset{\cdots}{r}_{\tilde{\tau}_i}^o\right)\right)^{\tilde{\omega}_i} + \left(\varsigma-1\right) \prod\limits_{i=1}^{\tilde{l}+1} \left(\overset{\cdots}{r}_{\tilde{\tau}_i}^o\right)^{\tilde{\omega}_i}}} \end{array} \right)$$

Consequently, the aggregated value is also a $GO_{mno}$-TSFV. Hence, by (a) and (b), we know that $n = \tilde{l} + 1$, Eq.(24) holds for all $n$. We have completed the proof. ∎

$GO_{mno}$-TSFHWG operator exhibits the following properties.

**(1) Idempotency:** If $\tilde{\tau}_i \left(i=1,2,...,n\right)$ are identical, i.e. $\tilde{\tau}_i = \tilde{\tau}$ for all $i$, then

$GO_{mno}\text{-TSFHWG}_{\tilde{\omega}}\left(\tilde{\tau}_1, \tilde{\tau}_2, ..., \tilde{\tau}_n\right) = \tilde{\tau}.$

**(2) Boundedness:** Let $\tilde{\underline{\varsigma}}_i\,(i=1,2,...,n)$ be a collection of $GO_{mno}$-TSFVs, and let $\tilde{\underline{\varsigma}}^- = \min\limits_{1\le i\le n}\left(\tilde{\underline{\varsigma}}_i\right), \tilde{\underline{\varsigma}}^+ = \max\limits_{1\le i\le n}\left(\tilde{\underline{\varsigma}}_i\right).$ Then,

$$\tilde{\underline{\varsigma}}^- \le GO_{mno}\text{-TSFHWG}_{\tilde{\omega}}\left(\tilde{\underline{\varsigma}}_1,\tilde{\underline{\varsigma}}_2,...,\tilde{\underline{\varsigma}}_n\right) \le \tilde{\underline{\varsigma}}^+$$

.

**(3) Monotonicity:** Let $\tilde{\underline{\varsigma}}_i\,(i=1,2,...,n)$ and $\tilde{\underline{\varsigma}}_i'\,(i=1,2,...,n)$ be two set of $GO_{mno}$-TSFVs, if $\tilde{\underline{\varsigma}}_i \le \tilde{\underline{\varsigma}}_i',$ for all $i$, then,

$$GO_{mno}\text{-TSFHWG}_{\tilde{\omega}}\left(\tilde{\underline{\varsigma}}_1,\tilde{\underline{\varsigma}}_2,...,\tilde{\underline{\varsigma}}_n\right) \le GO_{mno}\text{-TSFHWG}_{\tilde{\omega}}\left(\tilde{\underline{\varsigma}}_1',\tilde{\underline{\varsigma}}_2',...,\tilde{\underline{\varsigma}}_n'\right)$$

.

## 5. $GO_{mno}$-TSF Hamacher MCGDM scheme with its Application

Using the $GO_{mno}$-TSFHWA and $GO_{mno}$-TSFHWG operators, we propose a model for MCGDM using $GO_{mno}$-TSF information.

Let $Þ = \{?_1, ?_2,...,?_m\}$ be a discrete set of alternatives, and $L = \{L_1, L_2,...,L_n\}$ will be a set of characteristics. A weighting vector will be assigned to each attribute $\tilde{\omega} = \left(\tilde{\omega}_1, \tilde{\omega}_2,...,\tilde{\omega}_n\right)^T$ is $L_1\,(i=1,2,...,n)$, where $\tilde{\omega}_i \in [0,1], \sum\limits_{i=1}^n \tilde{\omega}_i = 1.$

Suppose that $\tilde{F} = \left(\tilde{f}_{ji}\right)_{m\times n} = \left(\ddot{\ddot{\mathfrak{A}}}_{ji}, \ddot{\ddot{\Xi}}_{ji}, \ddot{\ddot{\mathfrak{M}}}_{ji}; \ddot{r}_{ji}\right)_{m\times n}$ is the $GO_{mno}$-TSF decision matrix (DM), where $\ddot{\ddot{\mathfrak{A}}}_{ji}$ represents the DoM to which alternative $Þ_j$ satisfies Characteristics $L_i$, as determined by the decision maker. Additionally, $\ddot{\ddot{\Xi}}_{ji}$ denotes the DoI, indicating that alternative $Þ_j$ neither satisfies nor dissatisfies Characteristics $L_i$. Finally, $\ddot{\ddot{\mathfrak{M}}}_{ji}$ signifies the DoN, showing that alternative $Þ_j$ does not satisfy Characteristics $L_i$, as determined by the decision maker.

The constraints are $\ddot{\ddot{\mathfrak{A}}}_{ji} \in [0,1], \ddot{\ddot{\Xi}}_{ji} \in [0,1], \ddot{\ddot{\mathfrak{M}}}_{ji} \in [0,1], \ddot{r}_{ji} \in [0,1],$ and $0 \le \ddot{\ddot{\mathfrak{A}}}_{ji}^m + \ddot{\ddot{\Xi}}_{ji}^o + \ddot{\ddot{\mathfrak{M}}}_{ji}^n \le 1$, $i=1,2,...,n,$ and $j=1,2,...,m.$

Next, we utilize the GO$_{mno}$-TSFHWA and GO$_{mno}$-TSFHWG operators to address MCGDM problems that involve vague, indeterminate and imprecise information.

**Step 1.** By using the evaluation values presented in the matrix $\tilde{F}$ along with the GO$_{mno}$-TSFHWA operator, we have

$$\tilde{\underline{\xi}}_i = \text{GO}_{mno}\text{-TSFHWA}\left(\tilde{f}_{j1}, \tilde{f}_{j2}, ..., \tilde{f}_{jn}\right) = \overset{n}{\underset{i=1}{\oplus}}\omega_i \tilde{f}_{ji}$$

$$= \left( \begin{array}{cc} \sqrt[m]{\dfrac{\prod_{i=1}^{n}\left(1+(\varsigma-1)\ddot{\mathfrak{A}}_{ji}^m\right)^{\tilde{\omega}_i} - \prod_{i=1}^{n}\left(1-\ddot{\mathfrak{A}}_{ji}^m\right)^{\tilde{\omega}_i}}{\prod_{i=1}^{n}\left(1+(\varsigma-1)\ddot{\mathfrak{A}}_{ji}^m\right)^{\tilde{\omega}_i} + (\varsigma-1)\prod_{i=1}^{n}\left(1-\ddot{\mathfrak{A}}_{ji}^m\right)^{\tilde{\omega}_i}}}, & \dfrac{\sqrt[o]{\varsigma}\prod_{i=1}^{n}\dddot{\Xi}_{ji}^{\tilde{\omega}_i}}{\sqrt[o]{\prod_{i=1}^{n}\left(1+(\varsigma-1)\left(1-\dddot{\Xi}_{ji}^o\right)\right)^{\tilde{\omega}_i} + (\varsigma-1)\prod_{i=1}^{n}\left(\dddot{\Xi}_{ji}^o\right)^{\tilde{\omega}_i}}}, \\[30pt] \dfrac{\sqrt[n]{\varsigma}\prod_{i=1}^{n}\ddot{\mathfrak{M}}_{ji}^{\tilde{\omega}_i}}{\sqrt[n]{\prod_{i=1}^{n}\left(1+(\varsigma-1)\left(1-\ddot{\mathfrak{M}}_{ji}^n\right)\right)^{\tilde{\omega}_i} + (\varsigma-1)\prod_{i=1}^{n}\left(\ddot{\mathfrak{M}}_{ji}^n\right)^{\tilde{\omega}_i}}}, & \sqrt[o]{\dfrac{\prod_{i=1}^{n}\left(1+(\varsigma-1)\ddot{r}_{yji}^o\right)^{\tilde{\omega}_i} - \prod_{i=1}^{n}\left(1-\ddot{r}_{yji}^o\right)^{\tilde{\omega}_i}}{\prod_{i=1}^{n}\left(1+(\varsigma-1)\ddot{r}_{yji}^o\right)^{\tilde{\omega}_i} + (\varsigma-1)\prod_{i=1}^{n}\left(1-\ddot{r}_{yji}^o\right)^{\tilde{\omega}_i}}}, \end{array} \right), i=1,2,...,n.$$

Or $\tilde{\underline{\xi}}_i = \text{GO}_{mno}\text{-TSFHWG}\left(\tilde{f}_{j1}, \tilde{f}_{j2}, ..., \tilde{f}_{jn}\right) = \overset{n}{\underset{i=1}{\otimes}}\tilde{f}_{ji}^{\omega_i}$

$$= \left( \begin{array}{cc} \dfrac{\sqrt[m]{\varsigma}\prod_{i=1}^{n}\ddot{\mathfrak{A}}_{ji}^{\tilde{\omega}_i}}{\sqrt[m]{\prod_{i=1}^{n}\left(1+(\varsigma-1)\left(1-\ddot{\mathfrak{A}}_{ji}^m\right)\right)^{\tilde{\omega}_i} + (\varsigma-1)\prod_{i=1}^{n}\left(\ddot{\mathfrak{A}}_{ji}^m\right)^{\tilde{\omega}_i}}}, & \sqrt[o]{\dfrac{\prod_{i=1}^{n}\left(1+(\varsigma-1)\dddot{\Xi}_{ji}^o\right)^{\tilde{\omega}_i} - \prod_{i=1}^{n}\left(1-\dddot{\Xi}_{ji}^o\right)^{\tilde{\omega}_i}}{\prod_{i=1}^{n}\left(1+(\varsigma-1)\dddot{\Xi}_{ji}^o\right)^{\tilde{\omega}_i} + (\varsigma-1)\prod_{i=1}^{n}\left(1-\dddot{\Xi}_{ji}^o\right)^{\tilde{\omega}_i}}}, \\[30pt] \sqrt[n]{\dfrac{\prod_{i=1}^{n}\left(1+(\varsigma-1)\ddot{\mathfrak{M}}_{ji}^n\right)^{\tilde{\omega}_i} - \prod_{i=1}^{n}\left(1-\ddot{\mathfrak{M}}_{ji}^n\right)^{\tilde{\omega}_i}}{\prod_{i=1}^{n}\left(1+(\varsigma-1)\ddot{\mathfrak{M}}_{ji}^n\right)^{\tilde{\omega}_i} + (\varsigma-1)\prod_{i=1}^{n}\left(1-\ddot{\mathfrak{M}}_{ji}^n\right)^{\tilde{\omega}_i}}}, & \dfrac{\sqrt[o]{\varsigma}\prod_{i=1}^{n}\ddot{r}_{yji}^{\tilde{\omega}_i}}{\sqrt[o]{\prod_{i=1}^{n}\left(1+(\varsigma-1)\left(1-\ddot{r}_{yji}^o\right)\right)^{\tilde{\omega}_i} + (\varsigma-1)\prod_{i=1}^{n}\left(\ddot{r}_{yji}^o\right)^{\tilde{\omega}_i}}}, \end{array} \right), i=1,2,...,n.$$

to determine the whole preference values $\tilde{\underline{\xi}}_i \left(i=1,2,...,n\right)$ of the alternatives $Þ_j$.

**Step 2.** Compute the overall GO$_{mno}$-TSFVs' $\tilde{\underline{\xi}}_i \left(i=1,2,...,n\right)$ scores $\overset{\Rightarrow}{\Psi}(\hbar_i)\left(i=1,2,...,n\right)$ to established the ranking of all alternatives $Þ_j \left(j=1,2,...,m\right)$ and identify the best option(s). If two scores $\overset{\Rightarrow}{\Psi}(\hbar_i)$ and $\overset{\Rightarrow}{\Psi}(\hbar_j)$ are identical, calculate the accuracy degrees $\overset{\Rightarrow}{\xi}(\hbar_i)$ and $\overset{\Rightarrow}{\xi}(\hbar_j)$ of the overall $GO_{mno}$-TSFVs $\tilde{\underline{t}}_i$ and $\tilde{\underline{t}}_j$ for both alternatives. Rank the alternatives $Þ_i$ and $Þ_j$ according to these accuracy degrees $\overline{\overline{\xi}}(\hbar_i)$ and $\overline{\overline{\xi}}(\hbar_j)$.

**Step 3.** Determine the ranking of all the alternatives $Þ_j \left( j = 1, 2, ..., m \right)$ based on their scores $\overline{\overline{\Psi}}(\hbar_i)\left( i = 1, 2, ..., n \right)$ and select the best option(s).

**Step 4.** End.

**Example 2:** E-commerce online platforms (ECOP) have undoubtedly transformed the landscape of commerce, redefining how we shop, interact, and conduct business. They have become indispensable tools in the service of mankind, offering unparalleled convenience, accessibility, and choice to consumers worldwide. As digital marketplaces evolve and diversify, the significance of distinguishing and spotlighting the most effective platforms becomes ever more pronounced. In this dynamic environment, the criteria for evaluating e-commerce platforms assume a pivotal role, acting as guiding principles for both consumers and businesses seeking optimal solutions. These criteria serve as essential benchmarks, enabling stakeholders to assess the efficiency, dependability, and user-friendliness of online platforms. To evaluate these top five e-commerce online platforms effectively, three focus groups of field experts have been convened. These experts possess diverse qualifications and expertise, covering key areas such as Data Analytics and Data Science, Human-Computer Interaction, Cyber security and Information Security, Marketing and Consumer Behavior, E-commerce Technology and Development, Business Analytics and Strategy, User Experience Design, and Supply Chain Management. These experts were tasked with identifying and prioritizing key quality criteria essential for assessing the effectiveness and reliability of online platforms. Among the various important attributes considered, the experts initially highlighted the following four crucial quality criteria:

1. **User Experience:**

A seamless and intuitive user experience is essential for attracting and retaining customers in today's competitive online marketplace. A well-designed and user-friendly platform enhances customer satisfaction, encourages repeat visits, and ultimately drives conversions. By prioritizing user experience, businesses can differentiate themselves from competitors and foster long-term customer loyalty.

2. **Security and Trustworthiness:**

Security is paramount in e-commerce to protect sensitive customer data and maintain trust. A secure online platform instills confidence in customers, leading to increased sales and reduced cart abandonment rates. By investing in robust security measures and demonstrating a commitment to data protection, businesses can mitigate risks and safeguard their reputation.

3. **Scalability and Customization:**

As businesses grow and evolve, their e-commerce platforms must be able to scale and adapt accordingly. A scalable platform ensures that businesses can accommodate increasing website traffic, product catalog sizes, and transaction volumes without sacrificing performance. Additionally, customization options allow businesses to tailor the platform to their specific needs, branding, and customer preferences, enhancing flexibility and competitiveness in the market.

4. **Analytics and Reporting:**

Comprehensive analytics and reporting capabilities are essential for monitoring performance, identifying trends, and making data-driven decisions. By analyzing key metrics such as website traffic, conversion rates, and customer behavior, businesses can gain valuable insights into their audience and optimize their e-commerce strategies accordingly. Real-time reporting and forecasting tools enable businesses to stay agile and responsive to changing market dynamics, driving continuous improvement and growth.

To facilitate a clear understanding of the selection problem at hand, we denote the most rated top five e-commerce online platforms as $Þ_j (j = 1, 2, ..., 5)$ with a set notation. Similarly, we represent the set of criteria under consideration as $L_i (i = 1, 2, 3, 4)$. Additionally, we identify the focus groups involved in the evaluation process as $g_l (l = 1, 2, 3)$. This notation framework enables us to systematically analyze and compare the performance of the e-commerce platforms against the specified criteria, taking into account the perspectives and insights provided by each focus group. Through this structured approach, we aim to gain a comprehensive understanding of the selection problem and facilitate informed DM regarding the most suitable e-commerce platform for our intended purposes.

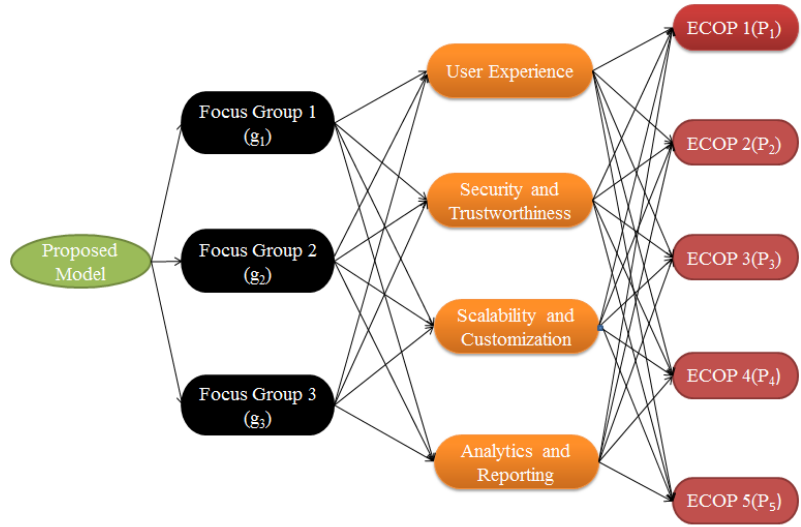

**Figure 4.** The proposed GO$_{mno}$-TSFHMCGDM model.

After extensive deliberations, experts within focus groups $g_l(l=1,2,3)$ conducted evaluation of the top five e-commerce online platforms, denoted by Þ$_j(j=1,2,...,5)$, based on the identified set of criteria $L_i(i=1,2,3,4)$, having weights represented in a weight vector as $w=\{0.2,0.1,0.3,0.4\}^T$. This process ensured alignment with industry standards and consumer expectations, guaranteeing a thorough and rigorous assessment of each platform's performance and suitability. Using the (m,n,o) T spherical fuzzy values, they assessed each platform's performance and suitability for consumers and businesses. The evaluation values for these platforms were presented in the form of (m,n,o) TSF-DM, shown in Table 1.

Table 1. (m,n,o)TSF-DM

| Options | Evaluators | $L_1$ | $L_2$ | $L_3$ | $L_4$ |
|---------|-----------|-------|-------|-------|-------|
| Þ$_1$ | $g_1$ | <0.25,0.45,0.60> | <0.36,0.48,0.57> | <0.25,0.39,0.84> | <0.92,0.21,0.15> |
| | $g_2$ | <0.34,0.62,0.41> | <0.51,0.63,0.32> | <0.75,0.28,0.33> | <0.44,0.50,0.64> |
| | $g_3$ | <0.74,0.34,0.28> | <0.57,0.26,0.54> | <0.33,0.48,0.65> | <0.45,0.54,0.38> |
| Þ$_2$ | $g_1$ | <0.29,0.65,0.54> | <0.80,0.40,0.36> | <0.80,0.25,0.30> | <0.94,0.20,0.10> |
| | $g_2$ | <0.15,0.82,0.37> | <0.25,0.32,0.76> | <0.40,0.60,0.50> | <0.32,0.80,0.26> |
| | $g_3$ | <0.47,0.56,0.55> | <0.60,0.18,0.40> | <0.15,0.45,0.57> | <0.72,0.18,0.43> |
| $Þ_3$ | $g_1$ | <0.55,0.07,0.30> | <0.75,0.08,0.24> | <0.71,0.24,0.22> | <0.77,0.50,0.70> |
| | $g_2$ | <0.60,0.22,0.66> | <0.65,35,0.67> | <0.51,0.25,0.68> | <0.70,0.79,0.49> |
| | $g_3$ | <0.65,0.18,0.88> | <0.55,0.22,0.77> | <0.57,0.56,0.61> | <0.65,0.11,0.97> |
| $Þ_4$ | $g_1$ | <0.88,0.56,0.21> | <0.11,0.29,0.93> | <0.56,0.23,0.62> | <0.43,0.55,0.63> |
| | $g_2$ | <0.63,0.48,0.32> | <0.60,0.70,0.16> | <0.41,0.33,0.66> | <0.50,0.71,0.29> |
| | $g_3$ | <023,0.38,0.74> | <0.41,0.56,0.43> | <0.90,0.13,0.34> | <0.60,0.30,0.40> |
| $Þ_5$ | $g_1$ | <0.75,0.35,0.22> | <0.15,0.45,0.66> | <0.77,0.24,0.38> | <0.56,0.26,0.44> |
| | $g_2$ | <0.34,0.54,0.70> | <0.60,0.25,0.50> | <0.33,0.54,0.21> | <0.20,0.60,0.50> |
| | $g_3$ | <0.14,0.34,0.65> | <0.44,0.58,0.16> | <0.22,0.45,0.61> | <0.35,0.70,0.55> |

Utilizing the definitions of our proposed $GO_{mno}$-TSFVs in Theorem 1, that was also outlined in Eq. (3), we transform the values of TSFVs presented in Table 1 into their corresponding average values of DoM, DoI and DoN for each alternative. These average $GO_{mno}$-TSFVs' membership grads are elucidated in Table 2.

Table 2. Average TSF-DM of the evaluation values

| Options | Attributes | | | |
|---|---|---|---|---|
| | $L_1$ | $L_2$ | $L_3$ | $L_4$ |
| $Þ_1$ | <0.49,0.48,0.44> | <0.48,0.48,0.48> | <0.49,0.39,0.64> | <0.64,0.44,0.43> |
| $Þ_2$ | <0.33,0.68,0.49> | <0.59,0.31,0.53> | <0.52,0.45,0.47> | <0.70,0.48,0.29> |
| $Þ_3$ | <0.60,0.16,0.65> | <0.65,0.36,0.60> | <0.60,0.38,0.54> | <0.70,0.54,0.74> |
| $Þ_4$ | <0.63,0.47,0.48> | <0.42,0.54,0.59> | <0.65,0.24,0.55> | <0.51,0.54,0.46> |
| $Þ_5$ | <0.48,0.42,0.56> | <0.43,0.44,0.48> | <0.50,0.42,0.43> | <0.39,0.55,0.49> |

We calculate the radius of the sphere for every $GO_{mno}$-TSFV in each of the alternatives by utilizing their average values of DoM, DoI and DoN shown in Table 2 through our proposed method for computing the radius of the sphere $r$ as provided in Eq.(4). The maximum radii for each alternative, expressed in terms of $GO_{mno}$-TSFVs, are then presented in Table 3.

Table 3. Maximum radii of the spheres

| Radii | Attribute | | | |
|---|---|---|---|---|
| | $L_1$ | $L_2$ | $L_3$ | $L_4$ |
| $Þ_1$ | 0.525901 | 0.486752 | 0.622307 | 0.610162 |
| Þ$_2$ | 0.471404 | 0.57431 | 0.580409 | 0.665069 |
| Þ$_3$ | 0.66273 | 0.659122 | 0.598247 | 0.634019 |
| Þ$_4$ | 0.678004 | 0.672512 | 0.521133 | 0.569087 |
| Þ$_5$ | 0.602431 | 0.533878 | 0.567917 | 0.579599 |

Next, we integrate the average evaluation values of the alternatives outlined in Table 2 and the values of maximum radii presented in Table 3 to construct the GO$_{mno}$-TSF-DM. The resultant GO$_{mno}$-TSF-DM is provided in Table 4.

Table 4. GO$_{mno}$-TSF-DM

| Options | Attributes | | | |
|---|---|---|---|---|
| | $L_1$ | $L_2$ | $L_3$ | $L_4$ |
| Þ$_1$ | <0.49,0.48,0.44;0.52> | <0.48,0.48,0.48;0.48> | <0.49,0.39,0.64;0.62> | <0.64,0.44,0.43;0.61> |
| Þ$_2$ | <0.33,0.68,0.49;0.47> | <0.59,0.31,0.53;0.57> | <0.52,0.45,0.47;0.58> | <0.70,0.48,0.29;0.66> |
| Þ$_3$ | <0.60,0.16,0.65;0.66> | <0.65,0.36,0.60;0.65> | <0.60,0.38,0.54;0.59> | <0.70,0.54,0.74;0.63> |
| Þ$_4$ | <0.63,0.47,0.48;0.67> | <0.42,0.54,0.59;0.67> | <0.65,0.24,0.55;0.52> | <0.51,0.54,0.46;0.56> |
| Þ$_5$ | <0.48,0.42,0.56;0.60> | <0.43,0.44,0.48;0.53> | <0.50,0.42,0.43;0.56> | <0.39,0.55,0.49;0.57> |

To identify the optimal choice, we employ the proposed GO$_{mno}$-TSFHWA and GO$_{mno}$-TSFHWG operators to devise a methodology for tackling multiple attribute decision-making problems involving GO$_{mno}$-TSF data, which can be characterized as follows:

**Step 1.** We implement our proposed GO$_{mno}$-TSFHWA and GO$_{mno}$-TSFHWG operators to aggregate the GO$_{mno}$-TSF values presented in Table 4, thereby driving the overall GO$_{mno}$-TSFVs for each alternative. This process involves applying our innovative aggregation methods to capture a comprehensive evaluation of each option. The results of this aggregation, with $\varsigma$, are detailed in Table 5. These results provide valuable insights into the relative performance of the alternatives, highlighting the effectiveness of our proposed operators in handling and synthesizing complex data.

**Step 2.** We then apply our score function, as defined in Equation (5), to the aggregated values presented in Table 5, which were obtained from the previous step using our proposed aggregation operators. The resulting score values are depicted in Table 6. These scores play a vital role in facilitating the comparison and ranking of the top-rated ECOPs. By systematically assessing each option's overall performance, these scores provide a clear and objective basis for DM.

**Step 3.** After a comprehensive evaluation and comparative analysis of the numerical values of the score function for various alternatives, as shown in Table 6, we ranked the top-rated ECOPs. The ordering and ranking of these ECOPs are presented in Table 7. In this context, the notation "$\succ$ "signifies that "the alternative on the left side of the operator is superior to the one on the right side." Our results demonstrate that the rankings obtained from both the $GO_{mno}$-TSFHWA and $GO_{mno}$-TSFHWG operators are consistent, identifying $Þ_2$ as the optimal choice among the top-rated ECOPs. This consistency underscores the robustness of our proposed methodology in selecting the most suitable ECOP.

Table 5. The aggregated results of the top rated ECOPs using $GO_{mno}$-TSFHWA and $GO_{mno}$-TSFHWG operators.

| $Þ_i$ | $GO_{mno}$-TSFHWA | $GO_{mno}$-TSFHWG |
|---|---|---|
| $Þ_1$ | <0.56,0.45,0.47;0.56> | <0.54,0.45,0.48;0.55> |
| $Þ_2$ | <0.61,0.45,0.41;0.60> | <0.56,0.51,0.45;0.58> |
| $Þ_3$ | <0.66,0.36,0.66;0.64> | <0.65,0.44,0.67;0.64> |
| $Þ_4$ | <0.54,0.49,0.51;0.62> | <0.52,0.51,0.52;0.61> |
| $Þ_5$ | <0.44,0.47,0.50;0.57> | <0.43,0.49,0.50;0.56> |

Table 6. The score functions of the top rated ECOPs.

| $Þ_i$ | $GO_{mno}$-TSFHWA | $GO_{mno}$-TSFHWG |
|---|---|---|
| $Þ_1$ | 0.599269459 | 0.579003508 |
| $Þ_2$ | 0.674296864 | 0.593614719 |
| $Þ_3$ | 0.638506119 | 0.59207529 |
| $Þ_4$ | 0.581091413 | 0.547233281 |
| $Þ_5$ | 0.515591876 | 0.504583715 |

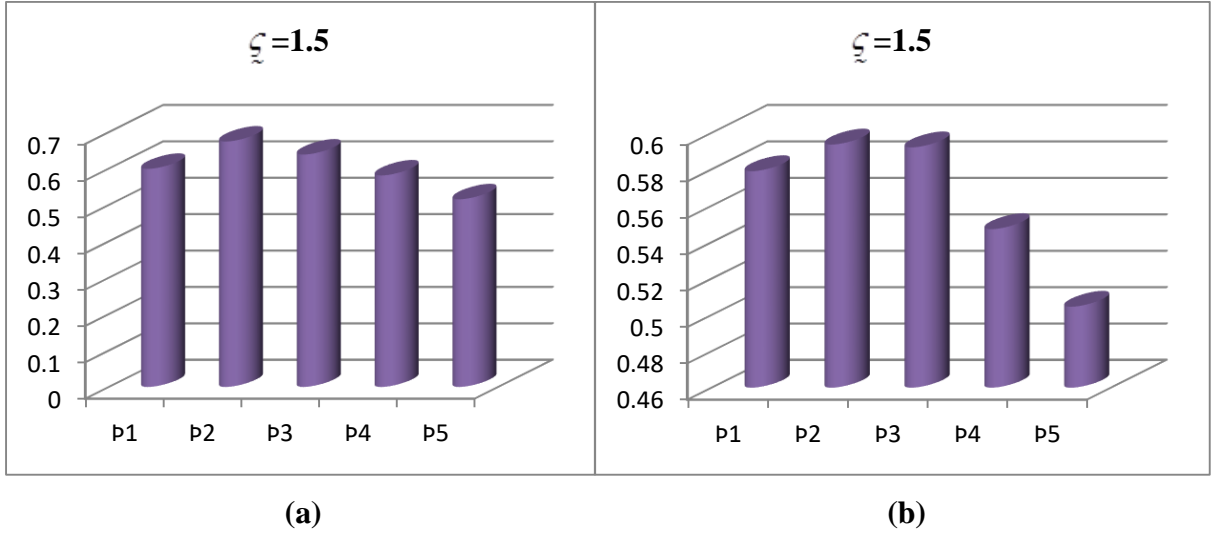

**(a)**                                   **(b)**

**Figure 5.** Score values of (a) GO$_{mno}$-TSFHWA and (b) GO$_{mno}$-TSFHWG for $\varsigma$ = 1.5

Table 7. Ranking of the top rated ECOPs.

| | Ranking |
|---|---|
| GO$_{mno}$-TSFHWA | $Þ_2 \succ Þ_3 \succ Þ_1 \succ Þ_4 \succ Þ_5$ |
| GO$_{mno}$-TSFHWG | $Þ_2 \succ Þ_3 \succ Þ_1 \succ Þ_4 \succ Þ_5$ |

### 5.1. Sensitivity analysis

To assess how the parameter $\varsigma$ affects DM outcomes in this scenario, we solve the above same numerical example using different values of the parameter $\varsigma$. Specifically, we consider $\varsigma$ values of 2, 2.5, 3, 3.5, and up to 6. The results, including the values of the GO$_{mno}$-TSFHWA operator and the GO$_{mno}$-TSFHWG operator for all alternatives, along with their respective ranking orders, are presented in Tables 8 and 9 separately. Upon reviewing the outcomes in Tables 8 and 9, it is obvious that even varying the values of the parameter $\varsigma$ produces consistent results, yielding the same ranking order for the alternatives. The ranking order (RO) of the top-rated five alternatives for $\varsigma$ values of 2, 2.5, 3, 3.5, and up to 6 is consistently $Þ_2 \succ Þ_3 \succ Þ_1 \succ Þ_4 \succ Þ_5$. This indicates that $Þ_2$ is the best alternative according to the proposed method, irrespective of the $\varsigma$ values used. Therefore, it is evident that the parameter $\varsigma$ does not affect the DM process and outcomes. This stability is attributed to the use of the GO$_{mno}$-TSFHWA and GO$_{mno}$-TSFHWG operators, which ensure that the aggregated values remain independent of the parameter $\varsigma$. Consequently, the results derived from the proposed methods are robust and independent of the parameter $\varsigma$.

Table 8. Ranking of the top rated ECOPs using GO$_{mno}$-TSFHWA for various values of $\varsigma$.

| $\varsigma$ | Þ$_1$ | Þ$_2$ | Þ$_3$ | Þ$_4$ | Þ$_5$ | Ranking |
|---|---|---|---|---|---|---|
| 1.5 | 0.599269 | 0.674297 | 0.638506 | 0.581091 | 0.515592 | Þ$_2$ ≻ Þ$_3$ ≻ Þ$_1$ ≻ Þ$_4$ ≻ Þ$_5$ |
| 2 | 0.598304 | 0.671857 | 0.637492 | 0.57967 | 0.515281 | Þ$_2$ ≻ Þ$_3$ ≻ Þ$_1$ ≻ Þ$_4$ ≻ Þ$_5$ |
| 2.5 | 0.597508 | 0.669943 | 0.636803 | 0.578556 | 0.515045 | Þ$_2$ ≻ Þ$_3$ ≻ Þ$_1$ ≻ Þ$_4$ ≻ Þ$_5$ |
| 3 | 0.596833 | 0.668366 | 0.6363 | 0.577641 | 0.514855 | Þ$_2$ ≻ Þ$_3$ ≻ Þ$_1$ ≻ Þ$_4$ ≻ Þ$_5$ |
| 3.5 | 0.596251 | 0.667024 | 0.635913 | 0.576868 | 0.514694 | Þ$_2$ ≻ Þ$_3$ ≻ Þ$_1$ ≻ Þ$_4$ ≻ Þ$_5$ |
| 4 | 0.595742 | 0.665859 | 0.635605 | 0.576202 | 0.514554 | Þ$_2$ ≻ Þ$_3$ ≻ Þ$_1$ ≻ Þ$_4$ ≻ Þ$_5$ |
| 4.5 | 0.595291 | 0.664831 | 0.635353 | 0.575619 | 0.514431 | Þ$_2$ ≻ Þ$_3$ ≻ Þ$_1$ ≻ Þ$_4$ ≻ Þ$_5$ |
| 5 | 0.59489 | 0.663914 | 0.635143 | 0.575104 | 0.514321 | Þ$_2$ ≻ Þ$_3$ ≻ Þ$_1$ ≻ Þ$_4$ ≻ Þ$_5$ |
| 5.5 | 0.59453 | 0.663087 | 0.634965 | 0.574644 | 0.514221 | Þ$_2$ ≻ Þ$_3$ ≻ Þ$_1$ ≻ Þ$_4$ ≻ Þ$_5$ |
| 6 | 0.594205 | 0.662337 | 0.634812 | 0.57423 | 0.514131 | Þ$_2$ ≻ Þ$_3$ ≻ Þ$_1$ ≻ Þ$_4$ ≻ Þ$_5$ |

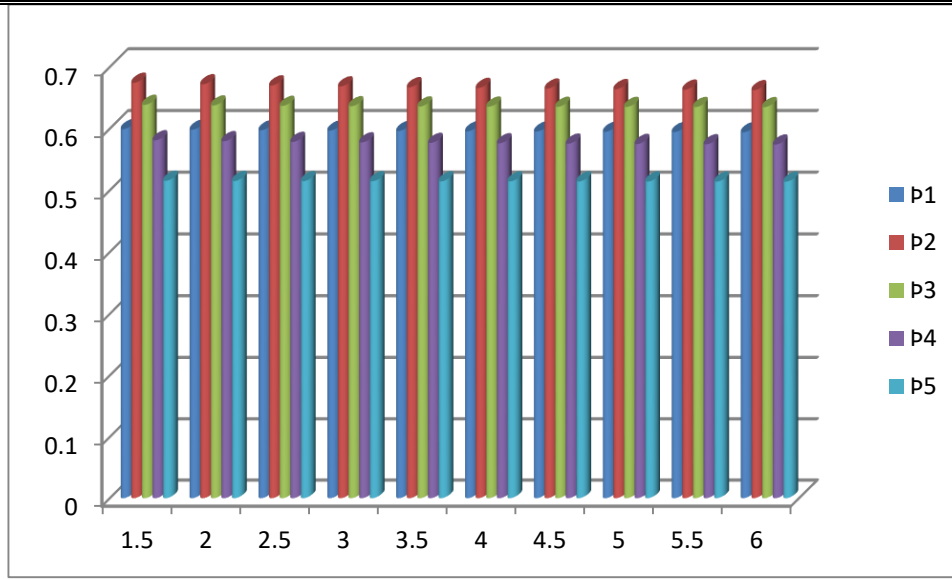

**Figure 6.** GO$_{mno}$-TSFHWA scores of ECOPs for various values of parameter $\varsigma$.

Table 9. Ranking of the top rated ECOPs using GO$_{mno}$-TSFHWG for various values of $\varsigma$.

| $\varsigma$ | $Þ_1$ | $Þ_2$ | $Þ_3$ | $Þ_4$ | $Þ_5$ | Ranking |
|---|---|---|---|---|---|---|
| 1.5 | 0.579004 | 0.593615 | 0.592075 | 0.547233 | 0.504584 | $Þ_2 \succ Þ_3 \succ Þ_1 \succ Þ_4 \succ Þ_5$ |
| 2 | 0.579861 | 0.59678 | 0.593571 | 0.548332 | 0.505032 | $Þ_2 \succ Þ_3 \succ Þ_1 \succ Þ_4 \succ Þ_5$ |
| 2.5 | 0.580509 | 0.599277 | 0.594841 | 0.54919 | 0.505414 | $Þ_2 \succ Þ_3 \succ Þ_1 \succ Þ_4 \succ Þ_5$ |
| 3 | 0.58103 | 0.601355 | 0.595949 | 0.5499 | 0.505748 | $Þ_2 \succ Þ_3 \succ Þ_1 \succ Þ_4 \succ Þ_5$ |
| 3.5 | 0.581463 | 0.603142 | 0.596935 | 0.55051 | 0.506045 | $Þ_2 \succ Þ_3 \succ Þ_1 \succ Þ_4 \succ Þ_5$ |
| 4 | 0.581832 | 0.604713 | 0.597823 | 0.551046 | 0.506312 | $Þ_2 \succ Þ_3 \succ Þ_1 \succ Þ_4 \succ Þ_5$ |
| 4.5 | 0.582152 | 0.606115 | 0.598632 | 0.551524 | 0.506554 | $Þ_2 \succ Þ_3 \succ Þ_1 \succ Þ_4 \succ Þ_5$ |
| 5 | 0.582434 | 0.607381 | 0.599375 | 0.551956 | 0.506774 | $Þ_2 \succ Þ_3 \succ Þ_1 \succ Þ_4 \succ Þ_5$ |
| 5.5 | 0.582685 | 0.608535 | 0.600062 | 0.552351 | 0.506976 | $Þ_2 \succ Þ_3 \succ Þ_1 \succ Þ_4 \succ Þ_5$ |
| 6 | 0.58291 | 0.609595 | 0.600701 | 0.552714 | 0.507162 | $Þ_2 \succ Þ_3 \succ Þ_1 \succ Þ_4 \succ Þ_5$ |

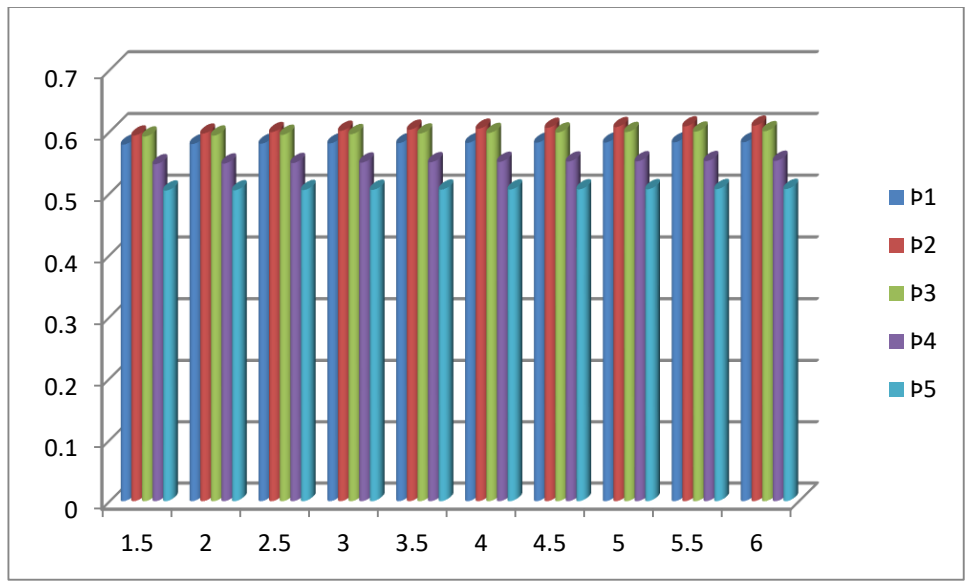

**Figure 7.** $GO_{mno}$-TSFHWG scores of ECOPs for various values of parameter $\varsigma$ .

### 5.2. Comparative analysis

The ranking results of our proposed methods are remarkably consistent, further verifying its effectiveness. As previously mentioned that the existing information aggregation operators based on Hamacher operations for FSs, IFS, C-IFSs, PyFSs, C-PyFSs, q-ROFSs, C-qROFSs, (p,q) QOFSs, PFSs, SFSs, C-SFSs, TSFSs, (p,q,r) SFSs, and G-TSFSs have limitations and cannot fully represent the situation. In the following, we furnish a tabular representation of the comparative analysis regarding the features of the proposed method and the existing frameworks.

Table 10. Comparative analysis

| S.# | Framework | Features | | | | | | |
|---|---|---|---|---|---|---|---|---|
| | | $\ddot{\mathfrak{A}}$ | $\ddot{\imath}$ | $\ddot{\mathfrak{M}}$ | $p$ | $q$ | $\ddot{r}$ | radius |
| 1 | FSs | Yes | No | No | No | No | No | No |
| 2 | IFSs | Yes | No | Yes | No | No | No | No |
| 3 | C-IFS | Yes | No | Yes | No | No | No | Yes |
| 4 | PyFSs | Yes | No | Yes | No | No | No | No |
| 5 | C-PyFSs | Yes | No | Yes | No | No | No | Yes |
| 6 | q-ROFs | Yes | No | Yes | No | Yes | No | No |
| 7 | C- qROFs | Yes | No | Yes | No | Yes | No | Yes |
| 8 | PFSs | Yes | Yes | Yes | No | No | No | No |
| 9 | SFSs | Yes | Yes | Yes | No | No | No | No |
| 10 | C-SFSs | Yes | Yes | Yes | No | No | No | Yes |
| 11 | TSFSs | Yes | Yes | Yes | No | Yes | No | No |
| 12 | G-TSFSs | Yes | Yes | Yes | No | Yes | No | Yes |
| 13 | (p,q) QOFSs | Yes | No | Yes | Yes | Yes | No | No |
| 14 | (p,q,r) SFSs | Yes | Yes | Yes | Yes | Yes | Yes | No |
| 15 | GO$_{mno}$-TSFS | Yes | Yes | Yes | Yes | Yes | Yes | Yes |

## 6. Conclusion

Effective decision-making is the cornerstone of success across various domains, from business ventures to academic pursuits and everyday challenges. In today's era of unprecedented complexity and uncertainty, the demand for robust decision-making tools has never been more evident. This research endeavors to address this critical need by introducing an innovative methodology aimed at refining the representation of uncertainty within decision-making processes. Through the introduction of Generalized Orbicular (m,n,o) T-Spherical Fuzzy Sets (GO$_{mno}$-TSFSs), we offer a versatile extension of existing fuzzy set models, providing decision-makers with a nuanced toolset for tackling ambiguity and imprecision in data analysis. By incorporating adjustable parameters m, n, and o into our framework, we empower decision-makers to finely tune the weighting of membership degrees, thereby facilitating more nuanced evaluations and yielding superior decision outcomes. The utilization of spheres to delineate membership, indeterminacy, and non-membership levels enhances the accuracy of data portrayal, enabling a more comprehensive understanding of complex datasets. Building upon the foundational concepts of $GO_{mno}$-TSFSs, we introduce essential set operations and algebraic operations for $GO_{mno}$-TSF Values ($GO_{mno}$-TSFVs). Additionally, the development of score functions, accuracy functions, and basic distance measures like Hamming and Euclidean distances further enriches the analytical capabilities of our framework. Furthermore, our research extends beyond mere theoretical propositions by proposing practical aggregation operators such as the $GO_{mno}$-TSFHWA and $GO_{mno}$-TSF HWG aggregation operators, which are tailored to suit the needs of our proposed sets. Through a case study involving MCGDM for selecting the most suitable e-commerce online shopping platform (ECOP), we have demonstrated the practical efficacy of our approach. Sensitivity analysis further validates the reliability and robustness of our results, reinforcing the confidence in the proposed methodology's suitability for addressing real-world decision-making challenges. This research opens new avenues for addressing uncertainty in decision-making processes, offering decision-makers a powerful toolkit for navigating complex and ambiguous decision landscapes. As the demand for sophisticated decision support systems continues to grow, the insights and methodologies presented in this study are poised to make a significant impact across various domains, driving innovation and fostering more informed decision-making practices.

**References**

[1]     Abid, M.N., Yang, M.S., Karamti, H., Ullah, K., & Pamucar, D. (2022). Similarity measures based on T-spherical fuzzy information with applications to pattern recognition and decision making. Symmetry, 14(2), 410.

[2]     Ashraf, S., Abdullah, S., Mahmood, T., Ghani, F., & Mahmood, T. (2019). Spherical fuzzy sets and their applications in multi-attribute decision-making problems. Journal of Intelligent & Fuzzy Systems, 36(3), pp. 2829-2844.

[3]     Ashraf, S., Abdullah, S., & Mahmood, T. (2020). Spherical fuzzy Dombi aggregation operators and their application in group decision-making problems. Journal of Ambient Intelligence and Humanized Computing, 11(7), pp. 2731-2749.

[4]     Atanassov, K.T. (1986). Intuitionistic fuzzy sets. Fuzzy Sets and Systems, 20(1), pp. 87-96.

[5]     Atanassov, K.T. (2020). Circular intuitionistic fuzzy sets. Journal of Intelligent & Fuzzy Systems, 39(5), pp. 5981-5986.

[6]     Bani-Doumi, M., Serrano-Guerrero, J., Chiclana, F., Romero, F.P., & Olivas, J.A. (2024). A picture fuzzy set multi-criteria decision-making approach to customize hospital recommendations based on patient feedback. Applied Soft Computing, 153, 111331.

[7]     Chaomurilige, C., Yu, J., & Yang, M.S. (2017). Deterministic annealing Gustafson-Kessel fuzzy clustering algorithm. Information Sciences, 417, pp. 435–453.

[8]     Cuong, B.C. (2014). Picture fuzzy sets. Journal of Computer Science and Cybernetics, 30(4), pp. 409-420.

[9]     Deschrijver, G., & Kerre, E.E. (2002). A generalization of operators on intuitionistic fuzzy sets using triangular norms and conorms. Notes on Intuitionistic Fuzzy Sets, 8(1), pp. 19-27.

[10]    Deschrijver, G., Cornelis, C., & Kerre, E.E. (2004). On the representation of intuitionistic fuzzy t-norms and t-conorms. IEEE Transactions on Fuzzy Systems, 12(1), pp. 45-61.

[11]    Dinh, N.V., Thao, N.X., & Xuan, N. (2018). Some measures of picture fuzzy sets and their application in multi-attribute decision making. Int. J. Math. Sci. Comput. (IJMSC), 4(3), pp. 23-41.

[12]    Garg, H. (2017). Some picture fuzzy aggregation operators and their applications to multicriteria decision-making. Arabian Journal for Science and Engineering, 42(12), pp. 5275-5290.

[13]    Garg, H., Munir, M., Ullah, K., Mahmood, T., & Jan, N. (2018). Algorithm for T-spherical fuzzy multi-attribute decision making based on improved interactive aggregation operators. Symmetry, 10(12), 670.

[14]    Hamacher, H. (1978). Über logische Verknüpfungen unscharfer Aussagen und deren zugehörige Bewertungsfunktionen. In Trappl, Klir, Riccardi (Eds.), Progress in Cybernetics and Systems Research, Vol. 3, Hemisphere, Washington DC, pp. 276–288.

[15]    Hwang, C.M., & Yang, M.S. (2013). New construction for similarity measures between intuitionistic fuzzy sets based on lower, upper and middle fuzzy sets. International Journal of Fuzzy Systems, 15(3), pp. 359−366.

[16]    Jafar, M.N., Saeed, M., Saqlain, M., & Yang, M.S. (2021) Trigonometric similarity measures for nutrosophic hypersoft sets with application to renewable energy source selection. IEEE Access, 9, pp. 129178-129187.

[17]    Jamkhaneh, E.B., & Nadarajah, S. (2015). A new generalized intuitionistic fuzzy set. Hacettepe Journal of Mathematics and Statistics, 44(6), pp. 1537-1551.

[18]    Li, D.F. (2005). Multiattribute decision making models and methods using intuitionistic fuzzy sets. Journal of Computer and System Sciences, 70(1), pp.73-85.

[19]    Mahnaz, S., Ali, J., Malik, M.A., & Bashir, Z. (2021). T-spherical fuzzy Frank aggregation operators and their application to decision making with unknown weight information. IEEE Access, 10, pp. 7408-7438.

[20]    Mahmood, T., Ullah, K., Khan, Q., & Jan, N. (2019). An approach toward decision-making and medical diagnosis problems using the concept of spherical fuzzy sets. Neural Computing and Applications, 31, pp. 7041-7053.

[21]    Rahim, M., Amin, F., Tag Eldin, E.M., Khalifa, A.E.W., & Ahmad, S. (2024). p, q-Spherical fuzzy sets and their aggregation operators with application to third-party logistic provider selection. Journal of Intelligent & Fuzzy Systems, 46(1), pp. 505-528.

[22]    Rahman, A.U., Saeed, M., & Mohammed, M.A. (2023) Fppsv-NHSS: Fuzzy parameterized possibility single valued neutrosophic hypersoft set to site selection for solid waste management, Applied Soft Computing, 140, 110273.

[23]    Roychowdhury, S., & Wang, B.H. (1998). On generalized Hamacher families of triangular operators. International Journal of Approximate Reasoning, 19(3-4), pp. 419-439.

[24]    Saqlain, M., Riaz, M., Saleem, M.A., & Yang, M.S. (2021) Distance and similarity measures for neutrosophic hypersoft set (NHSS) with construction of NHSS-TOPSIS and applications. IEEE Access, 9, pp. 30803-30816.

[25]    Singh, H., Gupta, M.M., Meitzler, T., Hou, Z.G., Garg, K.K., Solo, A.M., & Zadeh, L.A. (2013). Real-life applications of fuzzy logic. Advances in Fuzzy Systems, 2013, pp. 3-3.

[26]    Smarandache, F. (2005) Neutrosophic set, a generalization of the intuitionistic fuzzy sets. Inter. J. Pure Appl. Math., 24, pp. 287-297.

[27]    Smarandache, F. (2018) Extension of Soft Set to Hypersoft Set, and then to Plithogenic Hypersoft Set," Neutrosophic Sets and Systems, 22, pp. 168-170.

[28]    Ullah, K., Mahmood, T., & Jan, N. (2018). Similarity measures for T-spherical fuzzy sets with applications in pattern recognition. Symmetry, 10(6), 193.

[29]    Xu, Z., & Yager, R.R. (2006). Some geometric aggregation operators based on intuitionistic fuzzy sets. International Journal of General Systems, 35(4), pp. 417-433.

[30]     Yang, M.S., Akhtar, Y., & Ali, M. (2024). On Globular T-Spherical Fuzzy (G-TSF) Sets with Application to G-TSF Multi-Criteria Group Decision-Making. arXiv preprint arXiv:2403.07010.

[31]     Yue, C. (2024). A software trustworthiness evaluation methodology for cloud services with picture fuzzy information. Applied Soft Computing, 152, 111205.

[32]     Zadeh, L.A. (1965). Fuzzy sets. Information and Control, 8(3), pp. 338–356.

[33]     Zhao, H., Xu, Z., Ni, M., & Liu, S. (2010). Generalized aggregation operators for intuitionistic fuzzy sets. International Journal of Intelligent Systems, 25(1), pp. 1-30.